\documentclass[11pt,preprint]{emulateapj}

\newcommand{\mdot}{\mbox{$\dot M$}}
\newcommand{\lsun}{\mbox{L$_\odot$}}
\newcommand{\msun}{\mbox{M$_\odot$}}
\newcommand{\rsun}{\mbox{R$_\odot$}}
\newcommand\msunyr{\rm M_{\odot}\,yr^{-1}}
\newcommand\be{\begin{equation}}
\newcommand\en{\end{equation}}

\newcommand{\lsmm}{\mbox{$L_{\rm smm}$}} 
\newcommand{\lbol}{\mbox{$L_{\rm bol}$}} 
\newcommand{\tbol}{\mbox{$T_{\rm bol}$}} 
\newcommand{\trot}{\mbox{$T_{\rm rot}$}} 

\newcommand{\OI}{\mbox{[\ion{O}{1}]}}
\newcommand{\CI}{\mbox{[\ion{C}{1}]}}
\newcommand{\CII}{\mbox{[\ion{C}{2}]}}
\newcommand{\NII}{\mbox{[\ion{N}{2}]}}
\newcommand{\OIII}{\mbox{[\ion{O}{3}]}}
\newcommand{\NIII}{\mbox{[\ion{N}{3}]}}
\newcommand{\SIII}{\mbox{[\ion{S}{3}]}}
\newcommand{\Sitwo}{\mbox{[\ion{Si}{2}]}}
\newcommand{\jj}[2]{\mbox{$J = #1\rightarrow#2$}}
\newcommand{\funnyN}{\mbox{$\mathcal{N}$}}

\shorttitle{FUor Submm Environs}
\shortauthors{Green et al.}

\begin{document}

\title{An Analysis of the Environments of FU Orionis Objects with {\it Herschel}$^1$
}

\author{Joel D. Green\altaffilmark{1},
 Neal J. Evans II\altaffilmark{1}, 
 \'{A}gnes K\'{o}sp\'{a}l\altaffilmark{2},
    Gregory Herczeg\altaffilmark{3},
     Sascha  P. Quanz\altaffilmark{4},
       Thomas Henning\altaffilmark{5},
        Tim A. van Kempen\altaffilmark{6,7},
        Jeong-Eun Lee\altaffilmark{8}, 
        Michael M. Dunham\altaffilmark{9},
         Gwendolyn Meeus\altaffilmark{10}, 
         Jeroen Bouwman\altaffilmark{5},
          Jo-hsin Chen\altaffilmark{11}, 
         Manuel G\"udel\altaffilmark{12},
          Stephen L. Skinner\altaffilmark{13},
          Armin Liebhart\altaffilmark{12},
          \& Manuel Merello\altaffilmark{1}
}

\affil{
1.  The University of Texas at Austin, Department of
Astronomy, 2515 Speedway, Stop C1400,
Austin, TX 78712-1205, USA; joel@astro.as.utexas.edu \\
2. European Space Agency (ESA/ESTEC), Keplerlaan 1, 2200 AG, Noordwijk
The Netherlands \\
3. Kavli Institute for Astronomy and Astrophysics, Peking University, Beijing,
100871, PR China
4. Institute for Astronomy, ETH, Zurich, Switzerland \\
5. Max Planck Institute for Astronomy, Koenigstuhl 17, 69117 Heidelberg, Germany \\
6.  Leiden Observatory, Leiden University, PO Box 9513, 2300 RA Leiden, The Netherlands \\
7. Joint ALMA offices, Av. Alonso de Cordova 3107, Santiago, Chile \\
8. Department of Astronomy \& Space Science, Kyung Hee University, Gyeonggi 446-701, Korea  \\
9. Dept. of Astronomy, Yale University, New Haven, CT \\
10. Universidad Autonoma de Madrid, Dpt. Fisica Teorica, Campus Cantoblanco, Spain \\
11. Jet Propulsion Laboratory, Pasadena, CA \\
12. Dept. of Astronomy, University of Vienna, Austria \\
13. Center for Astrophysics and Space Astronomy (CASA), University of Colorado, Boulder, CO 80309-0389 \\
}

\begin{abstract}

We present {\it Herschel}-HIFI, SPIRE, and PACS 50-670 $\mu$m imaging and spectroscopy of six FU Orionis-type objects and candidates (FU Orionis, V1735 Cyg, V1515 Cyg, V1057 Cyg, V1331 Cyg, and HBC 722), ranging in outburst date from 1936-2010, from the ``FOOSH'' (FU Orionis Objects Surveyed with {\it Herschel}) program, as well as ancillary results from Spitzer-IRS and the Caltech Submillimeter Observatory.  In their system properties (\lbol, \tbol, line emission), we find that FUors are in a variety of evolutionary states.  Additionally, some FUors have features of both Class I and II sources: warm continuum consistent with Class II sources, but rotational line emission typical of Class I, far higher than Class II sources of similar mass/luminosity. Combining several classification techniques, we find an evolutionary sequence consistent with previous mid-IR indicators.  We detect \OI\ in every source at luminosities consistent with Class 0/I protostars, much greater than in Class II disks. We detect transitions of $^{13}$CO (J$_{up}$ of 5 to 8) around two sources (V1735 Cyg and HBC 722) but attribute them to nearby protostars.  Of the remaining sources, three (FU Ori, V1515 Cyg, and V1331 Cyg) exhibit only low-lying CO, but one (V1057 Cyg) shows CO up to \jj{23}{22} and evidence for H$_2$O and OH emission, at strengths typical of protostars rather than T Tauri stars.    Rotational temperatures for  ``cool'' CO components range from 20-81 K, for $\sim$ 10$^{50}$ total CO molecules. We detect \CI\ and \NII\  primarily as diffuse emission.
\end{abstract}

\section{Introduction}

FU Orionis-type objects (hereafter, FUors) are a class of low-mass pre-main
sequence objects named after the archetype FU Orionis (hereafter, FU Ori), which
produced a 6 magnitude outburst at $B$-band in 1936 and has remained close to peak
brightness ever since.   \footnote{{\it Herschel} is an
ESA space observatory with science instruments provided by European-led
Principal Investigator consortia and with important participation from NASA.} 

About five more sources have been observed to flare with lightcurves broadly resembling that of FU Ori (5 mag optical flare over $\sim$ 1 yr).  Similar spectral characteristics (broad blueshifted emission lines, IR excess, near-IR CO
overtone absorption) have been used to identify $\sim$ 10 additional
FUor-like objects \citep{hartmann96a, reipurth10}; a number of additional candidates have been  identified \citep[e.g.][]{quanz07b,semkov10a,reipurth12,fischer12,fischer13}.  \citet{paczynski76} first proposed that FUors are the result of a sudden cataclysmic accretion of material from a reservoir that had built up in the circumstellar disk surrounding a young stellar object (see also \citealt{lin85,hartmann85}).  Models of FUor outbursts indicate that over only a few months the accretion rate rises from the typical rate for a T Tauri star ($\dot{M}$ $\lesssim$
10$^{-7}$ M$_{\odot}$ yr$^{-1}$) up to 
10$^{-4}\,$M$_{\odot}$ yr$^{-1}$, and then decays 
over an e-fold time of $\sim$ 10-100 yr \citep{bell94}.   Over the entire outburst the star could accrete 
$\sim$ 0.01 M$_{\odot}$ of material, roughly the mass of a typical T
Tauri disk \citep{andrews05}.  Pre-outburst optical spectra available for two FUors (V1057 Cyg and HBC 722, below) are consistent with T Tauri stars \citep{herbig77,miller11}; they now have optical spectral types of F and K supergiants, respectively.  This change in spectral type is attributed to the large increase in continuum emission in the optical/near-IR, typically dominated by the newly heated inner disk annuli rather than the central star, during outburst.

Only in recent years, it has become possible to produce full spectral energy distributions (SEDs) of FUors, the closest of which are several times farther away than the best-studied low mass star forming regions, first with the Infrared Space Observatory \citep[ISO, 20-200 $\mu$m; ][]{lorenzetti00,abraham04}.  More recently, the Spitzer Space Telescope \citep{werner04} and the Herschel Space Observatory \citep{pilbratt10} cover an extended spectral range (5.3--670 $\mu$m) with a greater sensitivity due in part to {\it Herschel's} larger primary mirror and, at most wavelengths, higher spectral resolution. {\it Spitzer}-IRS \citep{houck04} observations reveal that some FUors resemble embedded protostars, in that their excess emission at wavelengths greater than $\sim$ 30 $\mu$m cannot be accounted for by circumstellar disk models alone, while others lack a strong mid-IR excess above that of a mildly flared disk \citep[e.g.][]{green06,quanz07a,zhu08}.  Thus a central question is whether FUors are all surrounded by envelopes -- and in a corresponding earlier stage of development than a T Tauri star -- or are drawn from both embedded and disk sources.   {\it Herschel} is particularly well-suited to distinguish between disk and envelope structure, as the SEDs of envelopes peak in the far-IR.

In \citet{green11b}, we presented {\it Herschel} observations of the region surrounding the newly outbursting FUor HBC 722. Here we present the first results of the ``FOOSH'' (FU Orionis Objects Surveyed with Herschel) Open Time program (PI: J. Green): improved PACS \citep[Photodetector Array Camera and Spectrometer, 50-210 $\mu$m;][]{poglitsch10} and SPIRE \citep[Spectral and Photometric Imaging REceiver, 194-670 $\mu$m;][]{griffin10} spectroscopy of HBC 722, and newly observed PACS and SPIRE spectroscopy, and imaging (70 -- 500 $\mu$m), of five older FUors: V1057 Cyg, V1331 Cyg, V1515 Cyg,  V1735 Cyg, and FU Ori.  Additionally we include HIFI  \citep[Heterodyne Instrument for the Far Infrared;][]{degraauw10} observations of CO \jj{5}{4} and \jj{14}{13}.  We re-reduce archival {\it Spitzer} observations of V1331 Cyg, originally presented in part by  \citet{carr11}.  Complementary to the HIFI data, we present velocity-resolved ground-based spectra of selected submillimeter lines.  We analyze the spatial structure of the line and continuum emission, and consider differences amongst the sample.  Lastly we compare FUors to Class 0/I embedded sources from the DIGIT Key Program \citep{green13b} and other protostar surveys (HOPS, \citealt{manoj12}; WISH, \citealt{karska13}) and consider the impact of the {\it Herschel} data on our understanding of the evolutionary state of FUors.

\section{Observations and data reduction}

The FOOSH program consisted of 21 hrs of Herschel observing time: V1057 Cyg, V1331 Cyg, V1515 Cyg, V1735 Cyg, and FU Ori were observed as part of FOOSH. In addition, HBC 722 was observed with Herschel as a Target of Opportunity (PI: J. Green) and the initial results were presented in \citet{green11b}. The region around each source was observed with the SPIRE-FTS and PACS spectrometers and imaged with the SPIRE bolometer and PACS cameras. Additionally, each source was observed in selected bands with HIFI.   Finally, all but FU Ori were observed in Sept. 2012 with the Caltech Submillimeter Observatory heterodyne receivers.
A list of the observations included in this work is presented in Table \ref{obslog}.

\subsection{The Sample}

The basic characteristics of the FUors in our sample are listed in Table \ref{fuors}.  Although all are classified as FUors, our sources were selected to span a variety of subtypes, with regard to local extinction, dust properties, observed eruption date, and SED shape \citep{green06,quanz07a,kospal11}.  Pre-{\it Herschel} observations of HBC 722 and FU Ori suggest that they fall toward the disklike end of the spectrum (based on their mid-IR colors and low extinction values), with low ($\sim$ 0.01-0.02 \msun) upper limits on envelope mass \citep{sandell01,dunham12b}.  V1735 Cyg is most similar to an embedded source with a highly extinguished  central star and a rising mid-IR SED.  V1331 Cyg, V1057 Cyg, and V1515 Cyg are intermediate between the FU Ori and V1735 Cyg cases; moderate extinction values, relatively blue SEDs but clear evidence for substantial envelope material. All but V1331 Cyg were identified as FUors by their optical flare, rather than by ancillary spectral characteristics.  Typically, FUors are classified based upon their optical and mid-IR spectral features; in this work we also classify the sources by their full SED properties.

{\it FU Ori.}  Located in Orion, FU Ori (600 pc) is the archetype of the FUor class, the earliest observed outburst and the most extreme, rising by 6 magnitudes in $B$ over the period of 3 months in 1936.  It has since faded by only $\sim$ 1 mag, at a rate of 14 mmag yr$^{-1}$ \citep{kenyon00}. FU Ori shows a relatively blue SED \citep{hartmann96a,green06,quanz07a}, with emission from pristine silicate dust in the Spitzer bands, and a relatively low extinction (A$_V$ $=$ 1.8 mag).  Models of the FU Ori system (with a central stellar mass of 0.3 $\msun$) suggest that the current accretion rate remains very high (10$^{-4}$ $\msunyr$) and it is the northern component in a close (0$\farcs$5) binary system \citep{reipurth04,wang04}.  The non-outbursting star, FU Ori S, may be a more massive companion ($>$ 0.5 $\msun$; \citealt{pueyo12}); the binary is unresolved by {\it Herschel}.

{\it V1735 Cyg.}  Also known as Elias 1-12, V1735 Cyg (950 pc) is the only source in our sample that does not show silicate emission features in its IRS spectrum, but instead shows absorption due to CO and H$_2$O ice.  It is the most extinguished source in the sample, with estimates of A$_V$ $>$ 6 
mag.  A second submillimeter source, V1735 Cyg SM1 is nearby ($\sim$ 20--24$\arcsec$ distant; \citealt{sandell01}).  The mid-IR SED suggests substantial envelope emission in the system.

{\it V1515 Cyg.} Gradually rising during the 1940s and 50s, V1515 Cyg (1000 pc) shows a flat mid-IR SED with weak silicate emission features, and A$_V$ $=$ 3.0 mag.  The region around V1515 Cyg shows arclike reflection nebula structure \citep{kospal11b}.

{\it V1057 Cyg.}  Erupting around 1969, V1057 Cyg (600 pc) is one of two sources in the FOOSH sample with a pre-outburst optical spectrum consistent with a T Tauri star \citep{herbig77}.  With A$_V$ $=$ 3.5 mag and a flat-spectrum SED, weak silicate emission, and a ring-like reflection nebula, V1057 Cyg is quite similar in character to V1515 Cyg in its optical/IR spectrum.

{\it V1331 Cyg.}  Classified as ``pre-outburst'' or between outbursts \citep{welin77}, V1331 Cyg (550 pc) was identified based on its spectral similarity to the pre-outburst optical spectrum of V1057 Cyg; no optical flare was observed and thus it is possible that V1331 Cyg is not an FUor 
\citep{sandell01,quanz07c, kospal11}.  V1331 Cyg is the least luminous of the Cygnus sources in our sample.  Hot water vapor was observed at K-band and was attributed to the inner disk \citep{najita09}.  We present archival IRS spectra of this source, which closely resemble the flat spectrum SEDs of V1515 Cyg and V1057 Cyg, with weak, pristine silicate emission features at 10 and 20 $\mu$m.  Even if V1331 Cyg is not an FUor, it represents an important historical comparison; the accretion rate would be very high for a non-outbursting T Tauri star, at $\sim$ 10$^{-6}$ $\msunyr$  \citep{najita09}.

{\it HBC 722.}  Also known as LkH$\alpha$ 188-G4, PTF10qpf, and V2493 Cyg, HBC 722 (520 pc) is a newly erupted FUor \citep{semkov10a,miller11} located in the ``Gulf of Mexico'' in the southern region of the North American/Pelican Nebula, among a closely-packed set of accreting young stars.  Pre-outburst observations \citep{cohen79} indicate that HBC 722 was a $\sim$ 0.5 \lsun\ K7-M0 spectral type T Tauri star prior to 2009.  Between 2009 and 2010, HBC 722 rose by $\sim$ 1 mag in $B$, and then suddenly flared between July and Sept. 2010, rising by 3.5 mag and in luminosity by a factor of $\sim$ 20 to 12 \lsun\ over the course of 3 months \citep{kospal11}.  After decreasing to 5.4 $\lsun$ between Sept. 2010 and May 2011, HBC 722 began to rise in brightness once more, re-achieving peak brightness by June 2012 and exceeding it in optical bands by 2013.  High cadence photometric observations reveal a multitude of periods and ``flickering'' that may be associated with a still-active inner accretion disk \citep{green13a}.  Resolved SMA observations detect little remnant envelope material close to the star \citep{dunham12b}.

\subsection{Imaging}

SPIRE imaging was gathered in single-cycle integration times in the small map mode, a single (349 s) observation per source.  SPIRE observed simultaneously at 250, 350 and 500 $\mu$m. The on-orbit beam sizes are 18$\farcs$1, 25$\farcs$2, and 36$\farcs$6, respectively.   The {\it Herschel} data were reduced using HIPE \citep[{\it Herschel} Interactive Processing Environment;][]{ott10} pipeline.

PACS imaging was acquired using two pairs of single-cycle scan maps.  For each source a ``blue'' (70 and 160 $\mu$m) and ``green'' (100 and 160 $\mu$m) simultaneous observation was taken, for two different orientation angles (70$^{\circ}$ and 110$^{\circ}$), a total of four (279 s) observations per source.  The final 160 $\mu$m image was produced by coadding all four scans; the 70 and 100 $\mu$m images were averaged from their two respective scans each.  The on-orbit beam sizes for 70, 100, and 160 $\mu$m are 5$\farcs$5, 6$\farcs$7, and 10$\farcs$7, respectively.

\subsection{Spectroscopy}

\subsubsection{SPIRE-FTS}

Our SPIRE-FTS data were taken in a single pointing with sparse image sampling, high spectral resolution, in 1 hr of integration time.  The spectrum is divided into two orders covering the spectral ranges 194 -- 325 $\mu$m (``SSW''; Spectrograph Short Wavelengths) and 320 -- 690 $\mu$m (``SLW''; Spectrograph Long Wavelengths), with a resolution of $\lambda$/$\Delta\lambda$ $\sim$ 300--800, increasing at shorter wavelengths.  Each order was reduced separately within HIPEv9.0 using the standard pipeline for extended sources, including apodization.  SPIRE used an onboard calibration source for flux calibration.  Although SPIRE observed simultaneously along multiple spatial pixels, at this stage the off-source pixels are not yet flux-calibrated.  Thus we  used off-source pixels only to determine if a source is extended; all presented spectra were taken using data from the central pixel only.  No background subtraction was performed, and thus the spectra should be considered an upper limit to the true flux density; we note cases where extended emission was seen in the following sections.
The measured uncertainties in the continuum ranged from 0.05-0.5 Jy (an improvement over that reported in \citealt{green11b} due to pipeline improvements in the interim), or $\sim$ 0.5-5 10$^{-18}$ W m$^{-2}$ in linefluxes, increasing with observed frequency and several times higher in SSW compared to SLW.

\subsubsection{PACS}

PACS was a 5$\times$5 array of 9$\farcs$4 $\times$9$\farcs$4  spatial pixels (also referred to as ``spaxels'') covering the spectral range from 51 -- 210 $\mu$m with a resolution $\lambda$/$\Delta\lambda$ $\sim$ 1000--3000, divided into four orders, covering $\sim$ 50--75, 70--105, 100--145, and 140--210 $\mu$m, referred to as ``B2A'', ``B2B'', ``short R1'', and ``long R1''.  Each order was first reduced using a modified pipeline optimized for extended sources; a detailed description is provided in \citet{green13b}.  
For the FOOSH sample we utilized the full range of PACS 
(50-210 $\mu$m) in
two linked pointed chop/nod rangescans: a blue scan covering 50-75 and
100-150 $\mu$m (SED B2A + short R1); and a red scan covering 70-105 and
140-210 $\mu$m (SED B2B + long R1). We used 6 and 4 range repetitions
respectively, for integration times
of 3530 and 4620 seconds (a total of $\sim$ 8000 seconds per target and off-positions combined, for
the entire 50-210 $\mu$m scan; the on-source integration time  is $\sim$ 3000 seconds).  The telescope sky background was subtracted using two nod positions 6\arcmin\ from the source.  The absolute flux uncertainty was estimated at 30\%, typically used for multi-wavelength comparisons in extended sources \citep[e.g.][]{herczeg12}.  Further detail on systematic uncertainties can be found in the PACS Observer Manual.

\subsubsection{HIFI}

 {\it Herschel}-HIFI provided high spectral resolution, down to $\sim$ 0.03 km s$^{-1}$).  HIFI was used in single point mode, in band 6b (1578.2--1697.8 GHz) tuned to CO \jj{14}{13} (1611.8 GHz), and in band 1b (562.6--628.4 GHz) tuned to CO \jj{5}{4} (576.3 GHz).
The data were again reduced using HIPE.  The beamsize of HIFI is 43$\farcs$1 in band 1b, and 15$\arcsec$ in band 6b \citep{roelfsema12}.  The total integration time was 622 and 677 seconds per observation in band 1b and 6b, respectively, and the uncertainty 
in frequency was less than 100 kHz, or 0.05 km s$^{-1}$.  
The HIFI data were reduced in a manner similar to that described in 
\citet{kristensen12}. The spectra were reduced in HIPE and exported 
to the ``CLASS'' analysis package for further reduction and analysis. The 
reduction consisted of subtracting linear baselines and averaging data 
from the H- and V-polarizations, the latter only after visual inspection of 
the data from the two polarizations. A main beam efficiency of 0.75 was 
adopted to convert from antenna temperature ($T_A^*$) 
to main beam temperature ($T_{\rm MB}$), and integration over the line
yielded a value for the integrated intensity ($\int T_{\rm MB}$ d$v$) \citep{roelfsema12}.

\subsubsection{CSO}

Observations of V1735 Cyg, V1057 Cyg, V1331 Cyg, V1515 Cyg, and HBC 722 were obtained with the 230 GHz heterodyne receiver installed on the Caltech Submillimeter Observatory (CSO) between 8 and 22 September 2012.  We observed each source in HCO$^+$ \jj{3}{2} (267.557 GHz) achieving a minimum S/N of 5 on the detected line, with $\tau_{225}$ (the atmospheric opacity at 225 GHz) varying between 0.08 and 0.13.  Additionally we observed HBC 722 in CO \jj{4}{3} (461.041 GHz) using the 460 GHz receiver on 17 September 2012, with $\tau_{225}$ $=$ 0.070.  The beamsize is $\sim$ 28$\arcsec$ for HCO$^+$ and $\sim$ 16.5$\arcsec$ for CO \jj{4}{3}.  The data were then exported to the CLASS package for further reduction and analysis, similar to HIFI.

\subsubsection{IRS}

{\it Spitzer}-IRS (InfraRed Spectrograph) was a long-slit spectrograph with 2-5$\arcsec$ pixels, covering the spectral range 5.3 -- 38 $\mu$m, with a resolution $\lambda$/$\Delta\lambda$ $\sim$ 60--600.  V1331 Cyg (AORs 14547712 and 14549760) was processed using a custom point-source pipeline \citep{furlan06}, cleaned using an aggressive bad pixel mask generated from a ``grand rogue mask'', in which all pixels exhibiting greater than 4$\sigma$ deviations in dark frames (attributed to cosmic rays, permanent hot pixels and other effects) taken during more than one campaign were flagged and interpolated using a ``nearest neighbor'' technique.

\subsection{Absolute Flux Calibration and SEDs}

In order to produce a single SED for each source, we needed to calibrate the continuum flux density across {\it Spitzer}-IRS, {\it Herschel}-PACS, and {\it Herschel}-SPIRE, which range in spatial resolution from 2--40\arcsec.  At the shortest wavelengths of IRS, the spatial resolution was  1100 AU at a distance of 550 pc; the material emitting in the {\it Spitzer} bands remains unresolved.  To determine the continuum level, we took the final output of each instrument and scaled, first between modules, and then between instruments.  At shorter wavelengths with improved spatial resolution these sources are dominated by disk/envelope warm dust,  within the IRS beam as listed above; the FWHM of the mid-IR continuum measured from IRS was typical of YSOs, with no more than a 20\% flux mismatch between modules for the four sources observed in \citet{green06}.  Thus in each case we assumed that the source was essentially unresolved, and increased spatial resolution at shorter wavelengths provided a better absolute flux calibration.  Where possible we then correlated this with photometry, extrapolating between missing wavelength regions.  All of the scaling is described below.

The IRS data for V1515 Cyg, V1057 Cyg, and FU Ori were first presented in \citet{green06}, and the IRS spectra for V1735 Cyg was presented in \citet{quanz07a}; for this work, we reduced V1331 Cyg using the same technique (the SH reduction appeared in \citealt{carr11}).  ``SL'' (Short-Low module; 5.3-14 $\mu$m) data were multiplied by 1.03 and ``LH'' (Long-High module; 20-38 $\mu$m) multiplied by 1.05 to match ``SH'' (Short-High module; 10-20 $\mu$m) to produce the final spectrum.   Additionally we note that in V1735 Cyg, the SL spectrum was scaled by 1.35 to match SH (SL was not included in the \citealt{quanz07a} data).  HBC 722 was not observed by IRS (pre- or post-outburst).  As described in \citet{green06}, the spectra were dereddened using an analytical extinction law \citep{mathis90}, using an R$_V$ of 3.1, in the cases of FU Ori, V1515 Cyg, and V1057 Cyg.

For all six sources, the PACS data were scaled, module-by-module, using a polynomial factor (as a function of wavelength), a technique that maximized continuum S/N while determining the proper absolute flux calibration.    The full technique, as applied to the DIGIT dataset, is described in detail in \citet{green13b}.  We modified this procedure in one way:  in all but HBC 722 we included all flux in the 3$\times$3 spaxels as part of the FUor environment, for both continuum and lines, rather than using the full array for continuum and the smaller 3$\times$3 for lines.   

For HBC 722, the SPIRE flux density was much greater than the PACS continuum.  This is likely because, as noted in \citet{green11b}, the continuum flux density near HBC 722 was contaminated by an embedded source, 2MASS 20581767+4353310, 17$\arcsec$ to the southeast.   SPIRE could not resolve these sources, but {\it Spitzer} and PACS photometry could do so.  We produced a PACS spectrum of the entire 3$\times$3 central region summed together, which included the embedded source and was similar in spatial extent and measured flux density to that of the SPIRE beam.  Thus the composite ``HBC 722'' spectrum is likely dominated by the embedded source rather than HBC 722 itself.  In addition, a second epoch of observations of the same field but with a different orientation angle were taken in this program, and the absolute flux matched to within 10\% for both extractions. For the HBC 722 composite spectrum, we multiplied B2B by 1.5, and both R1 modules by 0.67.  In all other cases the standard DIGIT procedure (scaling to the 3$\times$3 spaxels) produced a smooth PACS continuum.  No scaling relative to IRS was performed, but the final PACS spectrum was a reasonable match to the extrapolated IRS continuum at 60 $\mu$m.

V1735 Cyg also suffered from confusion with V1735 Cyg SM1 (24$\arcsec$ to the northeast), at $\lambda$ $>$ 100 $\mu$m.  The spectroscopic flux density using our standard  central 3$\times$3 spaxel aperture exceeded the PACS photometry by 50\% at 70 and 100 $\mu$m, and by 30\% at 160 $\mu$m.  We infer that an annulus of $\sim$ 17$\arcsec$ or less is required to avoid contamination.  However, the PACS spectroscopic data are not sufficiently sampled to extract the spectrum from a 17$\arcsec$ aperture at all wavelengths, and thus we use the larger aperture and assume contamination at these levels.
  
The SPIRE data were extracted using the ``extended source'' calibration pipeline, as this produced a smoother continuum between modules, better S/N, and fewer spectral artifacts than the ``point source'' pipeline. 
 First we multiplied SLW to match the SSW flux density in the overlap region of the two modules (300-320 $\mu$m), and then matched the flux at 210-220 $\mu$m to an extrapolation of the final PACS spectrum.  HBC 722 and V1735 Cyg required no additional scaling, either between modules, or between instruments.  For both FU Ori and V1515 Cyg, we multiplied SSW by 0.9 to match PACS.  For both V1331 Cyg and V1057 Cyg, we multiplied SLW by 1.55 to match SSW, with no additional scaling to match PACS.
  
 \subsection{Wavelength Calibration}

 For the DIGIT sample, we noted that the uncertainties in the PACS wavelength centroid were at least 30-50 km s$^{-1}$.  With shorter exposure times in FOOSH, we found slightly more erratic centering in our line positions, up to 100  km s$^{-1}$.  Small mispointings can masquerade as velocity shifts across the PACS IFU (see the PACS Observer Manual for further information); thus we cannot determine the gas velocities to better than 100 km s$^{-1}$ . The SPIRE spectra have even greater uncertainty due to lower spectral sampling and bigger beamsize.  Given the uncertainties, we did not observe any convincing velocity shifts in our data.

\section{Results}

\subsection{Imaging}

\begin{figure}
\begin{center}
\begin{tabular}{ll}
\includegraphics[width=8.0cm, angle=0]{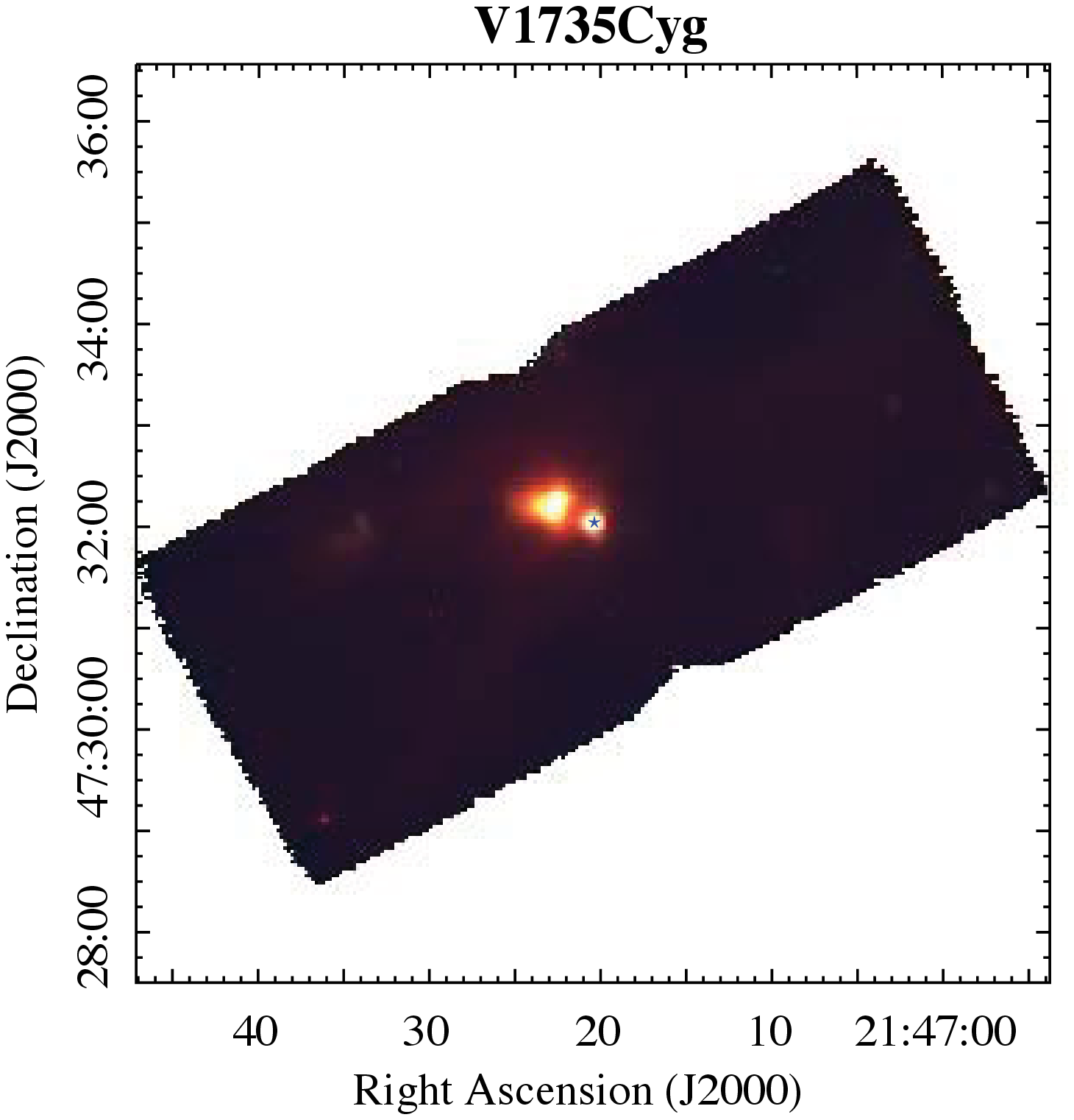} & \includegraphics[width=8.0cm, angle=0]{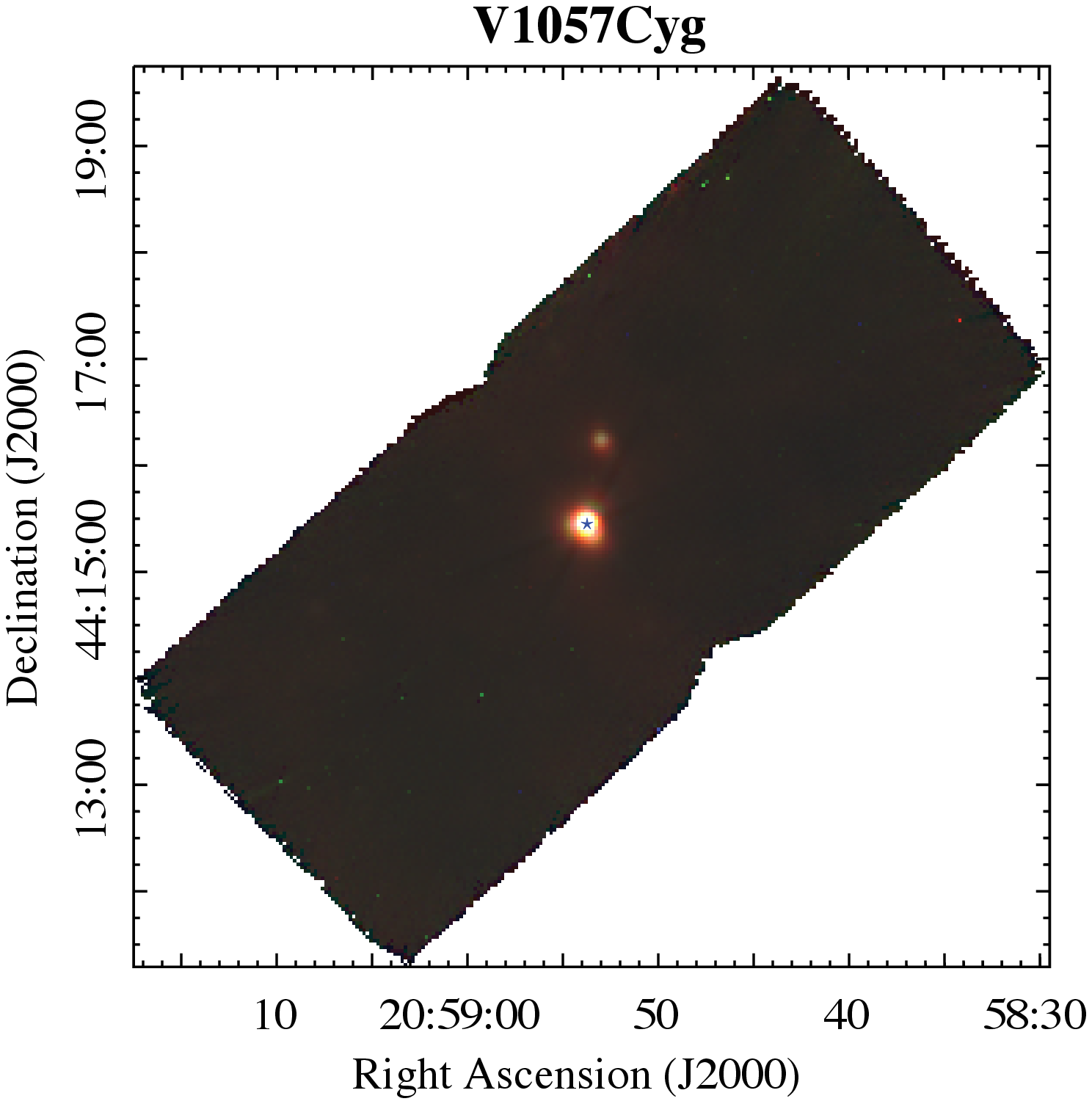} \\
\includegraphics[width=8.0cm, angle=0]{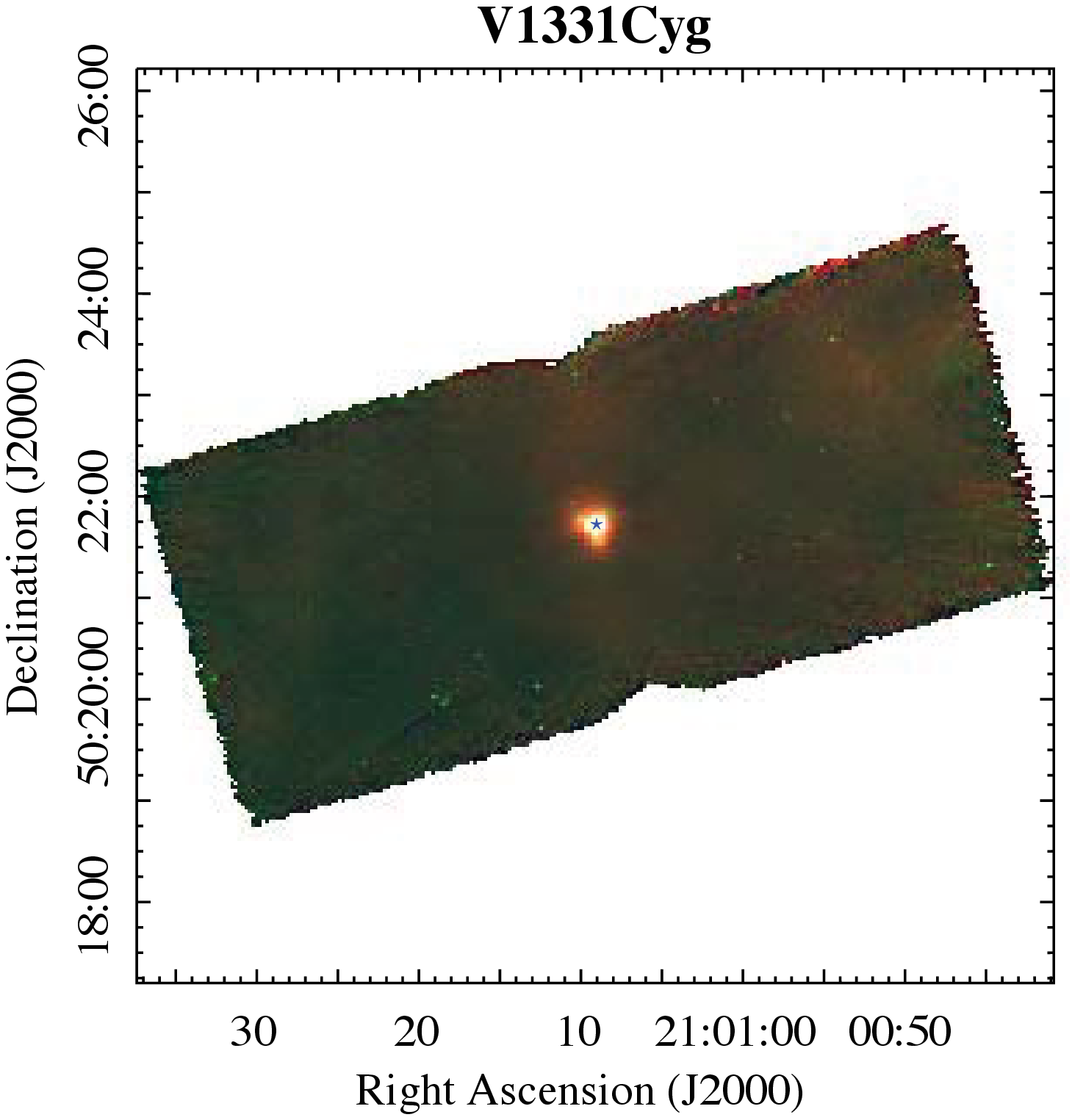}  & \includegraphics[width=8.0cm, angle=0]{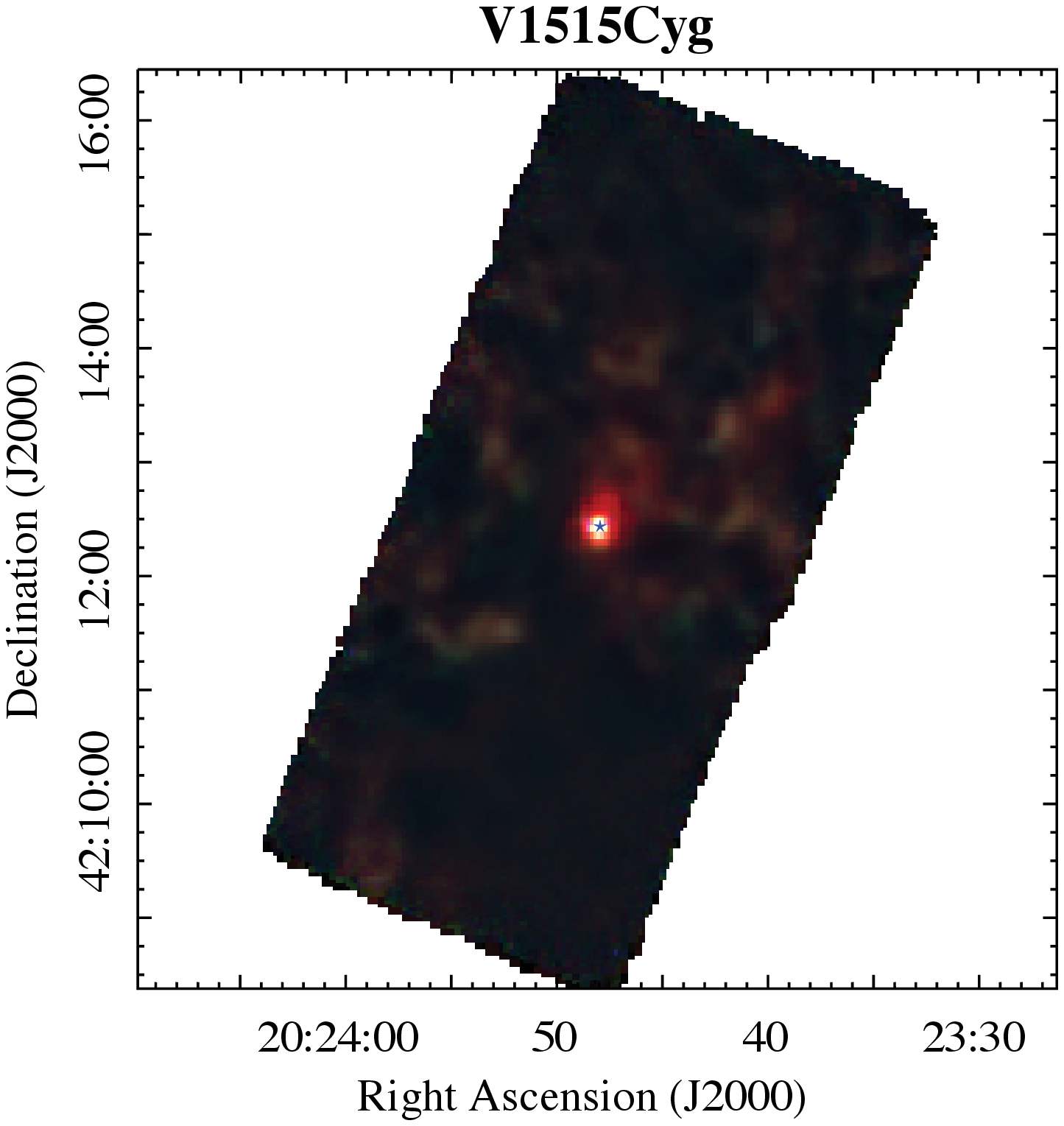} \\
\includegraphics[width=8.0cm, angle=0]{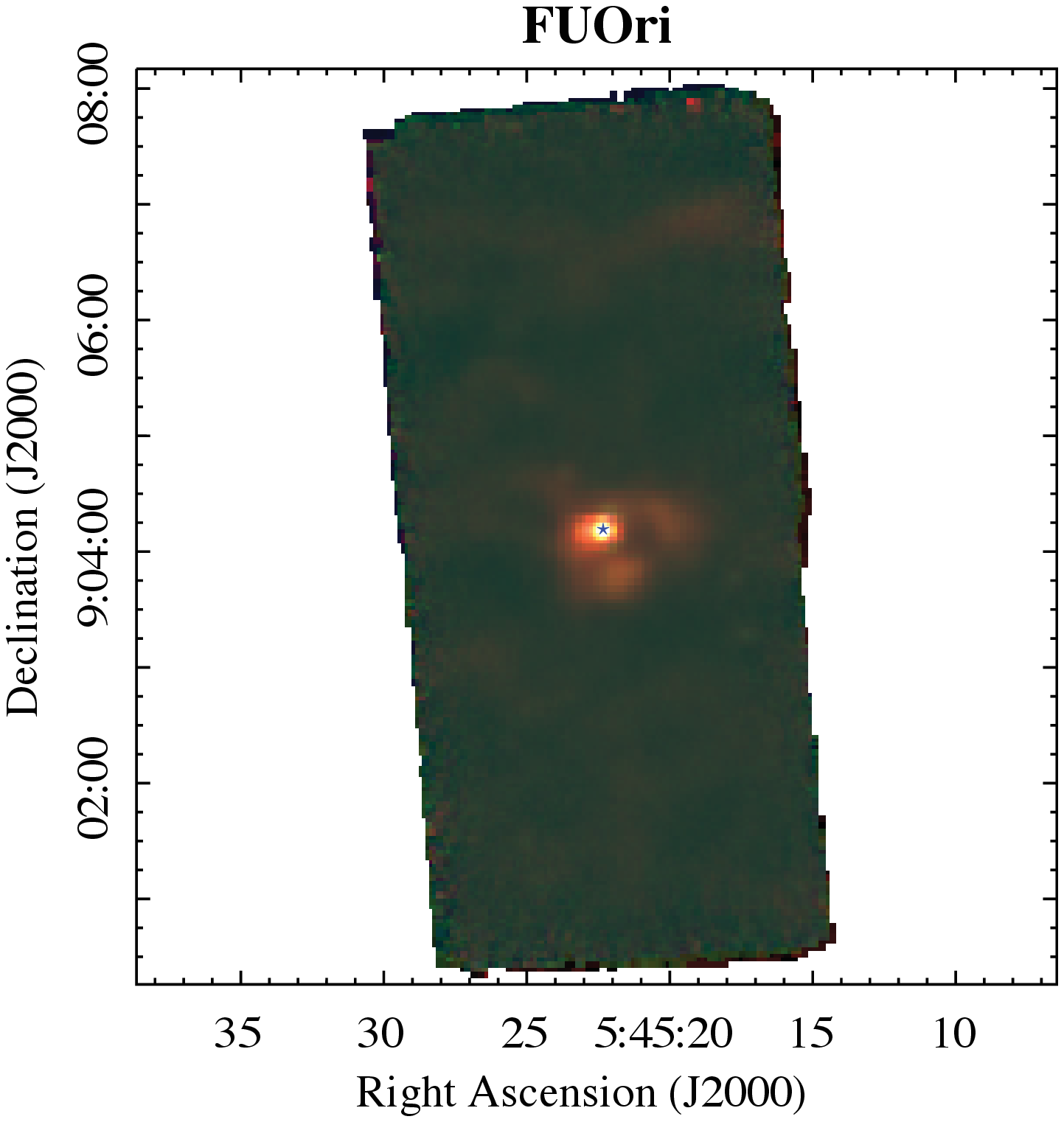}  \\
\end{tabular}
\caption{70 (blue), 100 (green), and 160 $\mu$m (red) composite PACS imaging of the FUors in our sample   The positions of the optical FUors are marked in each image with a blue *.  
}
\label{pacsphot}
\end{center}
\end{figure}

\begin{figure}
\begin{center}
\begin{tabular}{ll}
\includegraphics[width=8.0cm, angle=0]{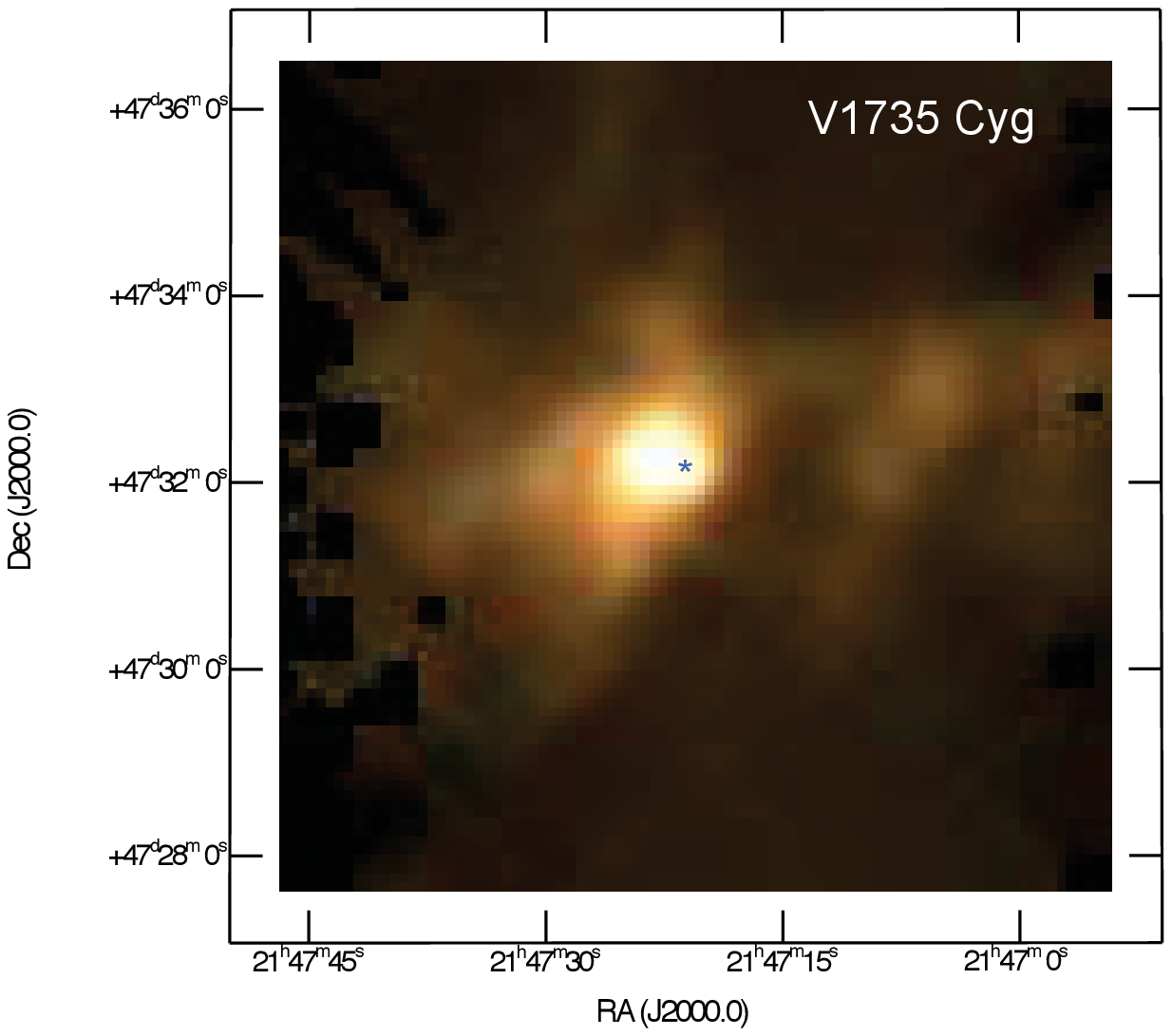} & \includegraphics[width=8.0cm, angle=0]{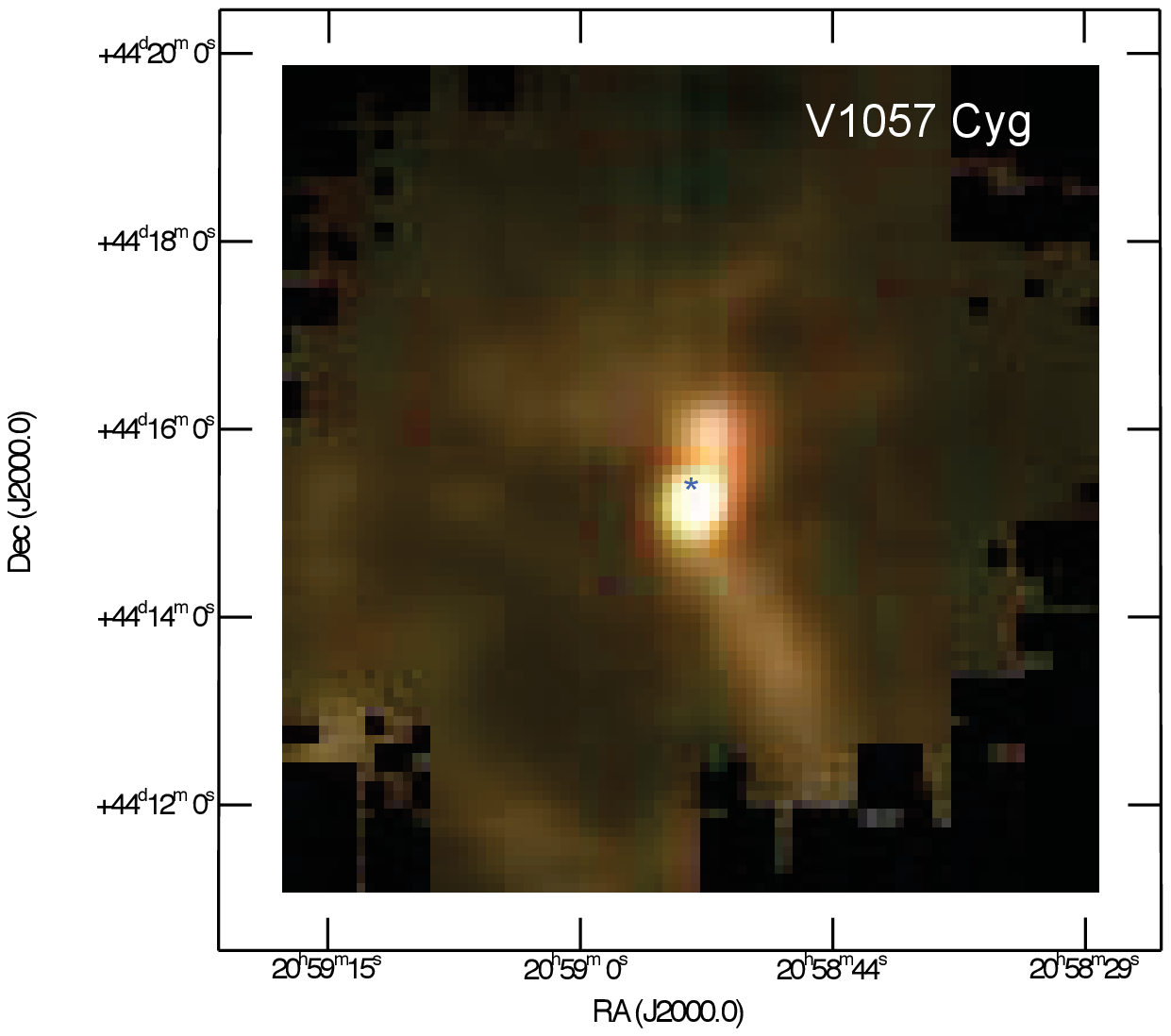} \\
\includegraphics[width=8.0cm, angle=0]{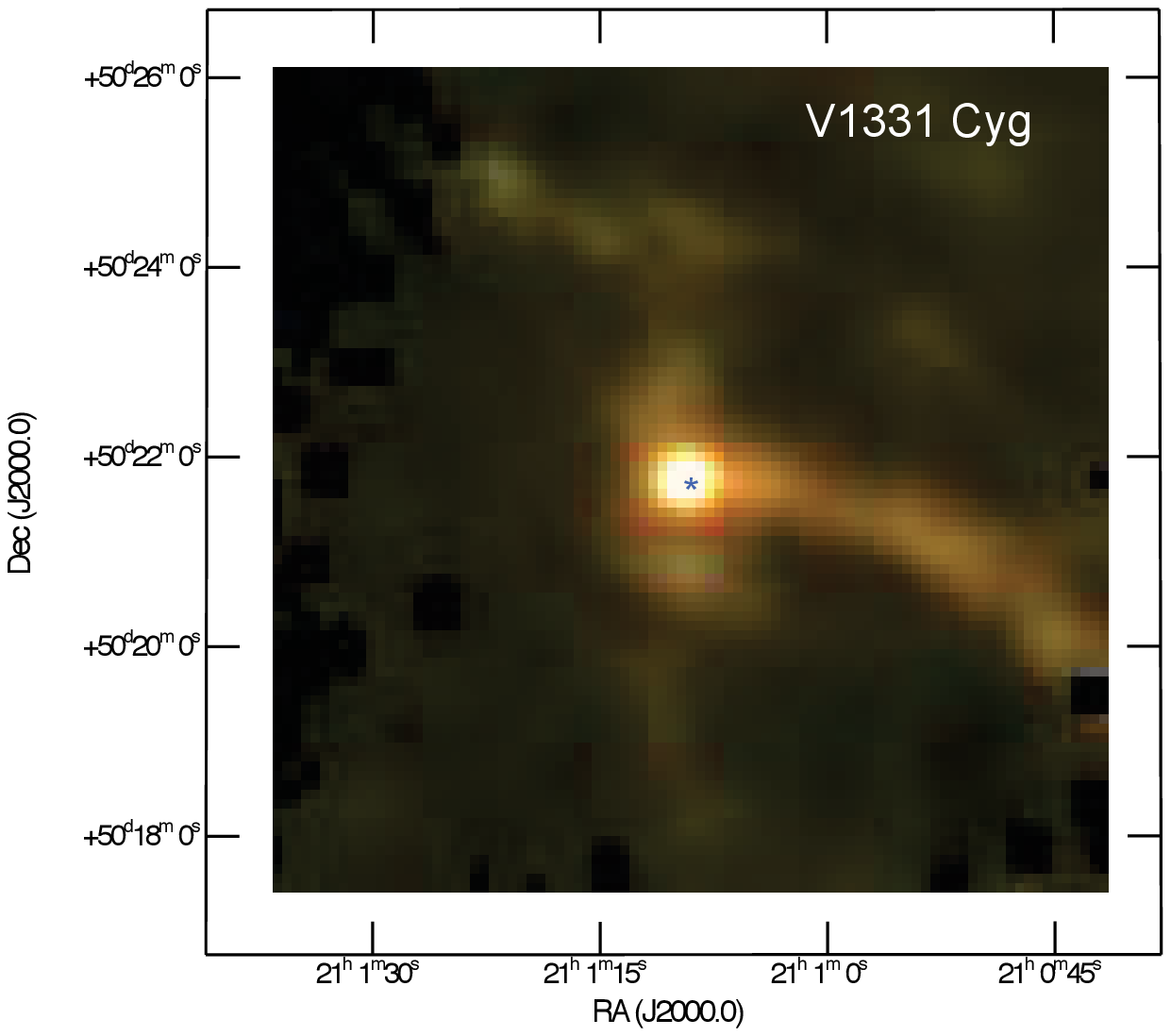} & \includegraphics[width=8.0cm, angle=0]{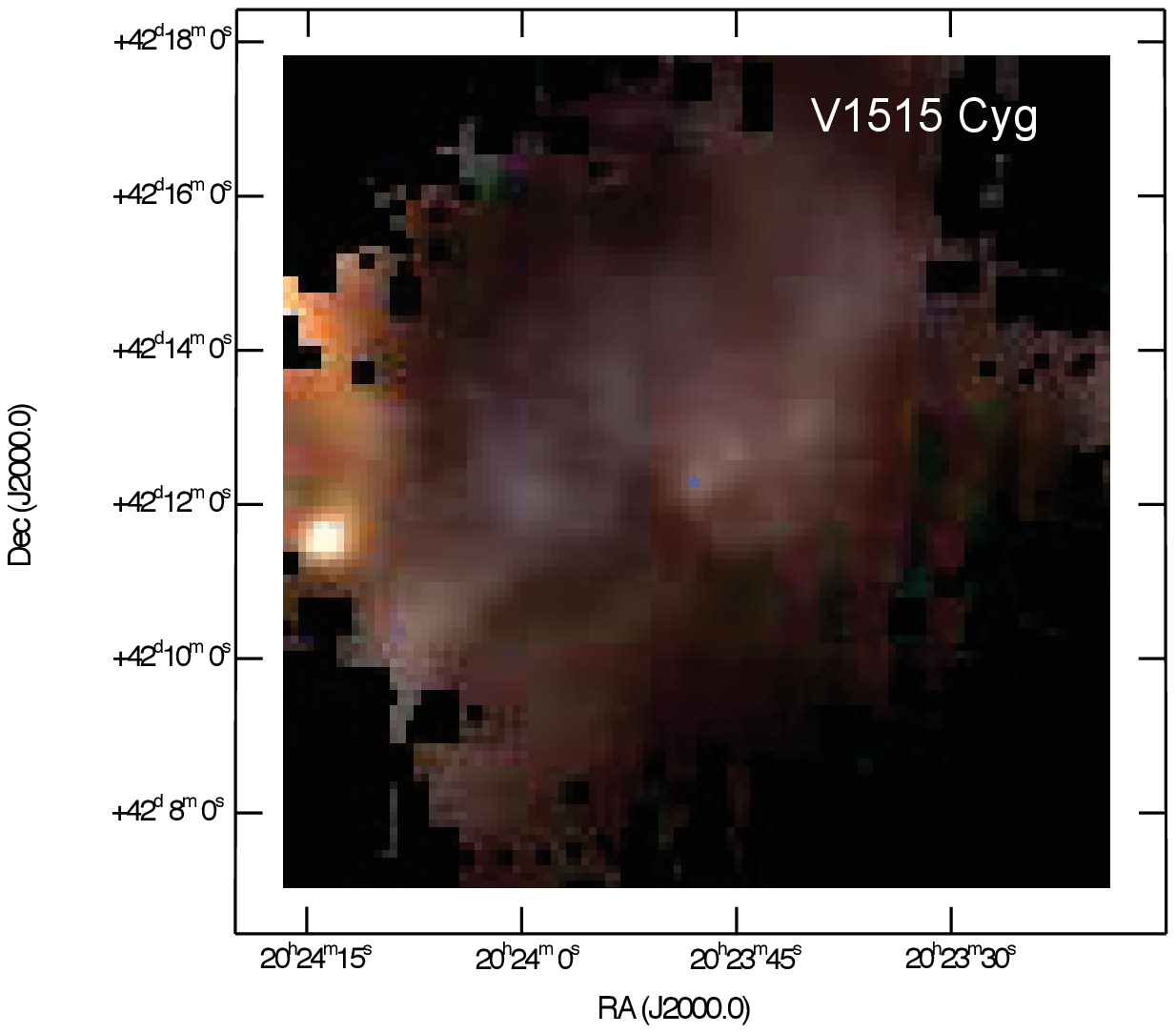} \\
\includegraphics[width=8.0cm, angle=0]{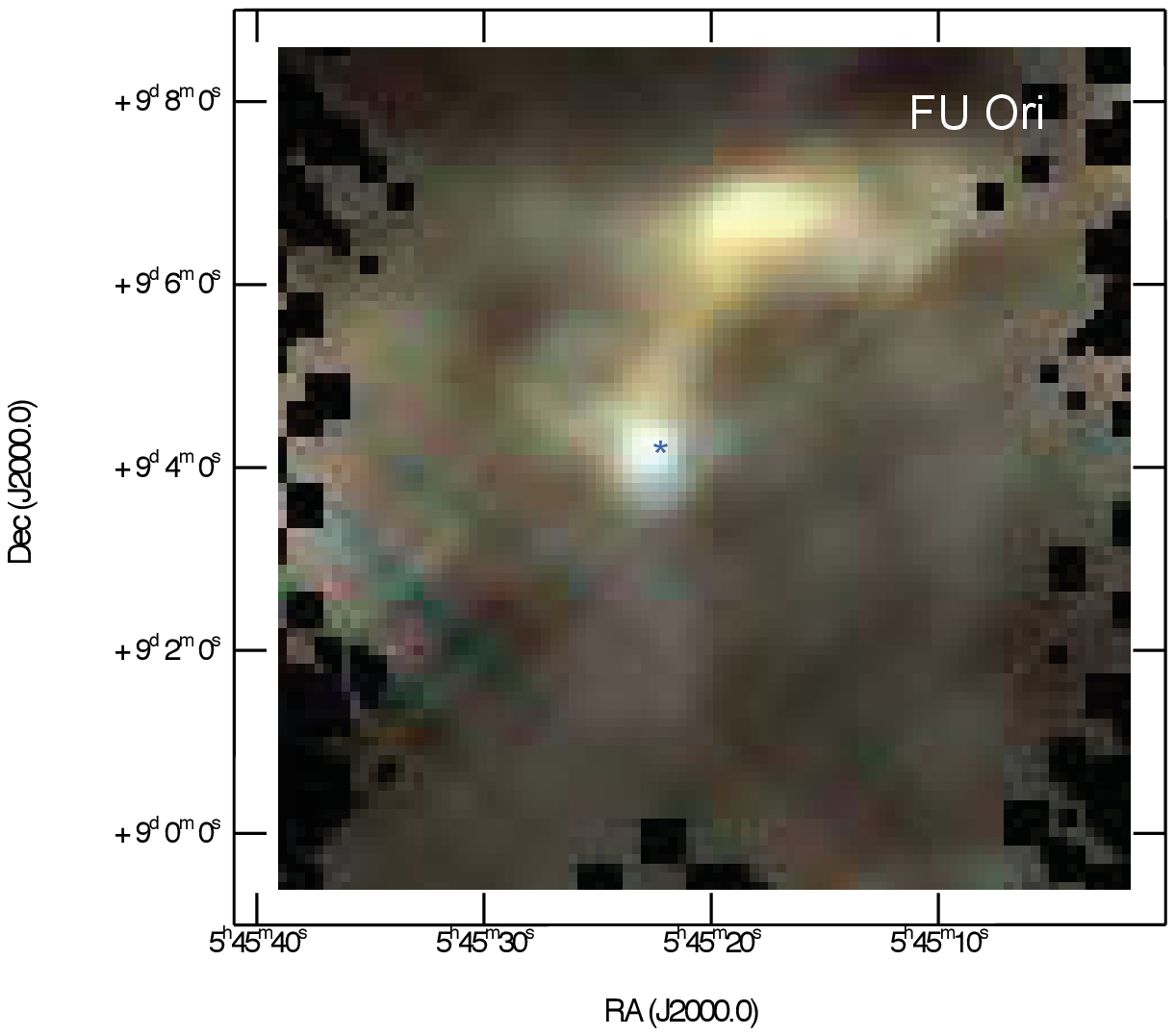}  \\
\end{tabular}
\caption{250 (blue), 350 (green), and 500 $\mu$m (red) composite SPIRE imaging of the FUors in our sample.  The positions of the optical FUors are marked in each image with a blue *.}
\label{spirephot}
\end{center}
\end{figure}

Figure \ref{pacsphot} shows three-color (70, 100, and 160 $\mu$m) PACS images, and Figure \ref{spirephot} shows three-color (250, 350, and 500 $\mu$m) SPIRE images for the five classical FUors in our sample.  We consider each source below.    Table \ref{phot} lists the detected flux in 10-15$\arcsec$ (PACS PSF corrected) and 30$\arcsec$ (SPIRE) apertures centered on the source coordinates.  In summary, most of the FUors appear pointlike at 70 $\mu$m, but gradually the extended filaments begin to dominate at longer wavelengths.

The HBC 722 images were first presented in \citet{green11b}.  All are surrounded by substantial emission from multiple sources in the far-IR which are detected within the SPIRE beam, especially at longer wavelengths.  The dominant continuum source in the HBC 722 field is actually 2MASS 20581767+4353310, an embedded protostar \citep{green11b}.   They note a separate 0.41 Jy source detected at PACS 70 $\mu$m at the position of HBC 722, below the sensitivity of the spectroscopic observations.

V1515 Cyg is detected in all bands, but does not show strong continuum sources at the stellar position at SPIRE wavelengths, relative to the surrounding nebulosity.  At 70 $\mu$m the FUor is easily distinguished from the background, but at 500 $\mu$m the arclike emission extending 2$\arcmin$ to the north becomes dominant.

Similarly, V1331 Cyg, V1057 Cyg, and FU Ori show clear detections with PACS but additional arclike structures generally correlating with previously analyzed mm data \citep[e.g.][]{kospal11}.   In each case, the FUor appears much bluer than the filamentary material.

In the case of V1735 Cyg, the optical outbursting source is the fainter western extension of the submillimeter peak; the peak emission comes from the embedded source V1735 Cyg SM1, consistent with observations by Spitzer-MIPS \citep{harvey08}.  The beamsize of MIPS at 70 $\mu$m is comparable to that of SPIRE at 250 $\mu$m, allowing marginal resolution of the two sources.  This is apparent from the PACS images where the sources are spatially resolved.  At longer wavelengths, they are inextricably blended.

\subsection{Continuum SEDs}

\begin{figure}
\begin{center}
\includegraphics[scale=0.4, angle=0]{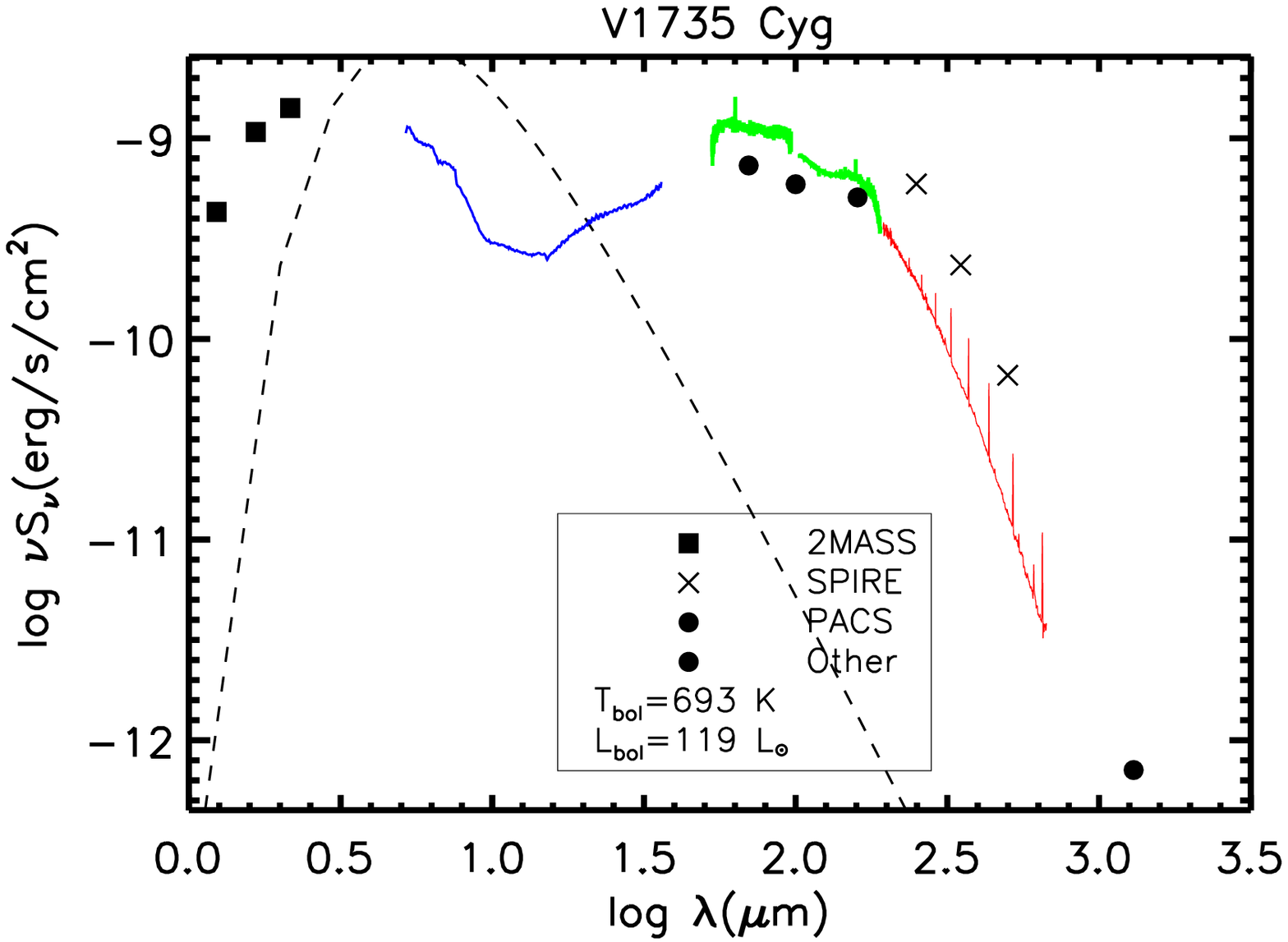} 
\includegraphics[scale=0.4, angle=0]{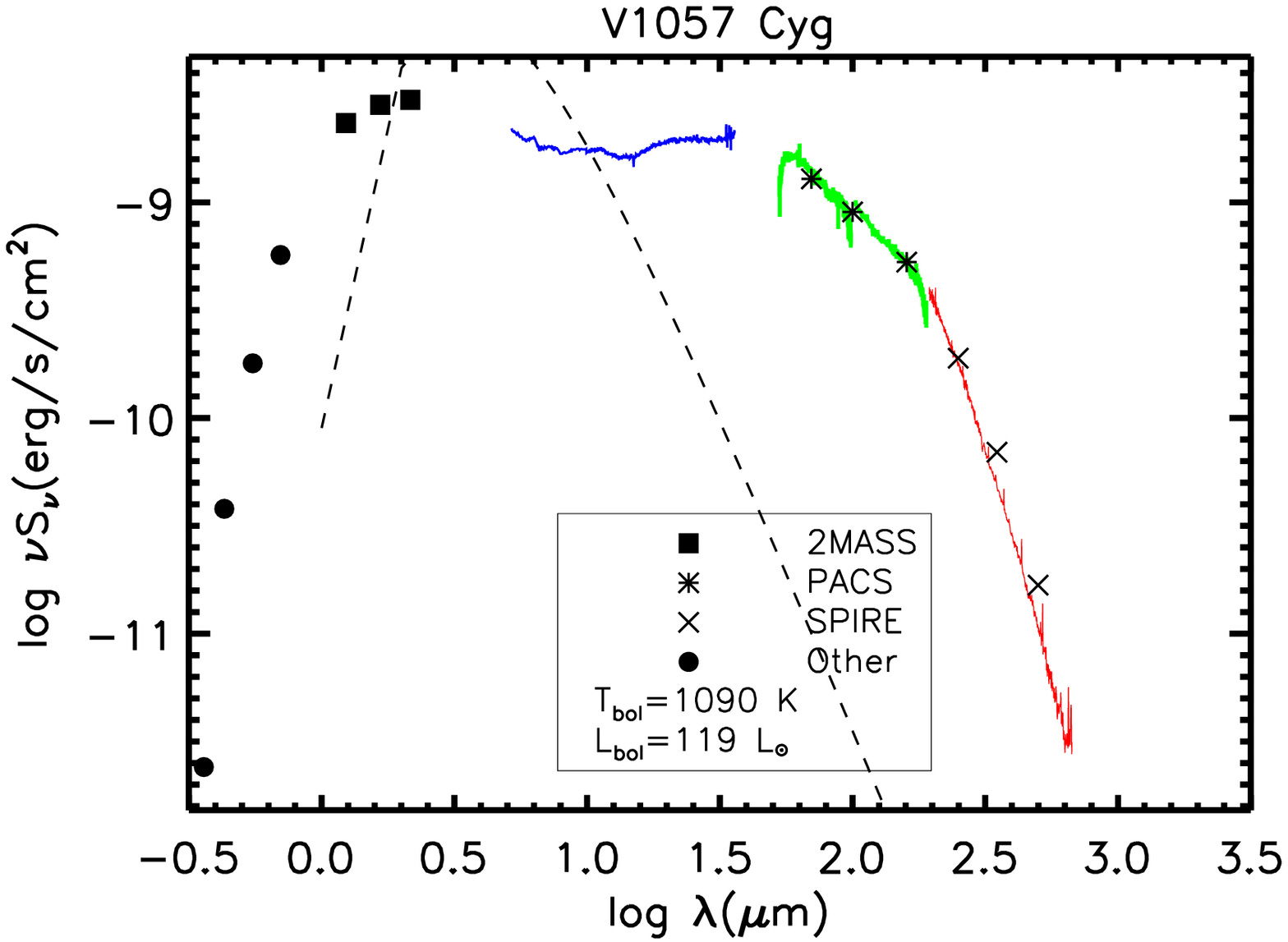} \\
\includegraphics[scale=0.4, angle=0]{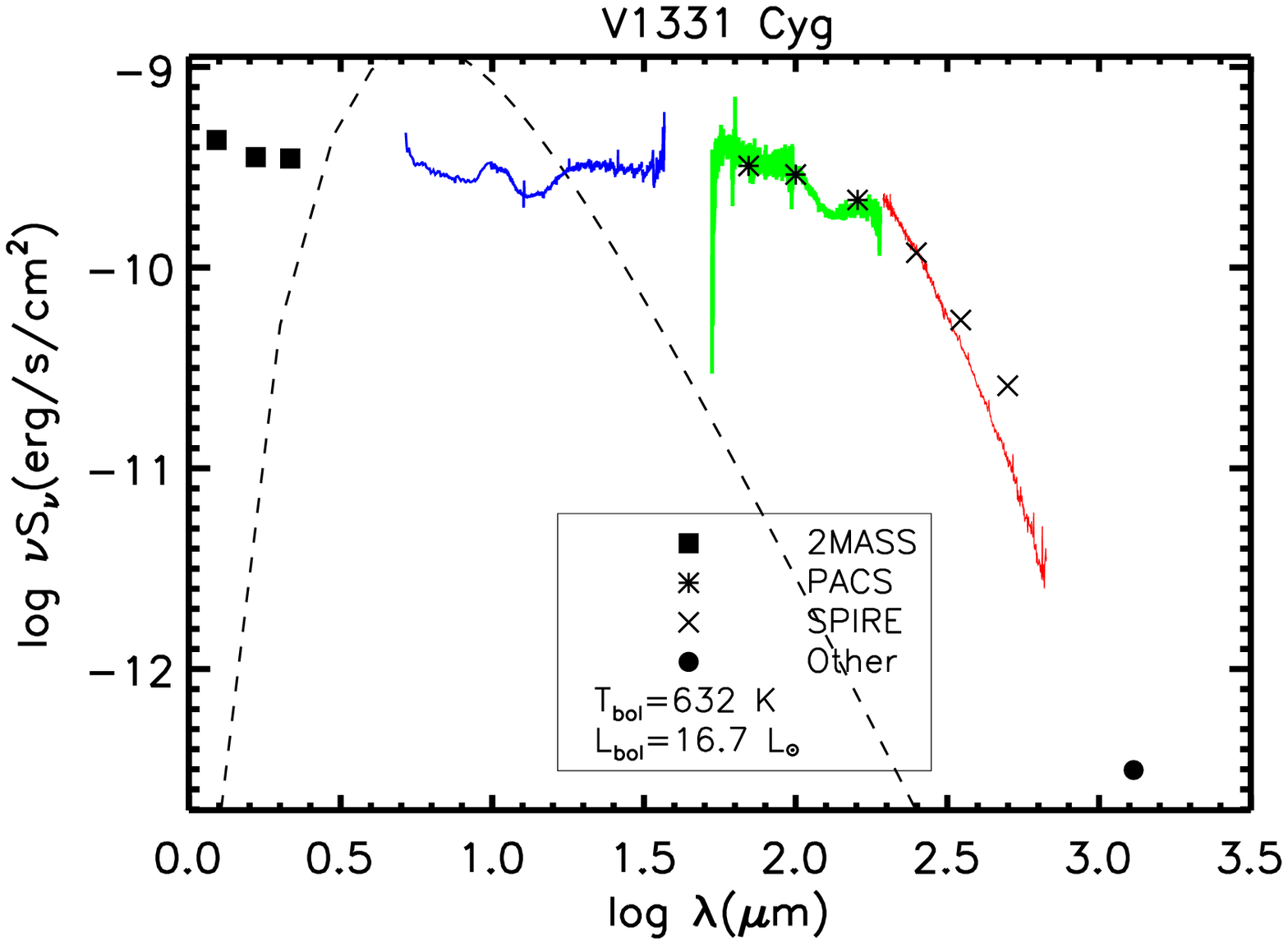} 
\includegraphics[scale=0.4, angle=0]{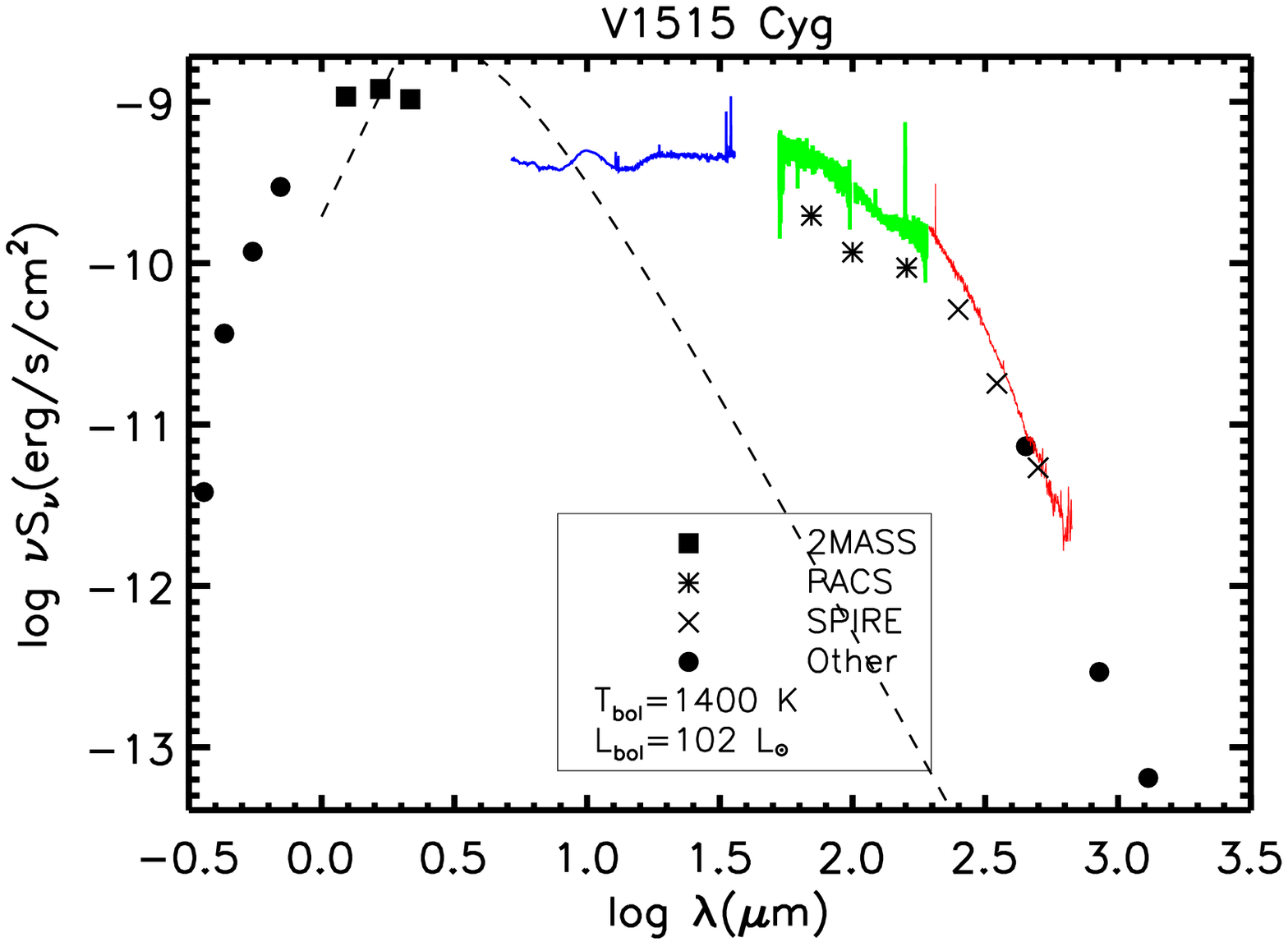} \\
\includegraphics[scale=0.4, angle=0]{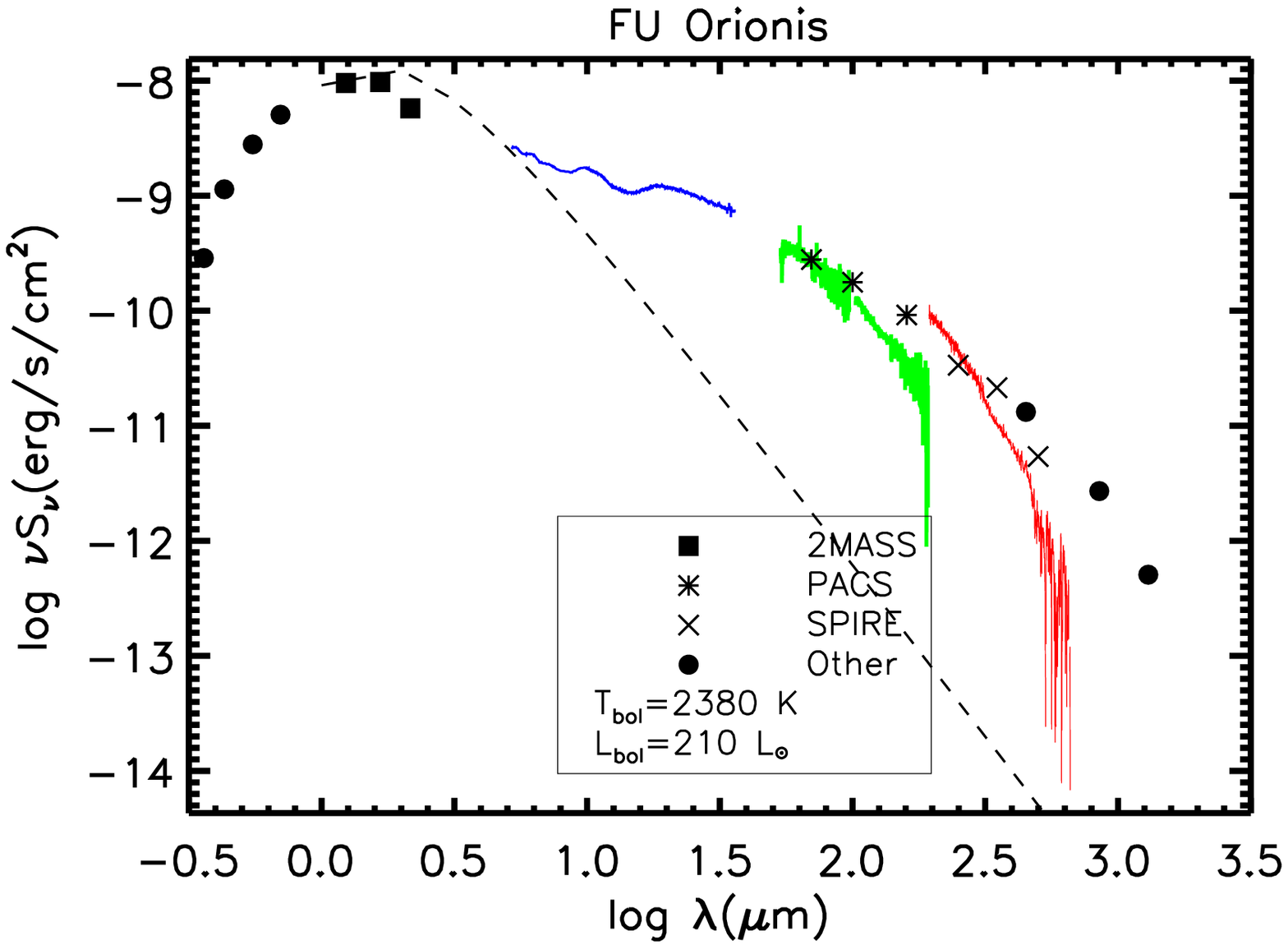} 
\includegraphics[scale=0.4, angle=0]{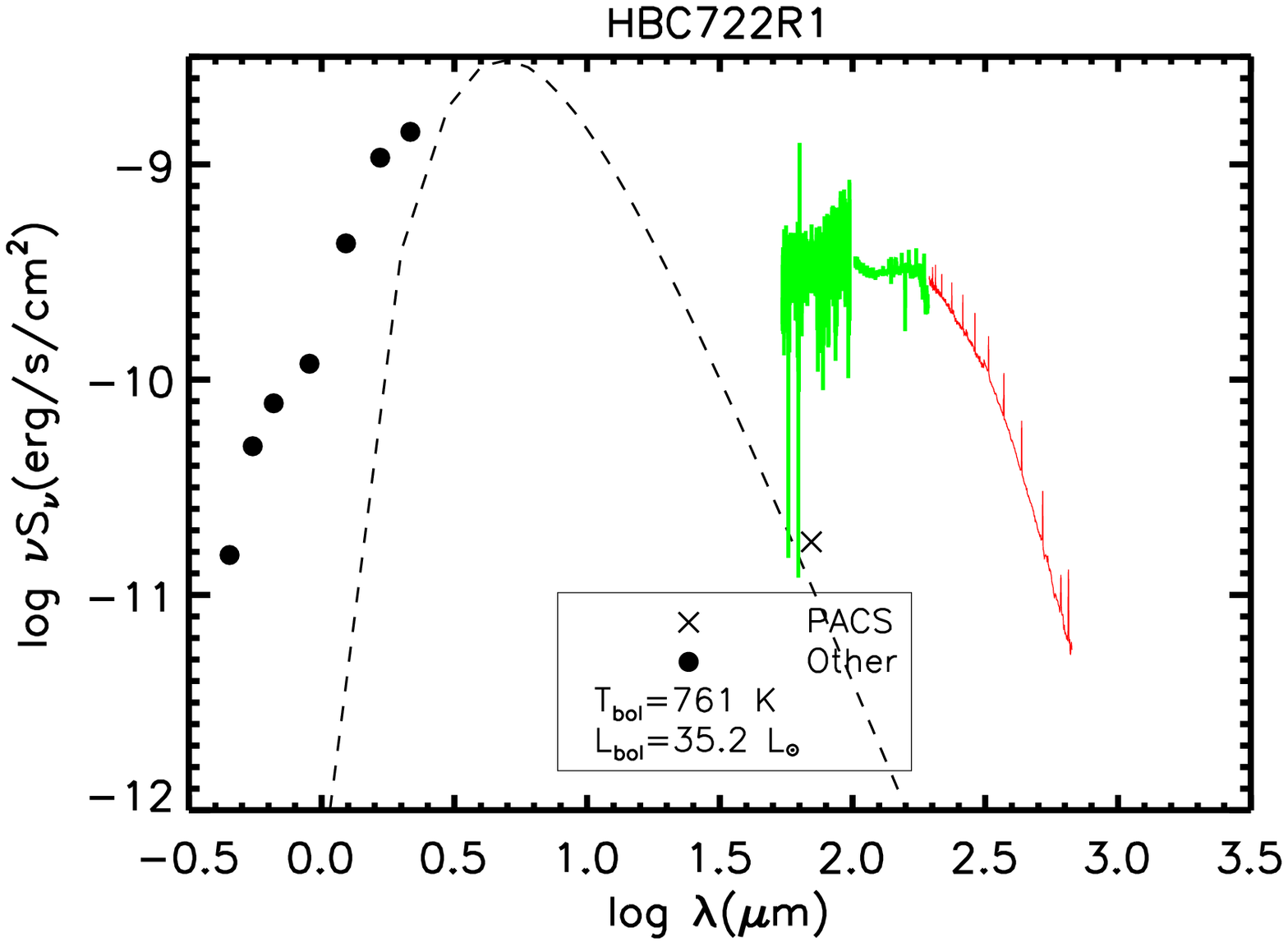} \\
\caption{0.5-1000 $\mu$m SEDs.
Both the UBVR (Maidanak Observatory)  photometric points and IRS spectra are circa 2005,  analytically dereddened \citep{mathis90} in the cases of FU Ori (A$_V$ $=$ 1.8), V1515 Cyg (A$_V$ $=$ 3.0) and V1057 Cyg (A$_V$ $=$ 3.5).  The other sources were not corrected for extinction.  The data is taken from \citet{green06} and \citet{quanz07a}.  The near-IR points are primarily from 2MASS (JHK; filled squares; circa 2000) and the dereddening correction was again applied.  The submm (`*') points are PACS 70, 100, and 160 $\mu$m, and the (`X') points are SPIRE 250, 350, and 500 $\mu$m (this work; circa 2011-12).  When available we include 0.45-1.3 mm SCUBA data.  The dashed line is a blackbody spectrum with the fitted bolometric luminosity and temperature.  We include the contaminated SED of the HBC 722 composite for comparison (BVRIJHK from \citealt{kospal11}, contemporaneous with our Herschel observations); note that only the optical/near-IR points and PACS 70 $\mu$m are attributed to HBC 722 itself.  The large discrepancy between the PACS photometry and spectroscopy indicates that the PACS spectra are dominated by nearby extended sources.}
\label{spirespec}
\end{center}
\end{figure}

Figure \ref{spirespec} shows continuum SEDs for all five FUors, including UBVR, JHK (2MASS, UKIDSS), Spitzer-IRS spectra (from \citealt{green06,quanz07a}, and this work) and PACS/SPIRE photometry (this work).   For comparison, we also show the incomplete SED of HBC 722 from \citet{green11b}; the PACS/SPIRE spectra for this source are clearly contaminated, as evidenced by the mismatch with the 70 $\mu$m PACS photometric datapoint, and the non-detection at 1.3 mm with the Submillimeter Array \citep{dunham12b}.  

Additionally, the PACS photometry falls below the PACS spectrum for V1735 Cyg and V1515 Cyg.  The former can be explained by contamination in the spectra from V1735 Cyg SM1, which is cleanly separated by the photometry but not by the spectroscopy.  A similar situation is likely in V1515 Cyg, where the spectroscopic data includes additional surrounding nebulosity.

The most notable slope discontinuity between PACS and SPIRE is in the FU Ori SED; the emission at SPIRE wavelengths is faint by comparison to the other sources.  In V1331 Cyg, the longest wavelength SPIRE photometry, at 500 $\mu$m, seems contaminated by extended emission.  In the other, brighter sources, the PACS and SPIRE photometry and spectroscopy are very well-aligned.

As FUors are variable and generally fading over time, we used photometry as close to contemporaneous as possible.  We then calculated \lbol\ and \tbol\ using the technique from \citet{dunham10}.  In summary, \tbol\ is the temperature of a blackbody with 
the same flux-weighted mean frequency as the measured spectrum, and \lbol\ is the total integrated luminosity, using trapezoidal interpolation between datapoints.

\subsubsection{Classification}

The original definition of Class I and II sources for young stellar objects is derived from the ratio of the 2 $\mu$m / 25 $\mu$m flux density \citep{andre93,greene94}, the spectral index $\alpha$.  For HBC 722, there is a pre-outburst measurement of $\alpha$ at -0.77, which would fall under the definition of Class II \citep{miller11}. \citet{quanz07a} used mid-IR spectral indices and found the FUors to be indistinguishable from Class II objects in Taurus \citep{furlan06}.  We compare the classifications of the non-confused FUors in our sample by three different classification methods, explained below, in Table \ref{class}.

In general, \tbol\ can be used as a proxy for evolutionary state \citep{chen95, robitaille06}: in this framework a Class I source has \tbol\ between 70 and 650 K; a Class II source has \tbol\ $>$ 650 K  .  Flat-spectrum sources can be characterized as having 350 K $<$ \tbol\ $<$ 950 K \citep{evans09,fischer13}.  \tbol\ is quite sensitive to the near-IR flux and is thus highly uncertain in variable sources such as FUors.  With that caveat in mind, we find that FU Ori, V1515 Cyg, and V1057 Cyg exhibit Class II SEDs, while V1735 Cyg and V1331 Cyg exhibit typical flat spectrum SEDs, near the Class I/II boundary.  This sequence agrees with the sequence from SED color, extinction, and strength of silicate emission found in \citet{green06} and \citet{quanz07a}.  The lone surprise is V1331 Cyg which might be expected to be the second-most blue/evolved source, given its silicate emission feature and flat/slightly declining mid-IR SED.   In summary, the \tbol\ values suggest that FUors are more evolved than Class I sources.  Unfortunately, the value of \tbol\ as an evolutionary proxy is debatable.  In the case of an FUor flare, models suggest that the far-IR luminosity will increase within a few months of outburst \citep{johnstone13}, although \citet{abraham04} find, in contrast, that  changes to the SED occur primarily in the optical/IR rather than the submillimeter on the 1-10 yr timescale.  If the far-IR lagged significantly behind, this would increase \tbol\ without any change in the submillimeter SED or derived envelope mass.  However, this scenario would require the dust to be 1-10 ly (0.3 - 3.0 pc) distant, and hence unrelated to the FUor.

Thus we also consider a criteria related to the physical stage of the object, using the Herschel spectra.  We calculate the envelope mass using the ``OH5'' opacity law \citep{ossenkopf94} as described in e.g., \citet{dunham11}.  First we determine an approximate dust temperature from the shape of the PACS and SPIRE continuum from 100-600 $\mu$m.  We compare dust at this temperature, assuming an emissivity $\alpha$ $\lambda^{-1.8}$, normalized to the 350 $\mu$m continuum, to confirm that the model slope is similar to the observed spectrum.  Then we calculate an envelope mass from the SPIRE photometry.  The results for the four isolated sources in our sample are shown in Table \ref{envmass}.    Using this criteria, \citet{crapsi08} consider 0.1 $\msun$ as the envelope mass cutoff between Stage I and II sources.   Their boundary value was derived from \citet{robitaille06} models, converting their boundary accretion rate value of 10$^{-6}$ $\msunyr$ to 0.07 $\msun$, with the additional consideration that disks greater than 0.1 $\msun$ were gravitationally unstable.  However, there are two dominant sources of uncertainty in this calculation.  First, the dust temperature is not perfectly constrained, and varies between 38 and 54 K, causing an uncertainty in the derived envelope mass of $\sim$ 35\%.  Second, the estimated mass increases significantly if derived using longer wavelengths, due to the inclusion of increasingly larger areas of colder dust from the more distant parts of the envelope, suggesting that values calculated at 350 $\mu$m generally underestimate the envelope mass.

The results for each classification method are summarized in Table \ref{class}.  The $\alpha$ and \tbol\ classifications agree very well, while the envelope mass method indicates an earlier evolutionary stage than the other methods in the cases of V1057 Cyg and V1515 Cyg. 
If we crudely average the classifications of the three different methods (assigning values of 1, 1.5, and 2 to Class/Stage I, Flat Spectrum, and Class/Stage II, respectively), we find the following sequence in increasing order of Class/Stage: V1057 Cyg (1.5), V1515 Cyg (1.67), V1331 Cyg (1.67), V1735 Cyg (1.75), and HBC 722 (2.0).  This ordering is broadly in agreement with the mid-IR dust feature sequence from \citet{quanz07a}.  Therefore, we find that the umbrella group of ``FUors'' includes both Stage I and II objects.

\subsubsection{Bolometric luminosity}

We also update \lbol\ for our sample.  First it is important to consider the observational epochs of the data in our SEDs: the optical/near-IR data has been updated to match the 2005-2006 {\it Spitzer} spectra; the {\it Herschel} data are from 2011.  Thus the computed \lbol\ values are upper limits to the current \lbol, if all are declining.  \citet{abraham04} found that the flux for $\lambda$ $>$ 20 $\mu$m held constant in time; thus the SEDs including the {\it Herschel} data may closely represent the SED from the 2005-6 timeframe.  However, \citet{johnstone13} find that while the wavelength-dependent variability during outburst is led on day timescales at mid-IR and shorter wavelengths, at wavelengths close to 100 $\mu$m the effect on the SED occurs over months, and the $\lambda$ $\lesssim$ 200 $\mu$m {\it Herschel} spectra could be sensitive to changes.

The addition of the Herschel data affects the derived \lbol\ and \tbol\ systematically, but this is overshadowed by a larger effect attributable to beamsize.  First, we compare the SEDs with and without the {\it Herschel} data: including PACS/SPIRE data along with ancillary photometry results in larger \lbol, and smaller \tbol, by 10-20\% for each source, compared to the values derived only from the ancillary photometry. The reason for this systematic increase in \lbol\ is that the broad extrapolation from Spitzer (35 $\mu$m) data to the 850 $\mu$m or 1.1 mm data, in the absence of the {\it Herschel} data, slightly understates the total flux.  

However, the beamsize of far-IR (IRAS) data used in older calculations of \lbol\ likely caused significant overestimation of the bolometric luminosity.  Comparing our measurements to earlier values of \lbol, V1057 Cyg peaked at 800 \lsun\ at peak and dropped to 250 \lsun\ \citep{kenyon91}, but we calculate an \lbol\ of 119 \lsun.  The large beamsize of the far-IR data used in the \citet{kenyon91} estimate causes a significant overestimate: we calculated 
\lbol\ for V1057 Cyg using only the optical/near-IR and IRAS fluxes from \citet{kenyon91} and found an \lbol\ of 215 \lsun, comparable to their derived value of 250 \lsun.  This is much greater than the 119 \lsun\ we calculate using the {\it Herschel} fluxes; thus we conclude that the  discrepancy is due to the greater fluxes at 25 and 60 $\mu$m from IRAS compared with PACS.  In the case of FU Ori, we observed a similar but smaller effect; although the IRAS fluxes are a factor of 2-3 higher than the PACS fluxes, \lbol\ only changes from 274 to 210 \lsun\ with the updated data.  FU Ori is a much bluer object than V1057 Cyg, and thus its \lbol\ is less sensitive to changes in submillmeter flux.   Whether this is a beamsize effect or an actual decrease in luminosity is not clear, in either source.

With these caveats in mind, we find that FU Ori is currently the brightest source at 210 \lsun.   V1057 Cyg, V1515 Cyg, and V1735 Cyg all show \lbol\ $\sim$ 100-120  \lsun, while V1331 Cyg is much fainter, at 16.7 \lsun.  In general, we find lower \lbol\ than previously reported values in the literature, but we cannot separate the large uncertainty in older measurements of far-IR/submm flux from an actual decrease in the far-IR/submm SED.  Resolving this ambiguity is beyond the scope of this work but can be addressed with multi-epoch submillimeter continuum measurements.

\subsection{Non-Detection of Dust and Ice Features}

We do not observe any evidence for cold crystalline Mg-rich or Fe-rich dust at 69 $\mu$m, which has been seen in some T Tauri and Herbig Ae/Be star disks \citep{sturm10,sturm13}.  However, our sensitivity is substantially lower than in sources with detections of this feature, due to shorter integration times, and greater distances to the source (lower S/N).  Furthermore, the sources in \citet{sturm13} typically show crystalline silicate dust in the IRS bands, whereas all of the FUor IRS spectra are pristine \citep{quanz07a}.  At this time, calibration uncertainties in spectral shape prevent a search for 60-70 $\mu$m H$_2$O crystalline ice features.

\subsection{Line Detections}

For the DIGIT embedded objects we used separate extractions to determine line spatial extent as distinct from continuum \citep{green13b}.  This was unnecessary for the FOOSH sources, which show no evidence of extended line emission, except in cases of contamination.  
Differences in the continuum and line extent resolved with {\it Herschel} in FUors is attributed to large scale structure, multiplicity, or outflows and is discussed in the following section.  For this work we follow a slightly simpler procedure than that used in \citet{green13b}, and assume the line and continuum emitting regions are the same, and apply the same scaling factors.

The list of detected lines in PACS and SPIRE for all six source regions is reported in Table \ref{pacsspirelines}.  In summary, the PACS observations show few lines overall, while the SPIRE spectra are richer.  The only lines  detected in all six sources are \OI\ 63 $\mu$m and \CI\ 370 and 610 $\mu$m.  Other observed fine structure lines include \OI\ 145 $\mu$m, \NII\ 205 and 122 $\mu$m, and \CII\ 158 $\mu$m.  We detect CO \jj{5}{4} and \jj{4}{3} in all but FU Ori (although CO \jj{5}{4} is detected only in the HIFI data, below the detection threshold of SPIRE due to the narrow feature width).   CO \jj{7}{6} and higher are seen only in V1057 Cyg, V1735 Cyg, and HBC 722; in V1057 Cyg we detect CO up to \jj{23}{22}.  We detect $^{13}$CO \jj{8}{7} to \jj{5}{4} in V1735 Cyg and HBC 722.  We detect H$_2$O 174.6 $\mu$m, and tentatively detect OH 84.41 $\mu$m, in one source (V1057 Cyg).  OH 119 $\mu$m, the lowest state of the OH 3/2 $\rightarrow$ 3/2 ladder, is the only line in absorption (seen in V1735 Cyg).

\begin{figure}
\includegraphics[scale=1.2]{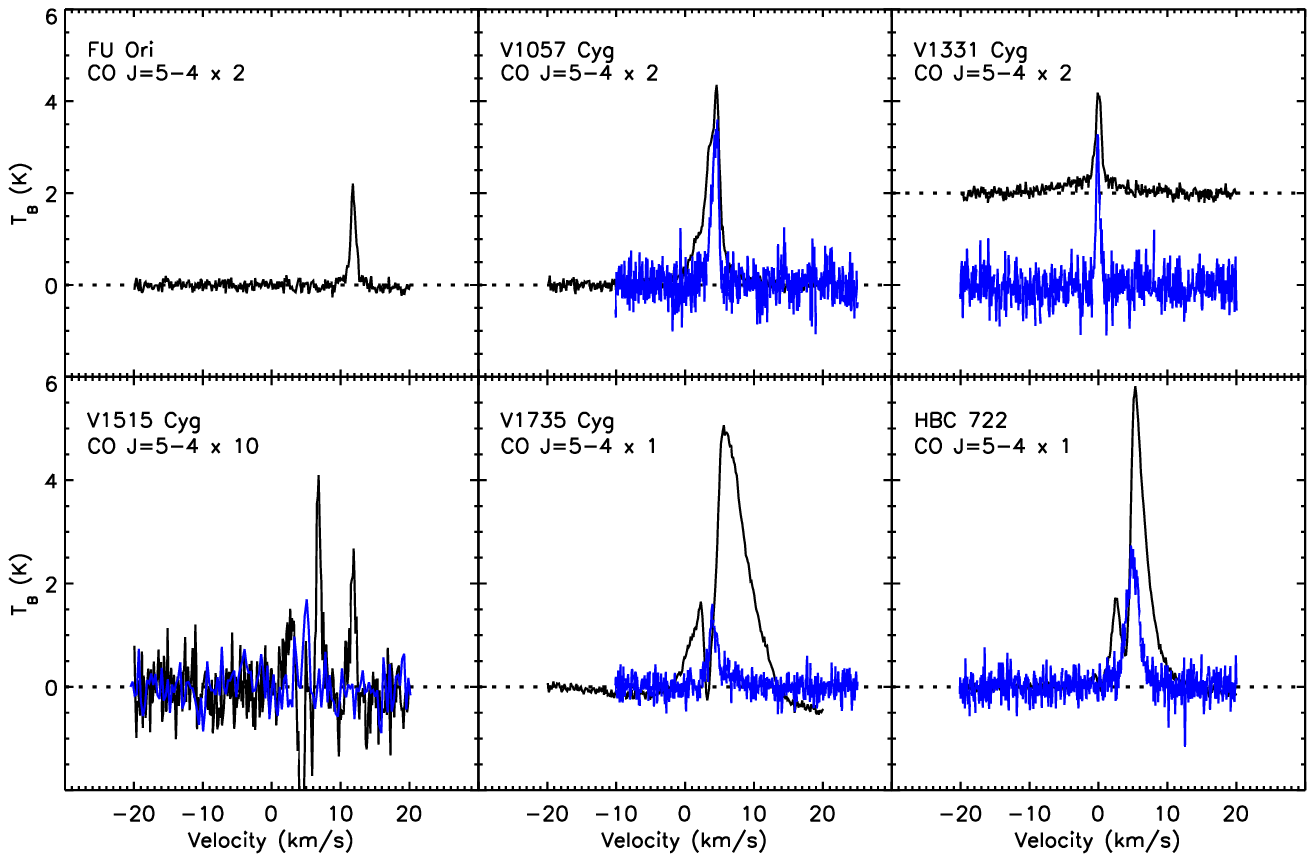}
\includegraphics[scale=1.2]{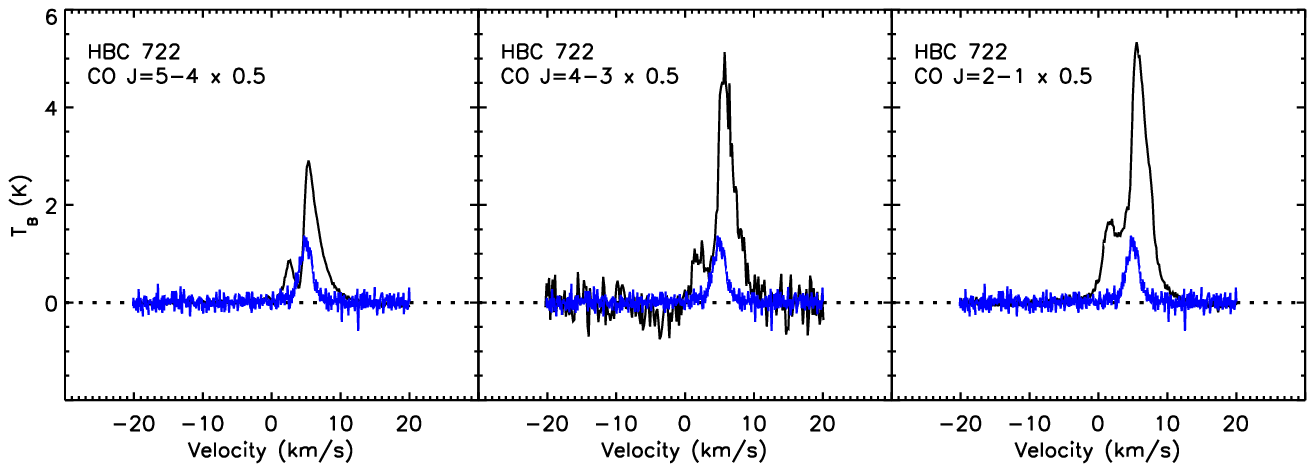}
\caption{{\bf Top:} HIFI observations of CO \jj{5}{4} (black) and HCO$^+$ \jj{3}{2} (blue).  The dashed horizontal line indicates the baseline offset; the CO \jj{5}{4} in V1331 Cyg is offset for clarity.  {\bf Bottom:}  CO \jj{5}{4} (HIFI),  \jj{4}{3} and \jj{2}{1} (CSO) vs. HCO$^+$ \jj{3}{2}, for HBC 722 .  Note that in each subfigure the black spectrum is scaled by the factors listed in its respective title.}
\label{co54}
\end{figure}

None of the sources was detected in CO \jj{14}{13} with HIFI, but all were detected in CO \jj{5}{4}.  Figure \ref{co54} (top) shows the CO \jj{5}{4} HIFI (black) compared with the HCO$^+$ \jj{3}{2} (267.558 GHz; blue) line observed with the CSO.  FU Ori was not observed in HCO$^+$ in our observing run; all of the others --  V1515 Cyg, V1735 Cyg, V1057 Cyg, V1331 Cyg, and HBC 722 -- were observed to emit in the HCO$^+$ \jj{3}{2}.  In the case of HBC 722, we also observed CO \jj{4}{3} and \jj{2}{1}.  The line profiles are compared in Figure \ref{co54} (bottom), placed on the same vertical scale for comparison.  (Note that the CO integrated intensity is scaled by the factors listed in each subfigure.)

\begin{figure}
\includegraphics[scale=0.75]{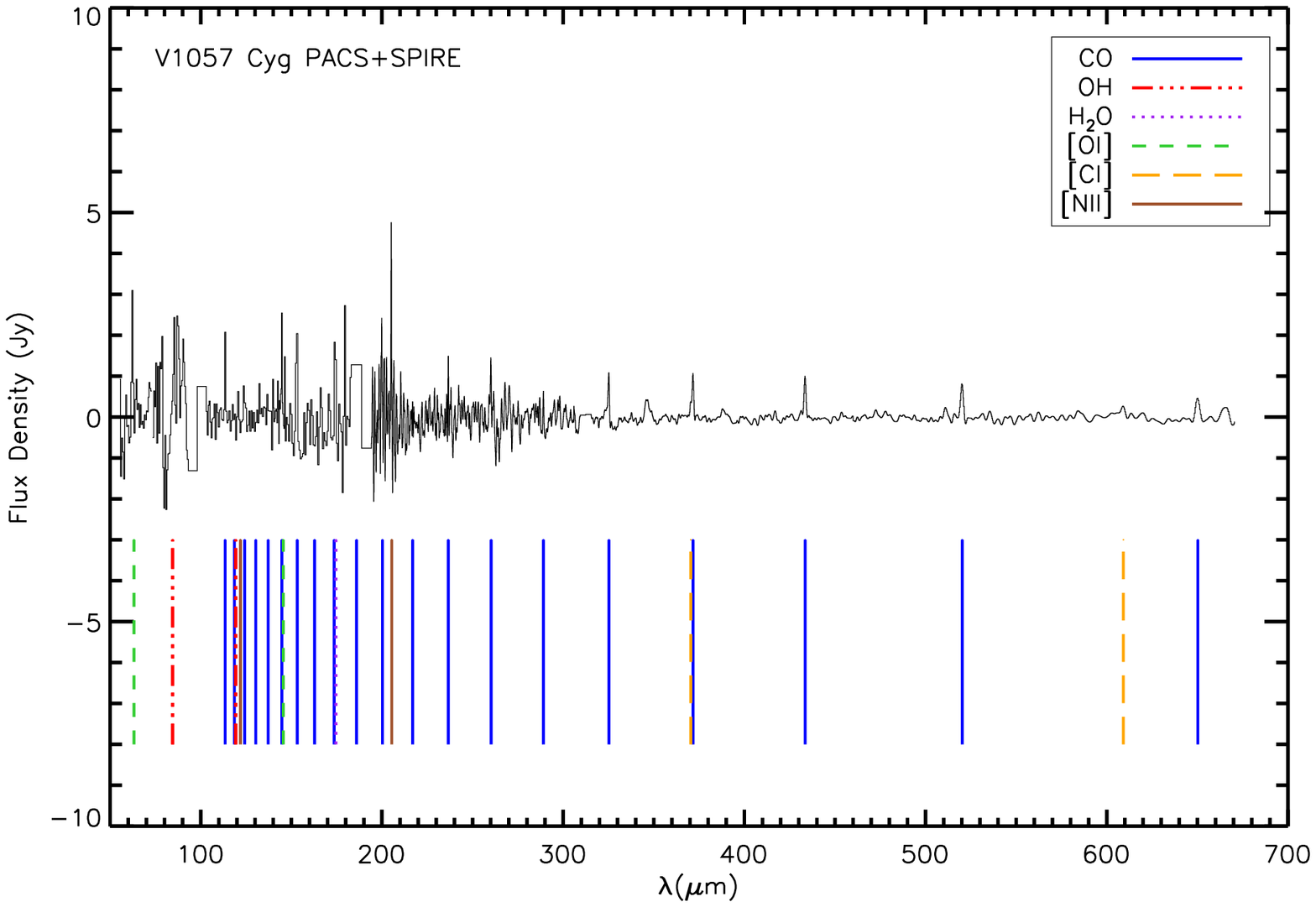}
\includegraphics[scale=0.75]{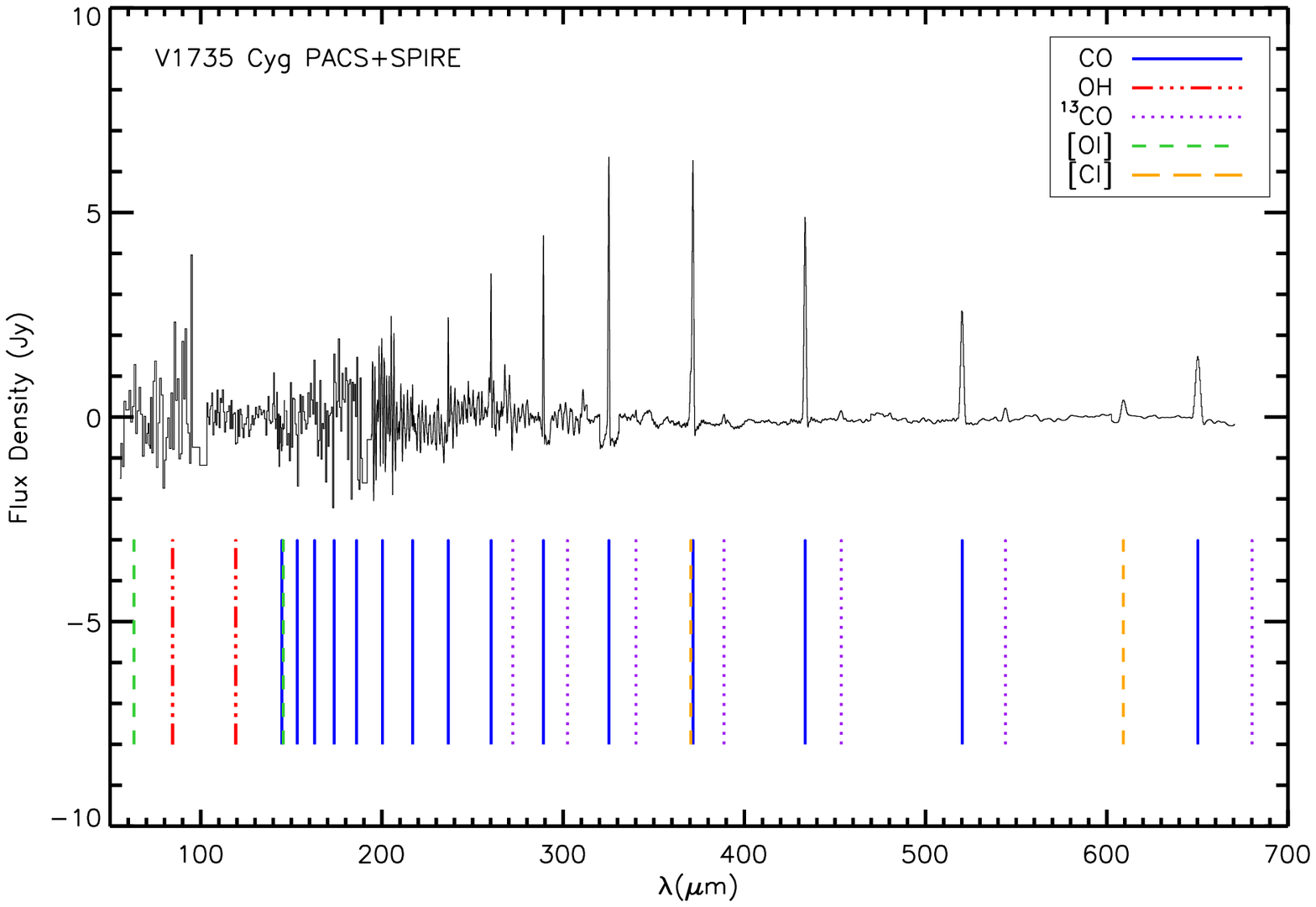}
\caption{{\bf Top:} Continuum-subtracted PACS/SPIRE spectrum of V1057 Cyg, rebinned to lower resolution for clarity.  {\bf Bottom}: Continuum-subtracted spectrum for V1735 Cyg, contaminated by V1735 Cyg SM1.}
\label{v1057}
\end{figure}

As an exemplar, in Figure \ref{v1057} we present the 50-670 $\mu$m continuum-subtracted spectrum of V1057 Cyg (top), rebinned to low resolution for clarity, vs. V1735 Cyg (bottom), which is partly contaminated by V1735 Cyg SM1.  In this form, a few details become apparent.  The peak CO lineflux appears at much longer wavelengths in V1735 Cyg, compared to V1057 Cyg, or to a  typical embedded source (e.g. Fig. 15, \citealt{green13b}).  The faint $^{13}$CO ladder appears from \jj{5}{4} to \jj{8}{7} only in the case of V1735 Cyg. The only detection of H$_2$O (at 174.6 $\mu$m) is in V1057 Cyg (the stronger nearby line is CO \jj{15}{14} at 173.6 $\mu$m).  V1735 Cyg shows a relatively weak \OI\ feature in comparison to the CO.

\begin{figure}
\includegraphics[scale=0.75]{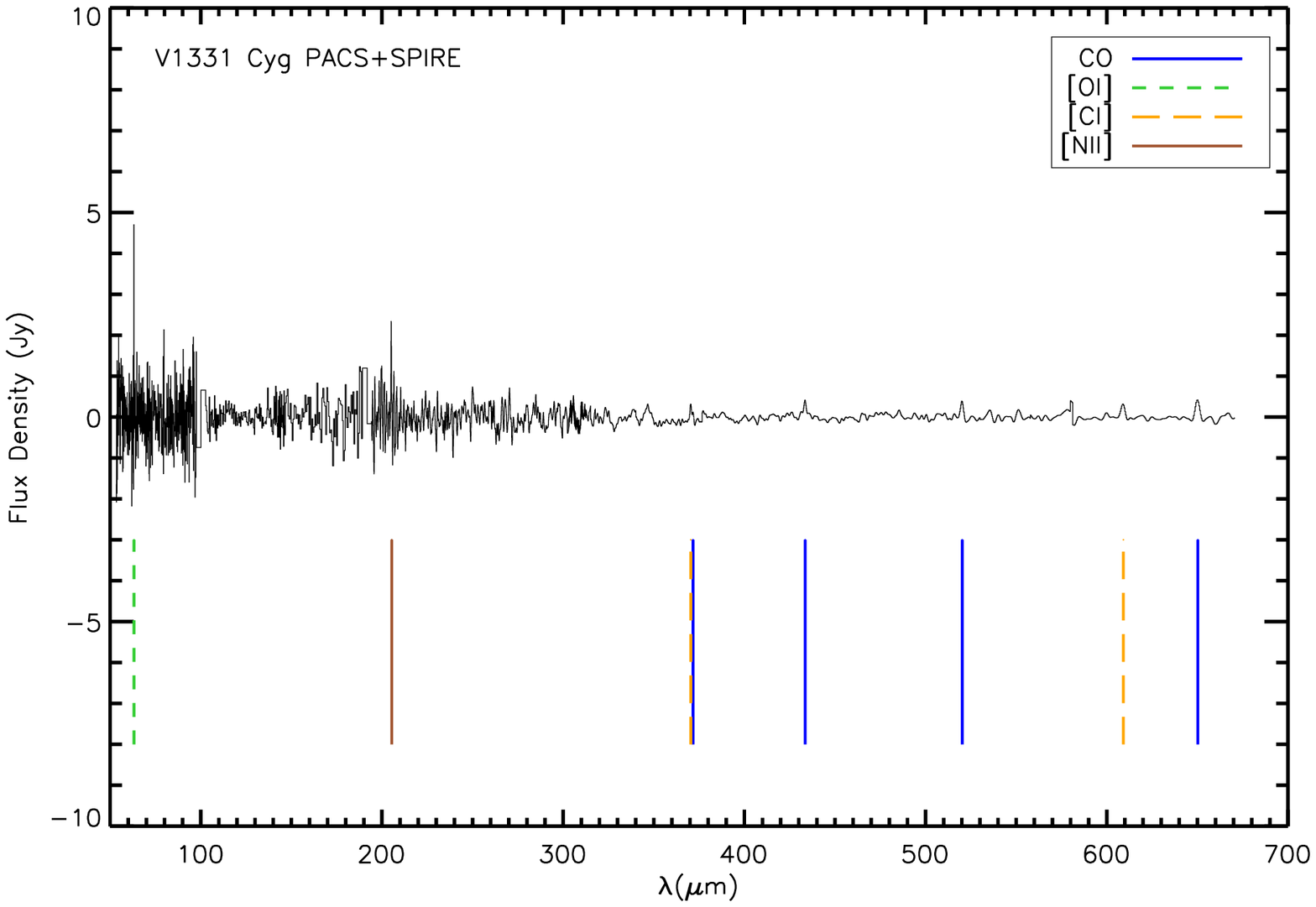}
\includegraphics[scale=0.75]{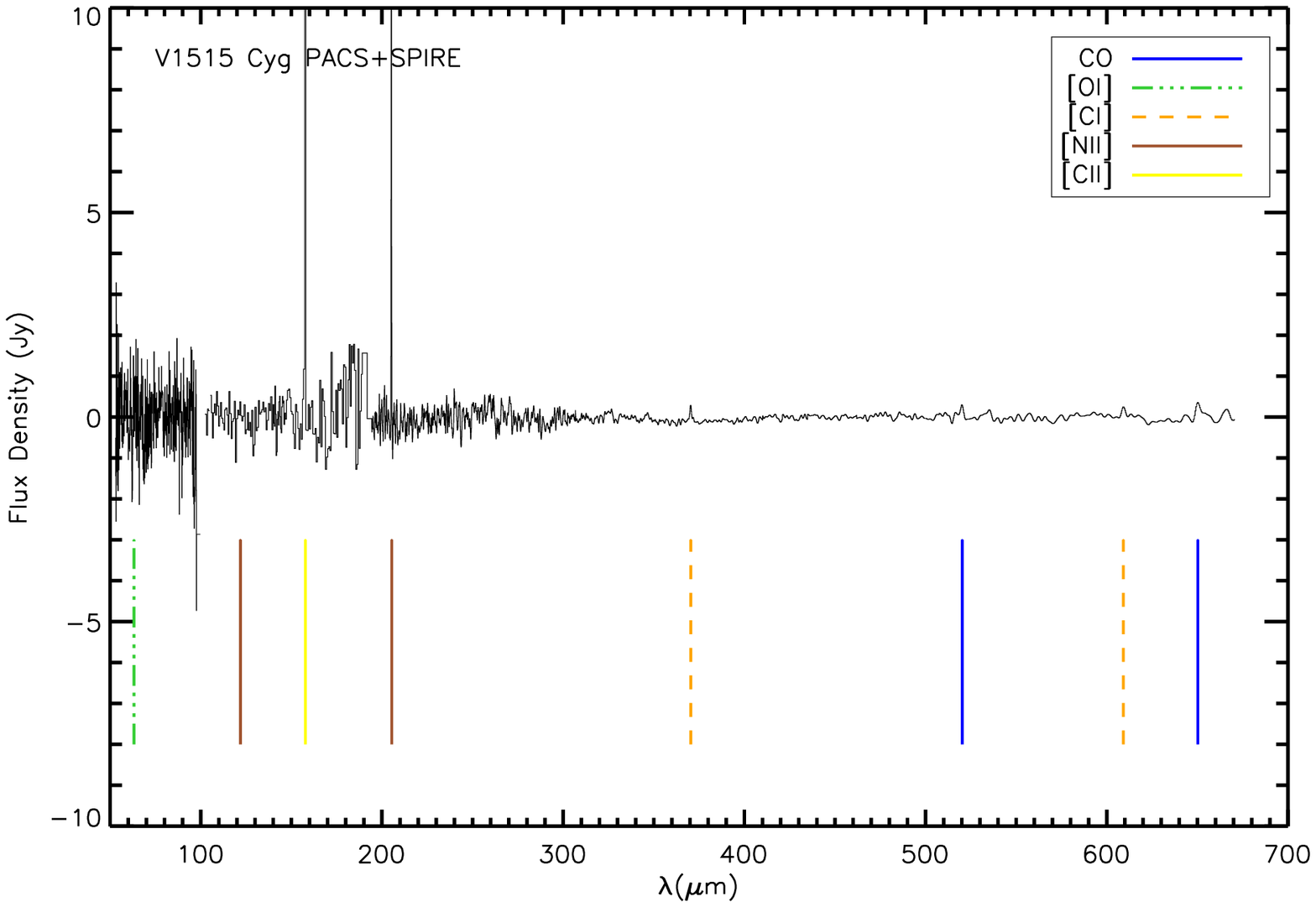}
\caption{{\bf Top:} Continuum-subtracted PACS/SPIRE spectrum of V1331 Cyg, rebinned to lower resolution for clarity.  {\bf Bottom}: Continuum-subtracted spectrum for V1515 Cyg.}
\label{flat2}
\end{figure}

\begin{figure}
\includegraphics[scale=0.75]{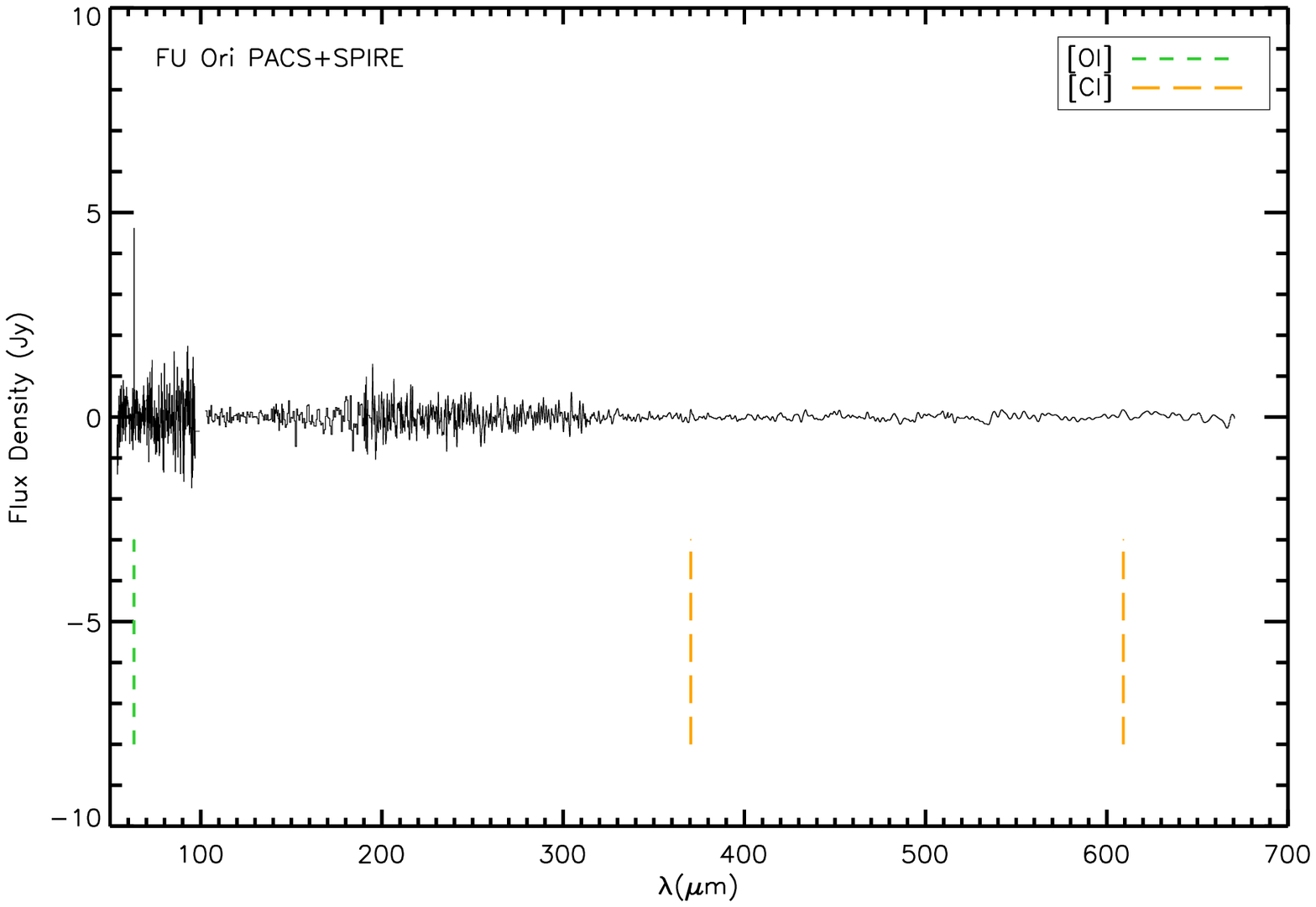}
\includegraphics[scale=0.75]{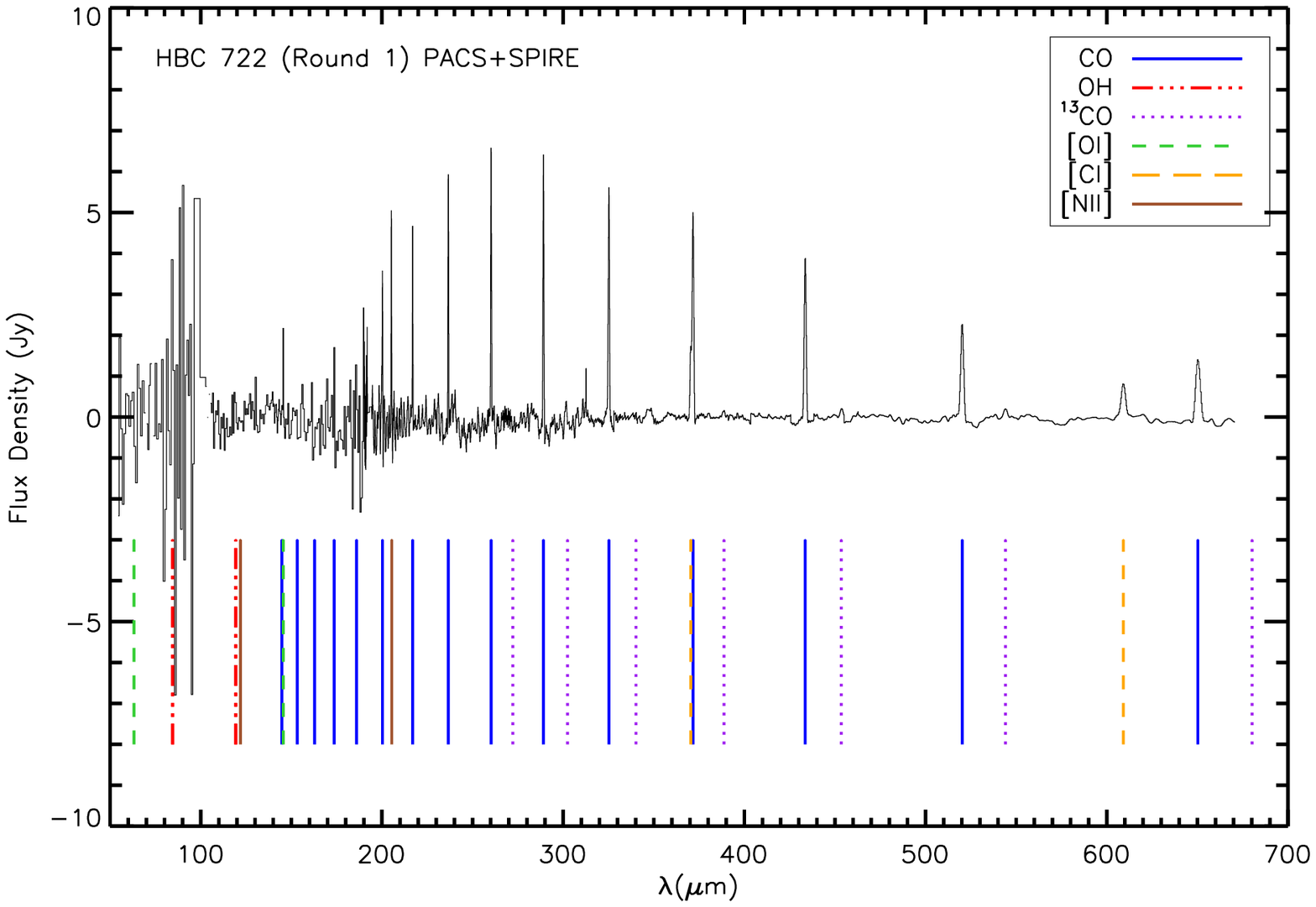}
\caption{{\bf Top:} Continuum-subtracted PACS/SPIRE spectrum of FU Ori, rebinned to lower resolution for clarity.  {\bf Bottom}: Continuum-subtracted spectrum for HBC 722 (taken during December 2010), contaminated by nearby Class 0/I protostars.}
\label{flat4}
\end{figure}

The remaining FUor spectra (including the contaminated HBC 722 from \citealt{green11b}) are shown in Figures \ref{flat2} and \ref{flat4}.

\subsection{Spatial Distribution of Continuum and Lines}

Although we detect substantial line emission from three sources (V1057 Cyg, V1735 Cyg, and HBC 722), we must consider whether the lines are local to each FUor, using the spatial information provided by the PACS array.  It is more difficult to determine this with SPIRE due to the larger beam, sparse spatial coverage, and lack of off-position data.

\subsubsection{PACS}

\begin{figure}
\includegraphics[scale=1.0]{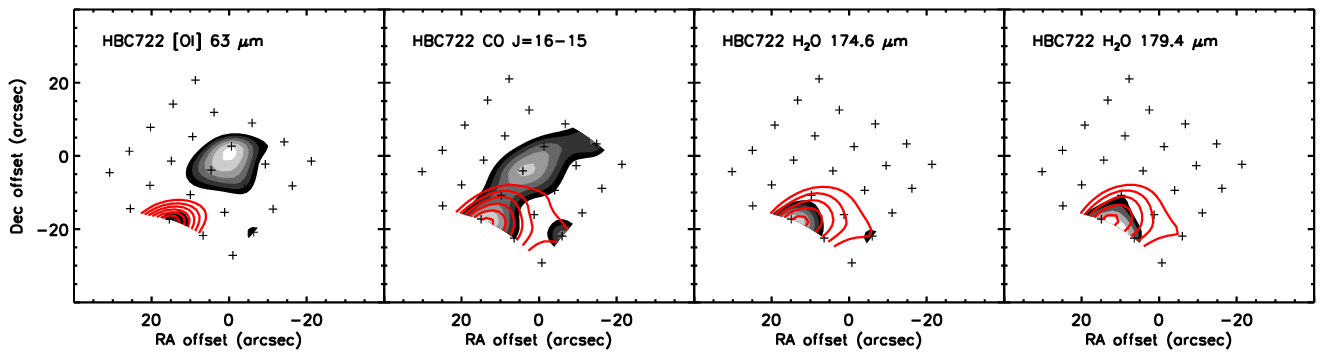}
\includegraphics[scale=1.0]{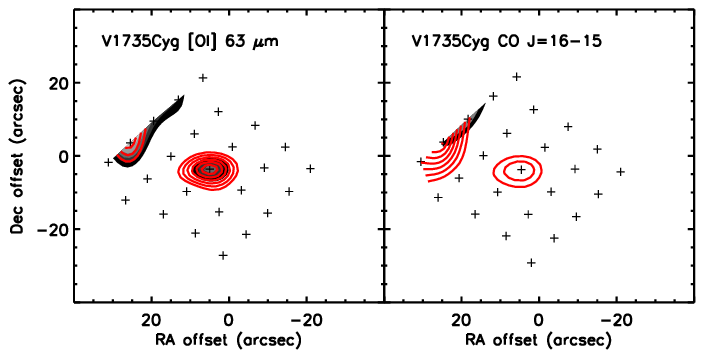}
\includegraphics[scale=1.0]{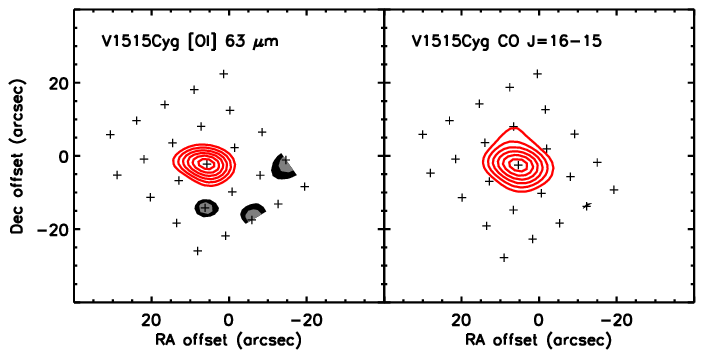}
\caption{Spatial distribution of PACS lines (grayscale) vs. local continuum (red); the `+' indicate the spaxel positions.  The 
contours are in increments of 10\% from the peak flux (usually in the central 
spaxel), plotted down to the noise limit.  The noise limit is computed as 2$\times$ 
the average flux of the outer ring of 16 spaxels, with the exception of the V1735 Cyg 
and HBC 722 maps, for which we avoid the edge spaxels containing a likely Class 0/I protostar (respectively, V1735Cyg SM1 and 2MASS 20581767+4353310).}
\label{spatial1}
\end{figure}

\begin{figure}
\includegraphics[scale=1.0]{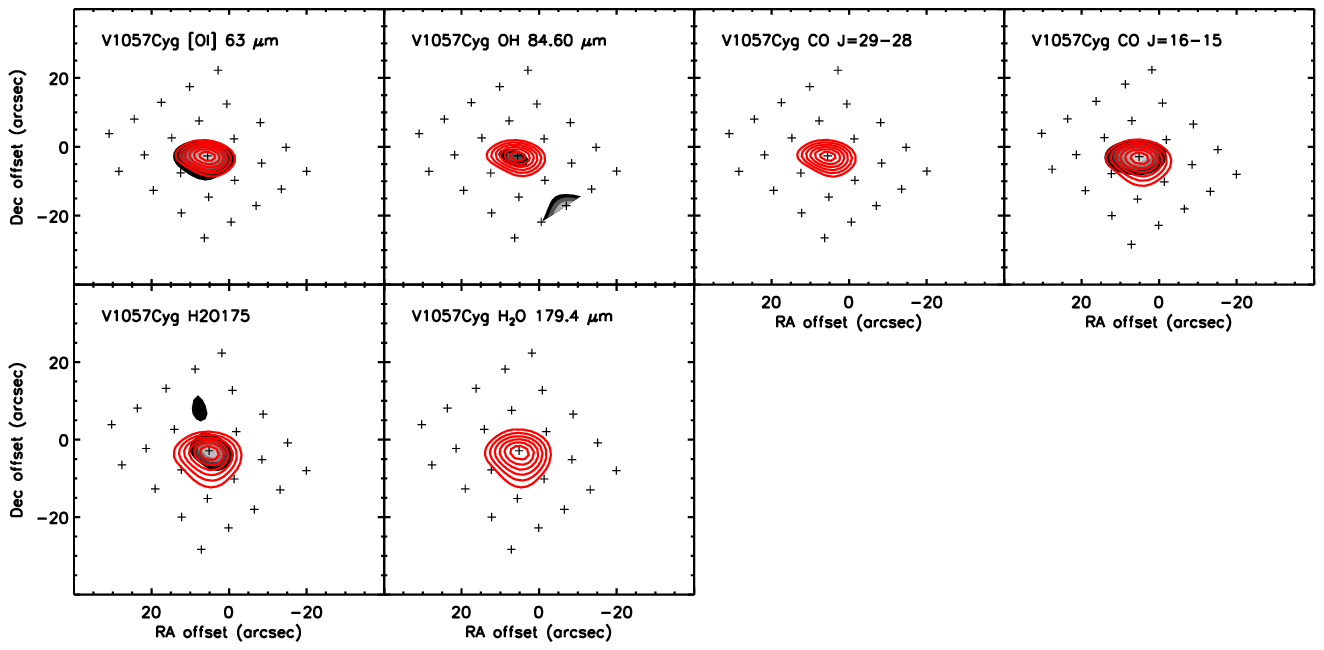}
\includegraphics[scale=1.0]{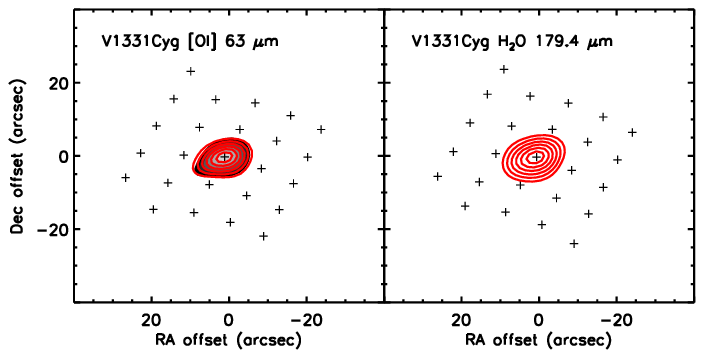}
\includegraphics[scale=1.0]{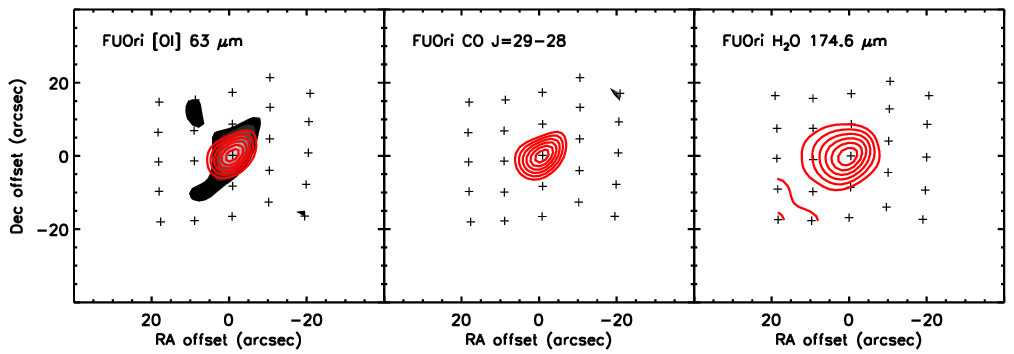}
\caption{See Figure \ref{spatial1} for a description of this figure.}
\label{spatial2}
\end{figure}

Figures \ref{spatial1} and \ref{spatial2} show 10\% contours for all detected lines (grayscale) and local continuum (red), with PACS.  The local continuum is selected from a line-free set of channels surrounding each line.  In HBC 722, we see a clear demarcation between the FUor (central spaxel) and the embedded source (two spaxels to the SE).  The submm continuum is strongly dominated at all wavelengths by the embedded source.   For V1735 Cyg, V1735 Cyg SM1 appears as a continuum source cleanly detected two spaxels to the east, at the edge of the array.  In the other four sources (V1515 Cyg, V1057 Cyg, V1331 Cyg, and FU Ori) the central spaxel dominates the continuum emission, and the pointing accuracy can be estimated as within 0.2 spaxels from the centroid at each wavelength.  The continuum footprint grows with wavelength, consistent with the point-spread function \citep{green13b}; thus there is no evidence for extended PACS continuum emission in these four sources.

The line emission differs from the continuum profiles in a few cases.  In the HBC 722 composite, the \OI\ 63.18 $\mu$m emission is co-spatial with the FUor, while the CO \jj{16}{15} emission extends along both sources in a SE-NW band roughly correlated with the CO \jj{2}{1} map \citep{green11b}.  The marginally detected H$_2$O emission at 179.53 $\mu$m is associated with the embedded object to the SE.   We attribute the \OI\ to the FUor, although we note that the \OI\ could be excited by shocks driven by other millimeter sources identified in SMA observations \citep[Fig. 1, MMS1 and MMS2;][]{dunham12b}.   In V1735 Cyg, the \OI\ emission is equally prominent around both the FUor and the SM1 source, but CO \jj{16}{15} is compact and associated only with SM1.  In V1057 Cyg, V1515 Cyg, V1331 Cyg, and FU Ori, the \OI\ and CO emission is compact and aligned with the FUor, when detected (although there is a hint of extended emission in FU Ori \OI).  There is very little line emission detected in V1515 Cyg at all.   \NII\ 122 $\mu$m is detected in off-center spaxels in V1515 Cyg, suggesting that the emission is diffuse.  \OIII\ (88.35 $\mu$m) and \NIII\ (57.0 $\mu$m) appear only in absorption, indicating emission lines in the off-position.  In HBC 722, the rotation of the off-position between our two data epochs dominated change in flux of the \CII, \OIII, and \NIII\ absorption features, clear evidence that they derive from the off-position and are not associated with the sources in the PACS field.

\subsubsection{SPIRE}

 We were able to utilize software within HIPE to determine the contributions from individual pixels of the SPIRE-FTS detector array.  The spacing between pixels in sparse mode is $\sim$ 33$\arcsec$ and 51$\arcsec$ for the short (SSW) and long wavelength (SLW) modules, with 37 and 19 hexagonal detectors respectively. A simple analysis of the spatial distribution of bright lines reveals some interesting trends.

The \NII\ 205.4 $\mu$m line appears to be diffuse and present in nearly every pixel surrounding the HBC 722, V1515 Cyg, and V1331 Cyg fields.  There is some hint of a central peak around V1057 Cyg.  \NII\ is absent in the FU Ori and V1735 Cyg fields.  As SPIRE uses an onboard calibrator, and no off-position, we attribute the \NII\ emission to diffuse background emission in Cygnus.  This is consistent with fine structure emission (\SIII, \Sitwo) seen in IRS spectra of these sources prior to off-position subtraction \citep{green06}.

The mid-$J$ CO emission is much more compact, with the low-lying ($J_{\rm up} < 9$) states of CO originating in the central pixel for V1515 Cyg, V1057 Cyg, V1331 Cyg, and FU Ori, the sources which do not suffer from confusion. 

\subsubsection{Summary of Results}

Both HBC 722 and V1735 Cyg are contaminated by nearby protostars at longer wavelengths.  In the case of HBC 722, the \OI\ line flux appears compact and local to the FUor; the CO is either extended in the foreground, or a chain of emission knots including HBC 722; the rest of the line emission in Table \ref{pacsspirelines} and all continuum emission originates from the spaxels containing 2MASS 20581767.  In V1735 Cyg, the contributions of the two sources are comparable at PACS wavelengths.  The \OI\ and PACS continuum are separable from V1735 Cyg SM1.  However, at SPIRE wavelengths they are blended.  Thus we {\it cannot} attribute the CO or $^{13}$CO emission to V1735 Cyg alone.  

In the other four FUors, we attribute all \OI, CO, H$_2$O, and continuum to the FUor itself.
V1057 Cyg is an outlier, as the only non-confused FUor with detectable high-$J$ CO, H$_2$O and OH emission; the other three (V1515 Cyg, V1331 Cyg, and FU Ori) show only strong \OI\ and weak detections of CO up to \jj{6}{5} with no H$_2$O or OH observed.

The final calibrated line fluxes appear in Table \ref{pacsspirelines}.

\section{Analysis}

\subsection{\OI}

HBC 722 and V1735 Cyg are the sources with the strongest  \OI\ 63 $\mu$m lines. They are also the only two sources with \OI\ 145 $\mu$m detections.
The 63/145 $\mu$m \OI\ flux ratios are 21 and 11 respectively.  The ratios fall slightly outside the range of the DIGIT embedded sample (14--20), for HBC 722 and V1735 Cyg.  For the other sources, the lower limit to the 63/145 $\mu$m ratio is 15 (V1057 Cyg), 7.6 (FU Ori), and 1.5 (V1331 Cyg and V1515 Cyg).

\begin{figure}
\includegraphics[scale=0.85]{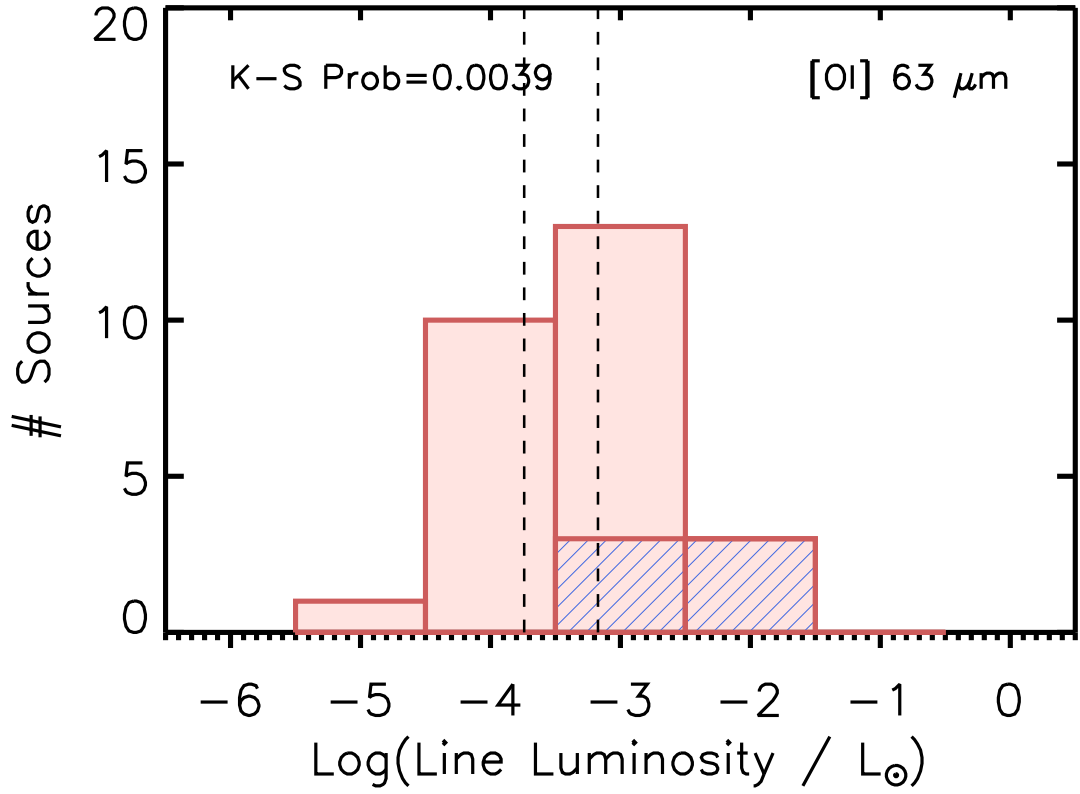}
\caption{Histogram of \OI\ 63 $\mu$m line luminosity for the full DIGIT embedded sample \citep{green13b}.  The FOOSH sample is superimposed in shaded bars.  The probability of the distributions arising from the same source function is included as the K-S probability, and is less than 1\%.  The dashed vertical lines are the detection thresholds for the closest (520 pc) and most distant (1000 pc) sources in the FOOSH sample.}
\label{o1hist}
\end{figure}

Figure \ref{o1hist} shows the distribution of  \OI\ 63 $\mu$m line luminosity for the DIGIT embedded and FOOSH samples.   A K-S test shows that the FUors are significantly different from the protostars as a group.  The average (median) \OI\ luminosity is 4.3 (3.1) $\times$ 10$^{-3}$ $\lsun$ for the FUors, compared to 1.6 (0.5) $\times$ 10$^{-3}$ $\lsun$ for the Class 0/I sources.  If we compare to more evolved sources, the strongest \OI\ emission in any Herbig Ae/Be star in the DIGIT sample is 1.4 $\times$ 10$^{-3}$ $\lsun$, and most Herbig/Class II sources fall below 10$^{-4}$ $\lsun$ (Meeus et al., Fedele et al., subm.), for cases where  the \OI\ line is detected at all.  

\begin{figure}
\includegraphics[scale=0.8]{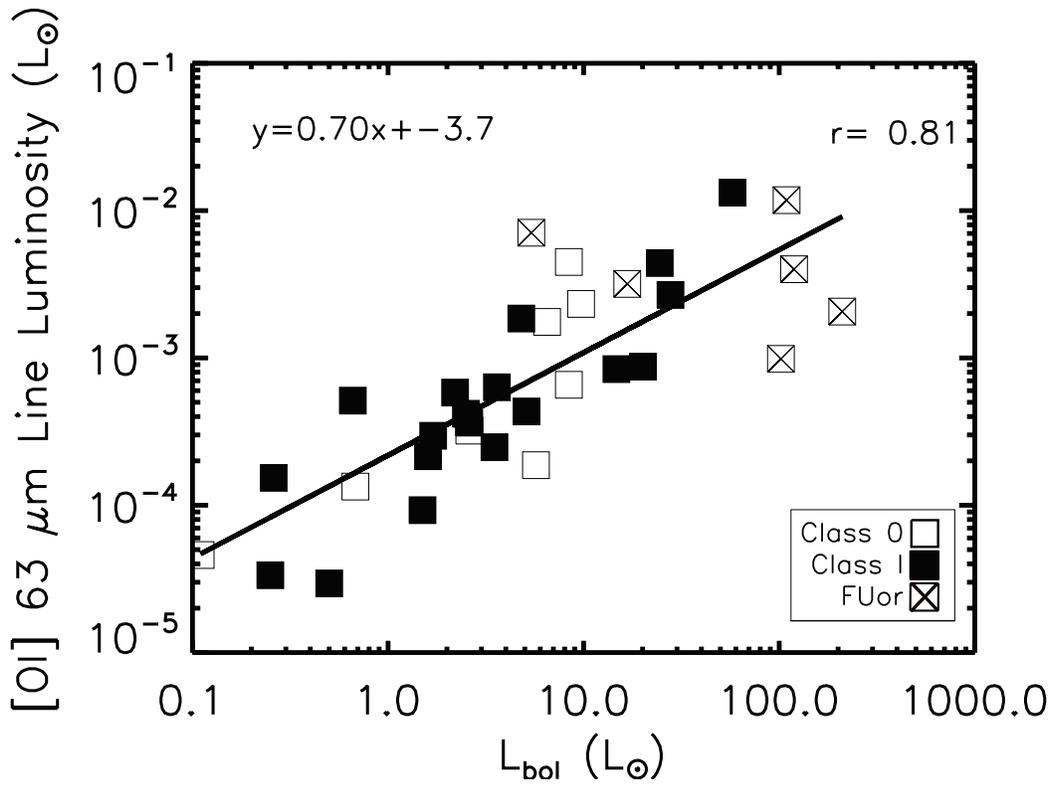}
\caption{\OI\ 63 $\mu$m vs. \lbol\ for the DIGIT and FOOSH samples.  The open squares are Class 0 sources, the filled squares Class I, and the X squares are the FUors. The line is a best-fit log-log  correlation characterized by the equation in the upper left portion of the plot, with r$=$0.81.}
\label{o1lbol}
\end{figure}

Figure \ref{o1lbol} shows the correlation between \OI\ line strength and \lbol\ across the DIGIT embedded and FOOSH samples.  The large \OI\ line strength in the FOOSH sample can be  attributed to greater \lbol, albeit with considerable scatter.  

We consider whether shocks contribute significantly to \OI\ emission in our sample.  The \OI\ can be excited by either UV radiation from the central star or shocks \citep{hollenbach85}. The earliest outburst in the FOOSH sample, FU Ori, occurred 75 years prior to the {\it Herschel} observations.  At 300 km s$^{-1}$, the wind-lofted material would have propagated $\sim$ 4800 AU, slightly less than the beamsize of $\sim$ 6000 AU for these sources.  The emitting area for \OI\ increases over time as the shock propagates outward, so a propagating shock model might predict higher \OI\ line fluxes
from the older outbursts, which we do not observe.

\begin{figure}
\includegraphics[scale=0.85]{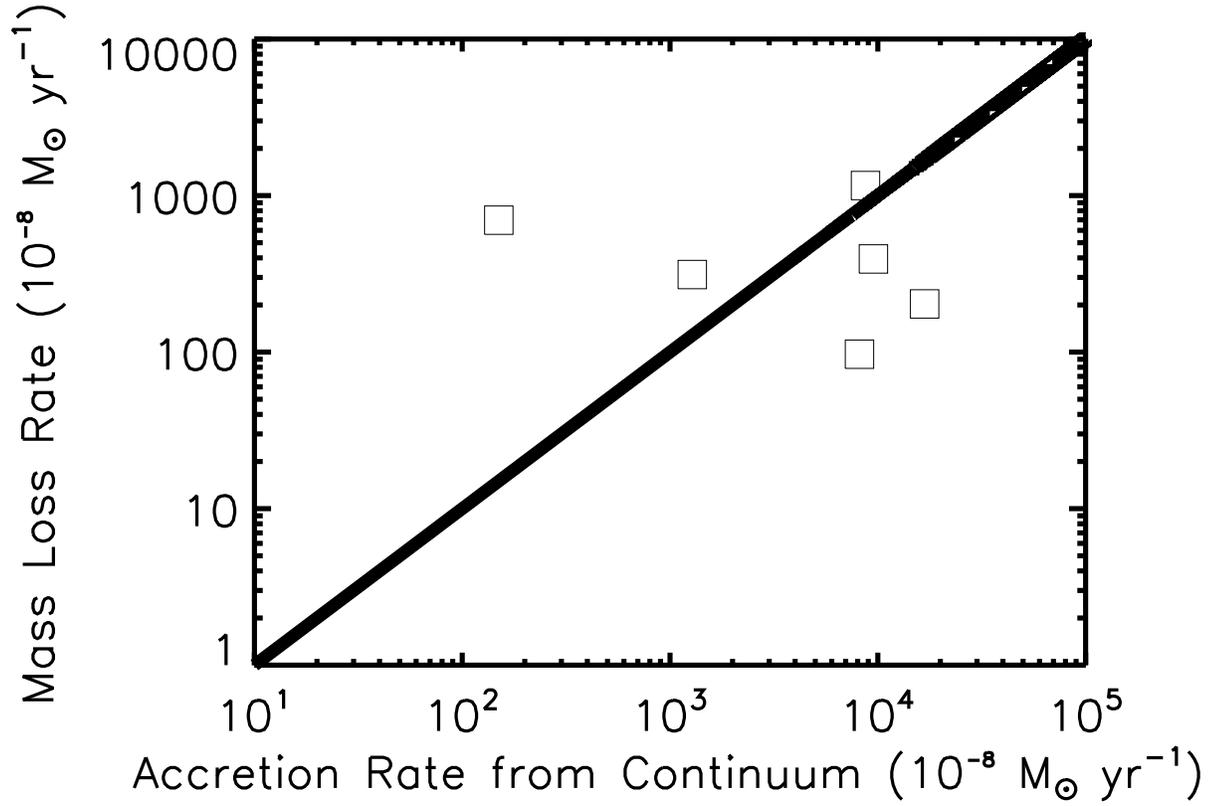}
\caption{Comparison of the accretion rate (calculated from the bolometric luminosity) and the instantaneous mass loss rate as derived from \OI, for the six FUors, assuming a wind velocity of 100 km s$^{-1}$ \citep{hollenbach85}.  The points would fall along the bold line if the \OI\ mass loss rate and accretion luminosity tracers were in perfect correlation, assuming a conversion rate of 10\% from accretion to outflow \citep[e.g.][]{kurosawa06}; the greatest outlier is HBC 722.}
\label{o1acc}
\end{figure}

Moreover, a slower wind speed of 100 km s$^{-1}$ is more consistent with the accretion rate in five of the six sources.  The accretion luminosity is calculated \citep[e.g.][]{hartmann98}:

\begin{equation}
L_{acc} = \frac{GM\mdot}{2R_{\star}}
\end{equation}

where R$_{\star}$ is taken as 2$\rsun$.  The \OI\ line luminosity is related to the instantaneous accretion rate, $\mdot=$10$^4$ L(\OI\ 63 $\mu$m, in \lsun), in units of 10$^{-5}$ $\msunyr$ \citep{hollenbach85}.    We compare these two mass loss indicators in Figure \ref{o1acc}.  The mass loss rates are consistent with a wind speed of 100 km s$^{-1}$, except in the case of HBC 722, in which the accretion luminosity {\it underestimates} the \OI\ luminosity.  The shock had only $\sim$ three months to propagate since peak outburst ($\sim$ 5 AU at a shock speed of 100 km s$^{-1}$), and thus the physical region of \OI\ enhancement due to shocks should be small, yet we observe the opposite effect (albeit at weak significance).   

In contrast, the \OI\ line flux from UV radiation should track \lbol.  Instead, the mass loss rate indicated by \OI\ is constant as a function of accretion luminosity,  between $\sim$ 10$^{-5}$ and 10$^{-6}$ $\msunyr$ for all six sources.  This suggests that either the \OI\ emission is dominated by the current UV field, or that contributions from outflow-driven \OI\ emission predate the current outburst.  This is particularly interesting given that HBC 722 appears to be a relatively evolved source with little remnant envelope to fuel multiple burst cycles at this stage.

\subsection{CO}

\begin{figure}
\begin{center}
\includegraphics[scale=0.4, angle=0]{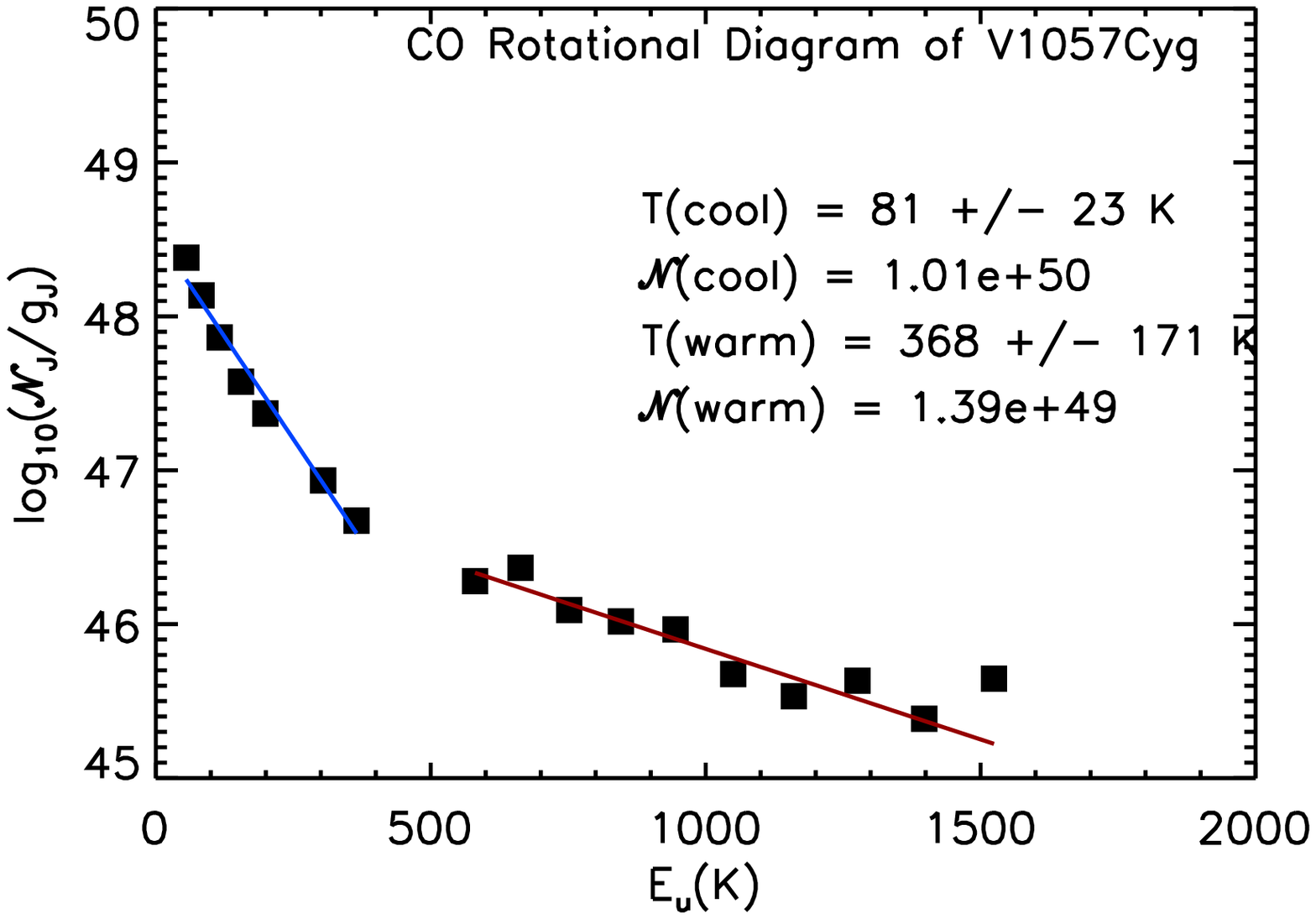}
\includegraphics[scale=0.4, angle=0]{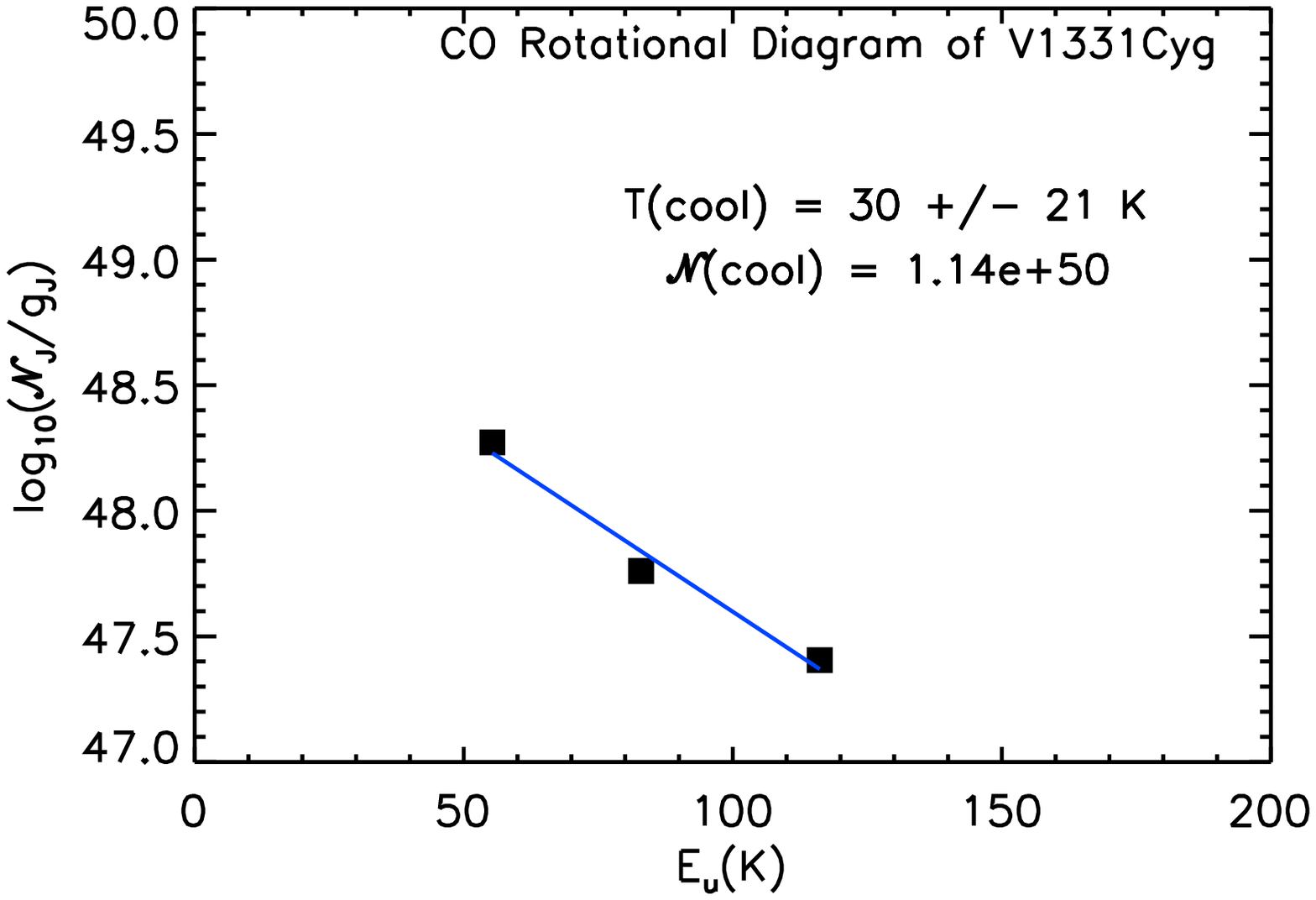} \\
\includegraphics[scale=0.4, angle=0]{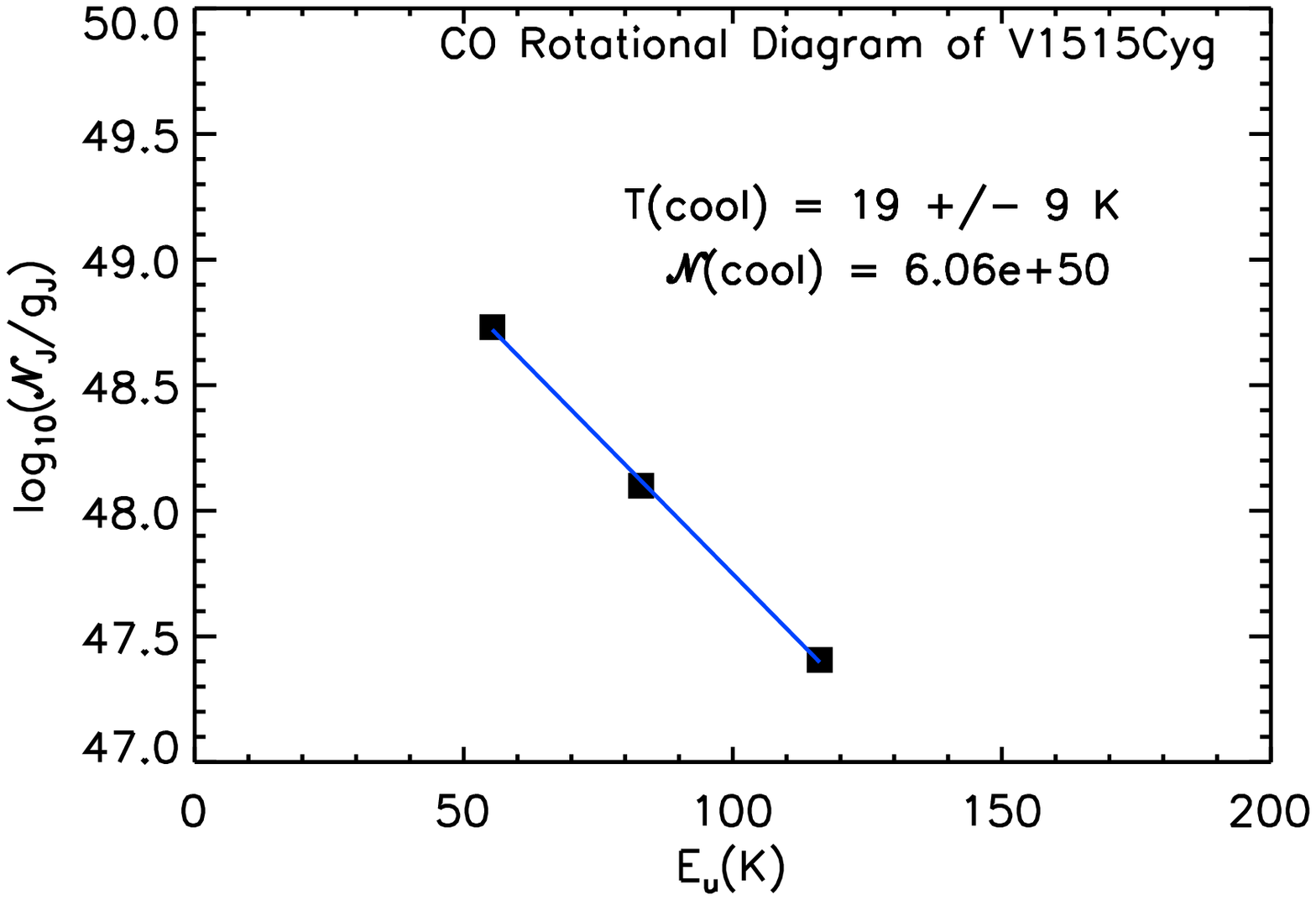}
\caption{CO rotational diagrams for the unconfused FUors in our sample. CO \jj{23}{22} is likely blended with H$_2$O and is not considered in the fit for V1057 Cyg.  FU Ori is not detected in any CO transitions with SPIRE; however a narrow CO \jj{5}{4} component was detected with HIFI, below the detection threshold of SPIRE.}
\label{spirerot}
\end{center}
\end{figure}

In addition to \OI, CO is detected in a few transitions in FU Ori, V1515 Cyg, and V1331 Cyg, and in 20 transitions in V1057 Cyg.  One useful tool in analyzing molecular emission is the rotational diagram; a detailed review can be found in \citet{goldsmith99}; a brief review in the context of {\it Herschel} spectroscopy is in \citet{green13b}.  As with samples of protostars, we can fit components to distinct ranges of CO $J_{\rm up}$: to reduce artifacts due to absolute flux calibration, we consider separately CO from SPIRE ($J_{\rm up} \leq$ 13), PACS R1 (14 $\leq J_{\rm up} \leq$ 24), and PACS B2A/B2B (25 $\leq J_{\rm up}$; no detections in this sample).  The results are presented in Figure \ref{spirerot}.

Most Class 0/I (embedded and non-outbursting) sources observed with {\it Herschel}-PACS exhibit CO emission up to \jj{24}{23} or even higher \citep{manoj12,karska13,green13b}.  These sources consistently show a 300-400 K component in this frequency range \citep{vankempen10b}, attributed to UV irradiation of the outflow cavity walls and C-shocks \citep{visser12}.  This component is usually referred to as ``warm''.  Many embedded sources also exhibit a ``hot'' component detected in CO transitions $J_{\rm up} \geq$ 25, ranging from 700-900 K.  We use the same definitions in this work, and add the ``cool'' component to identify fits to CO transitions $J_{\rm up} \leq$ 13; the source of the cool emission is usually attributed to passive heating (from gas particle collisions with dust grains heated by stellar radiation; \citealt{visser12}).

We do not observe a warm component in most of these systems.  Of the FUors, only V1057 Cyg shows robust cool and warm components at 81 and 368 K, respectively.  \citet{lorenzetti00} noted a detection of CO \jj{17}{16} for V1057 Cyg only, but no higher or lower-$J$ transitions, using ISO-LWS.  We also fit cool components to V1331 Cyg and V1515 Cyg and find \trot\ of 30 and 19 K, respectively.  We detect only one CO line in this range in FU Ori (\jj{5}{4}) from HIFI, and thus cannot compute \trot.  

We also fit a single component to the CO and $^{13}$CO rotational diagrams for V1735 Cyg.  We find \trot\ of 67 K and 45 K, respectively, but do not attribute the emission solely to the FUor.

As the highest observed rotational level of CO emission decreases, the rotational temperature of gas decreases as well, from 81 K in V1057 Cyg, to 19 K in V1515 Cyg.  The total amount of CO gas, \funnyN(CO), dominated in the FUors by the cool CO, does not follow this trend but instead varies more like \lbol.  This is consistent with \funnyN(CO) derived from the warm component in embedded sources \citep{manoj12,karska13,green13b}.  If the gas around V1057 Cyg is well-approximated by a two-temperature (cool/warm) fit, optical depth effects would be most apparent at low-$J$, where the line flux would fall below the fit, for certain density regimes \citep[][ Fig. 12]{goldsmith99}.  Instead, the CO continues to exhibit positive curvature to the longest SPIRE wavelengths.  This suggests that either the gas is populated at low densities \citep{neufeld12,manoj12}, that optical depth effects are not substantially affecting the V1057 Cyg line fluxes even at low-$J$, or that there is a larger reservoir of cooler gas.  

The best constraint on opacity arises from the ratio of $^{13}$CO to $^{12}$CO: in V1057 Cyg, the upper limit of the isotopic ratio of the \jj{5}{4} line intensity is $\sim$ 10, requiring an optical depth less than $\sim$ 7.  For comparison, the CO optical depth measured from the $^{12}$CO/$^{13}$CO ratio declines with upper state $J$ for both confused sources V1735 Cyg and HBC 722, from $\tau$ of $\sim$ 7 at \jj{5}{4} down to $\sim$ 3 at \jj{7}{6}.

In the limit of moderate or higher density, the observed curvature suggests that additional ``cold'' components, or even a power-law distribution of components may be needed to fit the rotational diagram, in order to produce sufficient low-$J$ emission.  It has already been suggested that a power law distribution is a good fit to the CO \jj{14}{13} up to \jj{49}{48} PACS lines detected in the Orion protostar sample \citep{manoj12}; whether this applies to the FUors is not clear.

\subsubsection{Velocity resolved mid-$J$ emission}

Using HIFI, CO \jj{14}{13} is weakly detected in HBC 722, likely contaminated by extended emission; it is not detected in any other object.
We also observed HCO$^+$ \jj{3}{2}, a dense gas tracer, using heterodyne observations at the CSO, resolving the line profiles (Figure \ref{co54}), for all sources except FU Ori.  We compared these profiles to CO \jj{5}{4} resolved lines from HIFI for each of V1057 Cyg, V1735 Cyg, V1331 Cyg, V1515 Cyg, and HBC 722.  Additionally, CO \jj{3}{2} and $^{13}$CO \jj{2}{1} profiles for these sources appear in \citet{mcmuldroch93}, \citet{evans94}, and \citet{mcmuldroch95}.
In the case of V1331 Cyg, the CO and HCO$^+$ exhibited similar profiles, with narrow features.  In the other four sources, the HCO$^+$ narrow feature peaked at the velocity of self-absorption in CO; the CO additionally showed wings much broader than those seen in HCO$^+$.  In fact, the self-absorption dip is the clearest indicator of the source velocity, and the HCO$^+$ peak is slightly redshifted in comparison to the absorption, indicating that some of the contributing gas in the red part of the wings is at the source velocity;  the wings are typical of embedded sources \citep[e.g.][]{jorgensen09}.

The sources with stronger high-$J$ CO also exhibit broader CO \jj{5}{4} profiles; the HCO$^+$ is always a narrow single peak in all sources.
In V1735 Cyg and HBC 722, the broad CO traces contaminating outflows, while the HCO$^+$ measures the velocity of the ambient dense gas.  To the extent that HCO$^+$ aligns with the absorption in CO, we presume that low excitiation ambient gas is present at that velocity.

In the case of V1515 Cyg, we see little excited gas, consistent with earlier observations \citep{evans94,kospal11b}.  In the cases of FU Ori and V1331 Cyg, we interpret the weak, single-peaked CO as tracing the source.  In the case of V1057 Cyg, we observe a narrower but still somewhat broad feature.  V1057 Cyg also has the richest PACS spectrum; in this case the CO emission may originate from material at the source, rather than contaminating outflows.  In this case, the outflow and the high-$J$ CO may be driven from V1057 Cyg itself.

Thus the mid-IR non-{\it Herschel} SEDs are not good predictors of which emission lines would be detected at longer wavelengths.  Despite similar mid-IR spectral features, shape, and extinction parameters, V1515 Cyg and V1057 Cyg are quite different from the {\it Herschel} perspective.

\subsubsection{Where is the Hot CO Emission in FUors?}

We might posit that CO emission from hot gas would be a transitory signature of heating from a recent outburst.  How then do we explain the lack of high-$J$ CO emission in these FUors?  

Analysis of the {\it non-detections} reveals that the absence of high-$J$ CO and all but one H$_2$O line in our sources is of only modest significance.  Figure \ref{gas} shows histograms of the line luminosities of the CO \jj{16}{15} (left) and H$_2$O 174.63 $\mu$m (right) for the DIGIT sample \citep{green13b}.  The vertical dashed lines show the strength of these lines in V1057 Cyg; the vertical dotted lines show 3$\sigma$ upper limits for the non-detections in the FOOSH sample.  In both cases, the detected CO \jj{16}{15} line (in V1057 Cyg) would be among the brightest in the DIGIT protostar sample, but not a notable outlier.  The upper limits from our sample for CO and H$_2$O are not low enough to rule out significant emission in any source.  

If the CO emission originates in a tenuous envelope, there may not be enough emitting material to produce a detectable signature in the warmer regions.  If the CO emission originates in a disk, then we require a thermal inversion to produce emission lines in the upper layer.  Models suggest that the midplane of FUors may become hotter than the upper disk layers during outburst, which would mask the signature of CO  \citep{zhu10}.  Our data do not sufficiently constrain the line luminosities to distinguish the FUors from ordinary protostars.

In the DIGIT sample, the CO emission from disks is much weaker.  The highest line luminosity from CO \jj{16}{15} detected in the DIGIT disks sample is in HD 100546, a Herbig Ae/Be star of similar luminosity to the FUors \citep{sturm10}.  The CO line is a factor of 20 weaker in luminosity than V1057 Cyg, at 3$\times$10$^{-5}$ \lsun\ (Meeus et al., subm.).  Thus it is unlikely that the disk is a significant contributor to the CO line luminosity in V1057 Cyg.

\subsection{Other Molecules}

We consider here a few other marginal detections.  We report the detection of a single line of H$_2$O at 174.63 $\mu$m in V1057 Cyg. The H$_2$O line luminosity in V1057 Cyg is 0.47 $\times$ 10$^{-3}$ $\lsun$, which would place this single line among the most luminous of the DIGIT sample.  The 174.63 $\mu$m line (excitation energy of 197 K) is not even the lowest excitation line within the PACS bands; in embedded sources the lowest energy detected H$_2$O line is at 179.53 $\mu$m (114 K); however the 174.63 $\mu$m line is typically among the most luminous H$_2$O lines detected in Class 0/I protostars .  We also do not detect the 557 GHz $1_{10}-1_{01}$ line in the SPIRE bands.  We do not detect H$_2$O in any other sources.  Although the detection in V1057 Cyg is robust, the upper limits on this line in the other FUors do not rule out substantial emission lines as well.

We tentatively detect OH 84.41 $\mu$m, half of a doublet in the 3/2 $\rightarrow$ 3/2 ladder, also in V1057 Cyg.   Although this is the most commonly detected line in the embedded sample, the 84.61 $\mu$m half of the doublet is typically just as strong.  More prominently, OH 119 $\mu$m is seen {\it in absorption} in V1735 Cyg.  Only V1735 Cyg and 2 of 30 Class 0/I sources in the DIGIT sample show this line in absorption.  There is also a hint of OH emission at 163 $\mu$m, possibly blended with a weak detection of the CO \jj{16}{15} line, but both detections are inconclusive.

No other molecules are detected in the FOOSH sample with {\it Herschel}.  In contrast, at least some embedded sources show rich SPIRE spectra (\citealt{goicoechea12}, Green et al., in prep.).
However, the upper limits for H$_2$O and OH line luminosity are consistent with the detections in embedded sources.

\subsection{Additional Lines}

\CII\ 157.74 $\mu$m is detected in V1735 Cyg, V1331 Cyg, and V1515 Cyg, but is not clearly associated with the sources, similar to the DIGIT embedded sources.

\CI\ is observed at 370 and 610 $\mu$m in all sources.  The \CI\ 370/610 $\mu$m flux ratio falls between 2.1 and 3.0 across the sample, consistent with ISM values \citep{jenkins01}, and the \CI\ emission is associated with cooler extended gas rather than the FUors themselves.  The SPIRE maps show that the lines are extended (at beamsize $\sim$ 40$\arcsec$); we do not attribute the emission to the FUors.
  
\NII\ 205 $\mu$m (SPIRE) is observed to emit strongly in HBC 722, V1331 Cyg, V1057 Cyg, and V1515 Cyg, but not V1735 Cyg or FU Ori.  The \NII\ 122 $\mu$m (PACS) line is detected along with the 205 $\mu$m line around V1515 Cyg; no other sources in our sample show the 122 $\mu$m  line.  This is in contrast with ISO-LWS results, in which \NII\ was marginally detected at 122 $\mu$m with ISO-LWS in V1735 Cyg, V1331 Cyg, V1057 Cyg \citep{lorenzetti05}.  As noted earlier, the difference in detection rates of \NII\ 122 $\mu$m is likely due to the use of an off-position for our PACS data, but an onboard calibration source for SPIRE.  

We detected both lines around V1515 Cyg, and examined the PACS and SPIRE maps to see if there was any hint of centrally peaked emission.   The flux of each line shows no significant variation ($<$ 50\%) across the entire PACS and SPIRE field-of-view, and is uncorrelated with proximity to the central position.  The \NII\ 122/205 $\mu$m flux ratio is diagnostic of the electron density \citep{hudson04,oberst11};  The ratio in the central pixel is 0.31, suggesting an extremely low electron density (n$_e$ $<$ 1 cm$^{-3}$.  We note that the observed ratio should be considered an upper limit, as the PACS spectra are observed using an off-position for sky subtraction, while the SPIRE spectra are not.

Further evidence of the diffuse nature of the \NII\ emission comes from the fact that there is no clear trend between \NII\  205 $\mu$m and the strength or position of the CO emission.  
Thus we conclude that among fine structure lines, only the \OI\ emission is localized to the FUor.

\section{Discussion}

\subsection{Evolutionary Stage}

It has been posited from occurrence rate \citep{hartmann96a} and from simulation \citep[e.g.][]{dunham12a} that FUors are Stage 0/I stars undergoing episodic outburst cycles with replenished material from an infalling envelope. However, FUors have much in common with Class II sources: in Spitzer-IRS bands all but V1735 Cyg show silicate emission and optical-IR continuum indices consistent with relatively low extinction \citep[e.g.][]{hartmann96a,green06, quanz07a}.  It has been suggested that the overall group of FUors fall into two different categories, those consistent with flared disks, and those requiring envelopes \citep[e.g.][]{kenyon91,zhu08}.  

\citet{zhu08} modeled the dust from Spitzer-IRS observations of FU Ori, V1057 Cyg, and V1515 Cyg and found that V1057 Cyg and V1515 Cyg required envelopes typical of protostars, while FU Ori did not require an envelope.  The {\it Herschel} spectra generally fall above their model predictions, indicating an underestimate of the reservoir of cold dust surrounding these systems.  Nonetheless, our estimates of the envelope mass from the {\it Herschel} dust emission broadly agrees with this classification; we find that V1057 Cyg and V1515 Cyg have envelopes greater than 0.1 $\msun$, while FU Ori has a much smaller envelope mass of $\sim$ 0.02 $\msun$. HBC 722 was considered a T Tauri star prior to outburst, and measurements of the envelope mass upper limit suggest a tenuous envelope at best.  

We noted some inconsistencies between the different classification methods for the continuum, but found broad agreement with the mid-IR sequence in \citet{quanz07a}.   However, FUors may not be well-characterized by the Stage I/II sequence, because the {\it Herschel} line observations provide a different picture from the continuum.  In general, the presence of weaker silicate features and larger mid-IR $\alpha$ is correlated with increased submillimeter dust and CO/H$_2$O emission, but is not well-correlated with \OI.  Class 0/I sources typically exhibit hotter and more excited CO rotational lines than Class II sources (cf. \citealt{karska13}, Meeus et al. subm.); by this standard FUors would {\it seem} to resemble disk sources; however, these lines may be too faint to detect in our observations.  The DIGIT (and HOPS) samples show a tight correlation between CO \jj{16}{15} line luminosity and derived \funnyN(warm).  In V1057 Cyg, the only FUor where the warm gas is detected, it falls in the DIGIT embedded source correlation.  Additionally the DIGIT sample showed a lack of correlation between \trot(warm) and \funnyN(CO); again, the lone FUor (V1057 Cyg) is consistent with the embedded sample. Finally, the DIGIT sample showed a (weaker) trend between \lbol\ and \funnyN(warm); once again V1057 Cyg fits this trend.  It is clear that the spectra of V1057 Cyg, as characterized by PACS, would not distinguish it as an outlier in the DIGIT Class 0/I sample, even though it has a Class II SED, suggesting it is a kind of hybrid object.  However, the high upper limits on line emission do not provide strong constraints on the evolutionary state of the other FUors.

\section{Conclusions}

We present analysis of six FUors with {\it Herschel}-PACS, SPIRE, and HIFI, {\it Spitzer}-IRS, and CSO heterodyne observations, from the ``FOOSH'' sample.

\begin{itemize}

\item We present PACS and SPIRE imaging of V1057 Cyg, V1331 Cyg, V1515 Cyg, V1735 Cyg, and FU Ori, and their surrounding regions.  Complementary to pre-existing near-IR and submillimeter ground-based imaging, all show filamentary morphologies.  In the case of V1515 Cyg, the optical FUor is only marginally detected at SPIRE wavelengths.    Most fall in complicated far-IR fields, typical of active star-forming regions, but only two (V1735 Cyg and HBC 722) are confused.  V1735 Cyg is blended with V1735 Cyg SM1, a nearby submillimeter source, at wavelengths greater than 100 $\mu$m, and HBC 722, as noted previously, is blended with 2MASS 20581767+4353310 at wavelengths greater than 70 $\mu$m.

\item The 1-1000 $\mu$m SEDs for the five older FUors indicate that FU Ori, V1515 Cyg, and V1057 Cyg are Class II, while V1735 Cyg and V1331 Cyg are borderline Class I/II, or flat spectrum sources, classified by \tbol.  We derive envelope masses of 0.07-0.3 $\msun$ from the long wavelength data, and find FU Ori and V1331 Cyg to be Stage II, V1515 Cyg to be borderline Stage I/II, and V1057 Cyg to be Stage I by the \citet{crapsi08} criterion (envelope mass $>$ 0.1 \msun).  Their derived \lbol\ values are somewhat lower than previously measured values, likely a complication of the replacement of the large IRAS beamsize with the smaller PACS/SPIRE beam in our SEDs.

\item All exhibit \OI\ 63 $\mu$m and \CI\ 370 and 610 $\mu$m lines.  \NII\ 122 and 205 $\mu$m and \CII\ 158 $\mu$m are detected as well; the latter two lines are detected in four of the sources while the former is only detected in one of those (V1515 Cyg).  Of these lines, only \OI\ (63 $\mu$m and 145 $\mu$m, when detected) is attributed to the FUors.  HBC 722 is detected only weakly in continuum at 70 $\mu$m, and cleanly detected in line emission only in \OI\ 63 $\mu$m.

\item Analysis of V1735 Cyg is complicated by the presence of a nearby submillimeter source (V1735 Cyg SM1), and the contributions from the two sources to gas lines other than \OI\ cannot be separated.   This includes four mid-$J$ transitions of $^{13}$CO as well as nine transitions of CO.  Mapping indicates that the detected CO emission is dominated by SM1.  The low-lying OH 119 $\mu$m doublet is observed in absorption toward V1735 Cyg; by comparison, this is seen in 2 of 30 protostars in the DIGIT sample.  The sources are separable in continuum emission out to 200 $\mu$m with {\it Herschel}, and we use ground-based observations to help constrain longer wavelengths.

\item V1057 Cyg is an outlier among the FOOSH sample.  We detect CO up to \jj{23}{22} and determine that the emitting source is compact.  We fit cool (81 K) and warm (368 K) rotational temperature components to the CO.  We detect H$_2$O in one transition at 174.63 $\mu$m, which would place this source among the most luminous protostars in line emission in the DIGIT sample, but not an extreme outlier.  V1057 Cyg also shows a tentative detection of OH 84.41 $\mu$m.

\item In the other three sources (V1331 Cyg, V1515 Cyg, and FU Ori), no H$_2$O or OH lines are detected, but upper limits are still consistent with the detections for protostars.  CO is detected up to \jj{6}{5} in these sources.  We fit cool components of 30 and 19 K to V1331 Cyg and V1515 Cyg, respectively; no component can be fit for FU Ori, which is detected in only a single CO transition (\jj{5}{4}).  Although V1057 Cyg and V1515 Cyg are very similar from the {\it Spitzer} perspective (in terms of their SED, extinction values, and dust properties), they are differentiated by their gas properties in the {\it Herschel} range: the former exhibits high-J CO, H$_2$O and luminous \OI; the latter shows no molecular emission and only weak \OI\ traceable to the source.

\item Dense gas is present at the source velocity in all four sources where it was observed, based on HCO$^+$ \jj{3}{2}.  In one case, V1331 Cyg, the HCO$^+$ profile matches the CO \jj{5}{4} from HIFI suggesting that the CO may originate on-source.  In V1057 Cyg, V1735 Cyg, and HBC 722, the CO \jj{5}{4} line is broader than the HCO$^+$, with hints of self-absorption at the velocity of the dense gas, suggesting that the CO may partially trace outflows in the region.  CO \jj{14}{13} is marginally detected in HBC 722 but contaminated by extended emission; it is not detected in any other objects.  Only in V1057 Cyg is the high-$J$ CO is suspected to originate at the source.

\item The FUors all exhibit large \OI\ line luminosities, which we attribute to increased UV radiation and outflow rates, via episodic accretion.  The mass loss rate derived from \OI\ assuming a 100 km s$^{-1}$ wind and the accretion luminosity from continuum are consistent (assuming a conversion factor of 10 to 1 from infall to outflow), except in the case of HBC 722, whose \OI\ implied mass accretion rate exceeds the rate derived from the accretion luminosity by two orders of magnitude.  This suggests that HBC 722 may have experienced previous outbursts, as the \OI\ could not have propagated far enough for a significant enhancement in first three months of outburst prior to observations.

\end{itemize}

From the {\it Herschel} perspective, FUors outbursts are associated with both Stage I and II sources; the evolutionary state of the sources is somewhat variable depending upon the criteria used.   They are invariably surrounded by considerable nebulosity and cold dust, typically associated with protostars rather than Class II sources.  Their continuum SEDs are not uniform, with some exhibiting ``flat-spectrum'' and others Class II features.  The sources with the lowest IR excess and most prominent mid-IR silicate dust features -- FU Ori, V1331 Cyg, and V1515 Cyg -- exhibit little or no excited CO gas, typical of most disk sources.  V1057 Cyg shows a cool and a warm gas component, consistent with embedded protostars, although not a detectable hot component (\trot\ $>$ 500 K).  The H$_2$O and OH detections in V1057 Cyg are typical for protostars but much more luminous than those seen in Herbig Ae/Be stars (Fedele et al., subm.) and T Tauri stars (Salyk et al., in prep.).  The defining characteristic from {\it Herschel} is that all FUors exhibit large \OI\ 63 $\mu$m line luminosities, indicative of high accretion rates.

\acknowledgements

Support for this work, part of the  {\it Herschel} Open Time Key
Project Program, was provided by NASA through an award issued by the Jet
Propulsion Laboratory, California Institute of Technology.  
The authors wish to acknowledge the Herschel Director, G. L. Pilbratt, for the timely approval and execution of the Target of Opportunity program, and Ivan Valtchanov and the {\it Herschel}  Helpdesk for its assistance in timely data reduction.  The authors also thank Michelle Rascati and Roderik Overzier for assistance in data reduction, and Colette Salyk, Amanda Heiderman, Isa Oliveira, Tom Megeath, Will Fischer, Manoj Puravankara, and Geoff Blake for helpful discussions, and the anonymous referee whose comments greatly improved the manuscript.  The research of J.-E. L. is supported by Basic Science Research Program 
through the National Research
Foundation of Korea (NRF) funded by the Ministry of Education, Science 
and Technology (No.
2012-044689).  The research of M.G. and A.L. has been supported by the Austrian Research Promotion Agency (FFG) through the ASAP initiative of the Austrian Federal Ministry for Transport, Innovation and Technology (BMVIT).  This material is based upon work at the Caltech Submillimeter Observatory, which is operated by the California Institute of Technology under cooperative agreement with the National Science Foundation (AST-0838261).  This publication makes use of data products from the Two Micron All Sky Survey, which is a joint project of the University of Massachusetts and the Infrared Processing and Analysis Center/California Institute of Technology, funded by the National Aeronautics and Space Administration and the National Science Foundation.

\bibliographystyle{apj}

\begin{center}
\begin{deluxetable}{l l l r}
\tabletypesize{\scriptsize}
\tablecaption{Herschel Observing Log \label{obslog}}
\tablewidth{0pt}
\tablehead{
\colhead{Object} & \colhead{Mode} & \colhead{AOR} & \colhead{Date}}
\startdata
HBC 722 & HIFI CO 14-13 & 1342210779 & 3 Dec 2010 \\
HBC 722 & HIFI CO 5-4 & 1342210806 & 4 Dec 2010 \\
HBC 722 & SPIRE Spec. & 1342210857 & 6 Dec 2010 \\
HBC 722 & SPIRE Phot. & 1342210915 & 8 Dec 2010 \\
HBC 722 & PACS Phot. Blue1 & 1342211094 & 13 Dec 2010 \\
HBC 722 & PACS Phot. Blue2 & 1342211095 & 13 Dec 2010 \\
HBC 722 & PACS Phot. Green1 & 1342211096 & 13 Dec 2010 \\
HBC 722 & PACS Phot. Green2 & 1342211097 & 13 Dec 2010 \\
HBC 722 & PACS Spec. Red & 1342211173 & 14 Dec 2010 \\ 
HBC 722 & PACS Spec. Blue & 1342211174 & 14 Dec 2010 \\ 
V1515 Cyg & SPIRE Phot. & 1342211353 & 20 Dec 2010 \\
V1057 Cyg & SPIRE Phot. & 1342211357 & 20 Dec 2010 \\
V1735 Cyg & SPIRE Spec. & 1342219560 & 24 Apr 2011 \\
V1735 Cyg & SPIRE Phot. & 1342219973 & 6 May 2011 \\
V1331 Cyg & SPIRE Phot. & 1342220630 & 8 May 2011 \\
V1735 Cyg & HIFI CO 14-13 & 1342220505 & 12 May 2011 \\
HBC 722 & HIFI CO 5-4 & 1342221418 & 20 May 2011 \\
V1515 Cyg & HIFI CO 14-13 & 1342221450 & 20 May 2011 \\
V1331 Cyg & HIFI CO 14-13 & 1342221451 & 20 May 2011 \\
V1057 Cyg & HIFI CO 14-13 & 1342221452 & 20 May 2011 \\
HBC 722 & HIFI CO 14-13 & 1342221453 & 20 May 2011 \\
V1515 Cyg & SPIRE Spec. & 1342221685 & 25 May 2011 \\
V1331 Cyg & SPIRE Spec. & 1342221694 & 25 May 2011 \\
V1057 Cyg & SPIRE Spec. & 1342221695 & 25 May 2011 \\
V1057 Cyg & PACS Phot. Green1 & 1342223184 & 23 Jun 2011 \\
V1057 Cyg & PACS Phot. Blue1 & 1342223185 & 23 Jun 2011 \\
V1057 Cyg & PACS Phot. Blue2 & 1342223186 & 23 Jun 2011 \\
V1057 Cyg & PACS Phot. Green2 & 1342223187 & 23 Jun 2011 \\
V1735 Cyg & PACS Phot. Green1 & 1342225246 & 22 Jul 2011 \\
V1735 Cyg & PACS Phot. Blue1 & 1342225247 & 22 Jul 2011 \\
V1735 Cyg & PACS Phot. Blue2 & 1342225248 & 22 Jul 2011 \\
V1735 Cyg & PACS Phot. Green2 & 1342225249 & 22 Jul 2011 \\
V1331 Cyg & PACS Phot. Green1 & 1342225252 & 22 Jul 2011 \\
V1331 Cyg & PACS Phot. Blue1 & 1342225253 & 22 Jul 2011 \\
V1331 Cyg & PACS Phot. Blue2 & 1342225254 & 22 Jul 2011 \\
V1331 Cyg & PACS Phot. Green2 & 1342225255 & 22 Jul 2011 \\
FU Ori & HIFI CO 5-4 & 1342228625 & 15 Sep 2011 \\
FU Ori Cyg & SPIRE Phot. & 1342219654 & 23 Sep 2011 \\
FU Ori & SPIRE Spec. & 1342230412 & 09 Oct 2011 \\
V1515 Cyg & PACS Phot. Blue1 & 1342232439 & 17 Nov 2011 \\
V1515 Cyg & PACS Phot. Green1 & 1342232440 & 17 Nov 2011 \\
V1515 Cyg & PACS Phot. Green2 & 1342232441 & 17 Nov 2011 \\
V1515 Cyg & PACS Phot. Blue2 & 1342232442 & 17 Nov 2011 \\
V1735 Cyg & HIFI CO 5-4 & 1342232695 & 23 Nov 2011 \\
V1057 Cyg & HIFI CO 5-4 & 1342232698 & 23 Nov 2011 \\
V1331 Cyg & PACS Spec. Red  & 1342233445 & 2 Dec 2011 \\
V1331 Cyg & PACS Spec. Blue & 1342233446 & 2 Dec 2011 \\
V1515 Cyg & PACS Spec. Red & 1342235690 & 28 Dec 2011 \\
V1515 Cyg & PACS Spec. Blue & 1342235691& 28 Dec 2011 \\
V1735 Cyg & PACS Spec. Red & 1342235848 & 28 Dec 2011 \\
V1735 Cyg & PACS Spec. Blue & 1342235849 & 31 Dec 2011 \\
V1057 Cyg & PACS Spec. Red  & 1342235852 & 1 Jan 2012 \\
V1057 Cyg & PACS Spec. Blue & 1342235853 & 1 Jan 2012 \\
FU Ori & PACS Phot. Blue1 & 1342242658 & 29 Mar 2012 \\
FU Ori & PACS Phot. Green1 & 1342242659 & 29 Mar 2012 \\
FU Ori & PACS Phot. Blue2 & 1342242660 & 29 Mar 2012 \\
FU Ori & PACS Phot. Green2 & 1342242661 & 29 Mar 2012 \\
V1515 Cyg & HIFI CO 5-4 & 1342244405 & 14 Apr 2012 \\
V1331 Cyg & HIFI CO 5-4 & 1342244406 & 14 Apr 2012 \\
FU Ori & PACS Spec. Blue & 1342250907 & 11 Sep 2012  \\
FU Ori & PACS Spec. Red & 1342250908 & 11 Sep 2012 \\
HBC 722 & HIFI CO 14-13 & 1342255774 & 22 Nov 2012 \\
\enddata
\tablecomments{ Each SPIRE image and spectrum includes a single observation ID (OBSID).  Each PACS spectral scan consists of a ``blue'' (50-75 and 100-150 $\mu$m) and ``red'' (70-100 and 140-200 $\mu$m) pair of observations.  Each PACS image consists of a pair of cross-scans (1 and 2) for each of the ``blue'' (70 and 160 $\mu$m) and ``green'' (100 and 160 $\mu$m) bands.}
\end{deluxetable}
\end{center}

\begin{center}
\begin{deluxetable}{l l l l l l l r}
\tabletypesize{\scriptsize}
\tablecaption{Sample Description \label{fuors}}
\tablewidth{0pt}
\tablehead{
\colhead{Source} & \colhead{RA (J2000)} & \colhead{Dec} & \colhead{Dist. (pc)}  & \colhead{Date of Eruption} &
 \colhead{Mid-IR Class} & \colhead{Binary?}  & \colhead{Ref.}
}
\startdata
FU Ori & 05:45:22.4 & +09:04:12 & 600 & 1936 & disk & few AU binary & \citet{herbig77} \\
V1515 Cyg & 20:23:48.0 & +42:12:26 & 1000 & $\sim$ 1950 & some env. & --- & \citet{herbig77} \\
HBC 722 & 20:53:17.0 & +43:53:43 & 520 & 2009-2010 & --- & --- & \citet{laugalys06} \\
V1057 Cyg & 20:58:53.7 & +44:15:29 & 600 & 1969 & some env. & --- & \citet{herbig77} \\
V1735 Cyg & 21:47:20.7 & +47:32:04 & 950 & 1952-1965 & env. & --- & \citet{elias78} \\
V1331 Cyg & 21:01:09.2  & +50:21:45 & 550 & --- & disk & --- & \citet{quanz07c}  \\
\enddata
\tablecomments{Sample description.  A portion of the table is from \citet{reipurth10}.  The Mid-IR class is described in the text.}
\end{deluxetable}
\end{center}

\begin{center}
\begin{deluxetable}{l r r r r r r}
\tabletypesize{\scriptsize}
\tablecaption{PACS/SPIRE Flux Density \label{phot}}
\tablewidth{0pt}
\tablehead{
\colhead{Object} & \colhead{70 $\mu$m} & \colhead{100 $\mu$m} & \colhead{160 $\mu$m} & \colhead{250 $\mu$m} & \colhead{350 $\mu$m} & \colhead{500 $\mu$m}}
\startdata
& Jy & Jy & Jy & Jy & Jy & Jy\\
& & & & & & \\
V1057 Cyg & 30.0 $\pm$ 3.0 & 30.1 $\pm$ 3.0 & 28.2 $\pm$ 2.8 & 15.8 $\pm$ 1.6 & 8.1 $\pm$ 1.2 & 2.8 $\pm$ 0.6 \\
V1331 Cyg & 7.5 $\pm$ 0.8 & 9.7 $\pm$ 1.0 & 11.6 $\pm$ 1.2 & 9.9 $\pm$ 1.1 & 6.4 $\pm$ 1.0 & 2.7$\pm$ 0.6 \\
V1515 Cyg & 4.6 $\pm$ 0.5 & 3.9 $\pm$ 0.4 & 5.0 $\pm$ 0.5 & 4.3 $\pm$ 0.8 & 2.1 $\pm$ 0.6 & 0.9 $\pm$ 0.4 \\
V1735 Cyg & 17.1 $\pm$ 1.7 & 19.7 $\pm$ 2.0 & 27.1 $\pm$ 2.7 & $<$ 49.4 $\pm$ 4.0 &  $<$ 27.3 $\pm$ 2.8 & $<$ 11.0 $\pm$ 1.4 \\
FU Ori & 6.5 $\pm$ 0.7 & 5.9 $\pm$ 0.6 & 4.9 $\pm$ 0.5 & 2.8 $\pm$ 0.5 & 2.5 $\pm$ 0.6 & 0.9 $\pm$ 0.3 \\
HBC 722$^1$ & 0.41 $\pm$ 0.02 & --- & --- & --- & --- & --- \\
\enddata
\tablecomments{PACS 70, 100, and 160 $\mu$m, and SPIRE 250, 350, and 500 $\mu$m photometry of the five FUors in our sample, with the extended source correction applied, in 30\arcsec apertures around the source coordinates.  The SPIRE uncertainty includes a 7\% calibration uncertainty, and sky subtraction from an annulus between 90-120$\arcsec$ from the source.  The PACS formal uncertainty is smaller, but we assume a 10\% minimum uncertainty to account for extended emission.  V1057 Cyg, V1331 Cyg, V1515 Cyg, and FU Ori were resolved as separate sources with PACS and were extracted using an annulus that included all flux.  V1735 Cyg includes a significant contribution from V1735 Cyg SM1 at SPIRE wavelengths (e.g. $>$ 200 $\mu$m), a nearby submm source to east of V1735 Cyg, and thus the flux densities for $\lambda > 200$ $\mu$m should be considered upper limits; in addition, there is some contamination at PACS wavelengths, where we find contributions from SM1 for annuli greater than 17\arcsec.     \\
$^1$HBC 722 \citep{green11b} was not detected as a separate continuum source with PACS/SPIRE except at 70 $\mu$m.}
\end{deluxetable}
\end{center}

\begin{center}
\begin{deluxetable}{l l l l}
\tabletypesize{\scriptsize}
\tablecaption{Classification \label{class}}
\tablewidth{0pt}
\tablehead{
\colhead{Object} & \colhead{$\alpha$} & \colhead{\tbol} & \colhead{Env. Mass}}
\startdata
& Class & Class & Stage \\
& & & \\
V1057 Cyg & FS & II & I \\
V1515 Cyg & II & II & I \\
V1331 Cyg & II & FS & I/II$^1$ \\
FU Ori & II & II  & II \\
V1735 Cyg & II & FS & -- \\
HBC 722 & II$^2$ & -- & -- \\
\enddata
\tablecomments{Classification of the FOOSH sources by three different techniques: mid-IR spectral index $\alpha$, \tbol, and envelope mass (see text).  ``FS'' indicates a flat-spectrum classification.  Contaminated measurements are reported as ``--''.  \\
$^1$ V1331 Cyg can be classified as Stage I or II depending upon the wavelength used to measure the envelope mass; the inferred mass is close to the boundary of 0.10 \msun\ defined in \citet{crapsi08}.\\
$^2$ $\alpha$ was measured pre-outburst for HBC 722 \citep{miller11}}
\end{deluxetable}
\end{center}

\begin{center}
\begin{deluxetable}{l r r r r r}
\tabletypesize{\scriptsize}
\tablecaption{Dust-Derived Envelope Mass \label{envmass}}
\tablewidth{0pt}
\tablehead{
\colhead{Object} & $\lbol$/$\lsmm$ & T$_{dust}$ & \colhead{Env. Mass (250 $\mu$m)} & \colhead{(350 $\mu$m)} & \colhead{(500 $\mu$m)} }
\startdata
& & K & $\msun$ & $\msun$ & $\msun$ \\
& & & & & \\
V1057 Cyg & 975 & 43 & 0.16 & 0.25 & 0.30 \\
V1515 Cyg & 505 & 42 & 0.13 & 0.19 & 0.28 \\
V1331 Cyg & 159 & 38 & 0.11 & 0.07 & 0.46 \\
FU Ori & 5858 & 54  & 0.02 & 0.05 & 0.07 \\
\enddata
\tablecomments{Ratio of bolometric over submillimeter luminosity, dust temperature, and envelope mass calculated at each of the three SPIRE photometric bands.  The large envelope mass derived at 500 $\mu$m in V1331 Cyg is attributed to extended/diffuse emission.}
\end{deluxetable}
\end{center}

\begin{center}
\begin{deluxetable}{l r r r r r r r r r r r}
\tabletypesize{\scriptsize}
\tablecaption{Line fluxes  \label{pacsspirelines}}
\tablewidth{0pt}
\tablehead{
\colhead{Line} & \colhead{Freq.} & \colhead{$\lambda$} & \colhead{HBC 722} & \colhead{V1735 Cyg} & \colhead{V1057 Cyg} & \colhead{V1331 Cyg} & \colhead{V1515 Cyg} & \colhead{FU Ori}}
 \startdata
 & & & 10$^{-18}$ &10$^{-18}$ & 10$^{-18}$ &10$^{-18}$ &10$^{-18}$ &10$^{-18}$ \\
 Units: & GHz & $\mu$m & W m$^{-2}$ & W m$^{-2}$ & W m$^{-2}$ & W m$^{-2}$ & W m$^{-2}$ & W m$^{-2}$ \\
 & & & & & & & & & \\
CO \jj{4}{3} & 461.0 & 650.25 & 33.99 & 31.54 & 9.44 & 6.73 & 7.58 & $<$ 1.72 \\
$[$C I] $^3$P$_1 \rightarrow ^3$P$_0$ & 491.4 & 610.9 & 20.13 & 11.15 & 4.81 & 6.46 & 4.56 & 3.75 \\
$^{13}$CO  \jj{5}{4} & 551.0 & 544.1 & 4.04 & 5.72 &  $<$ 1.65 &  $<$ 1.65 &  $<$ 1.65 &  $<$ 1.65 \\
CO  \jj{5}{4} &  576.3 & 520.23 & 50.84 & 58.27 & 16.30 & 4.50 & 5.40 &  $<$ 3.72$^1$ \\
$^{13}$CO  \jj{6}{5} & 661.4 & 453.5 & 5.14 & 5.46 &  $<$ 3.06 &  $<$ 3.06 &  $<$ 3.06 &  $<$ 3.06 \\
CO  \jj{6}{5} & 691.5 & 433.56 & 89.72 & 111.02 & 21.62 & 8.25 &  $<$ 2.85 &  $<$ 2.85 \\
$^{13}$CO  \jj{7}{6} & 771.3 & 388.74 & 4.78 & 8.54 &  $<$ 3.15 &  $<$ 3.15 &  $<$ 3.15 &  $<$ 3.15 \\
CO  \jj{7}{6} & 806.7 & 371.65 & 125.88 & 157.59 & 23.98 & 3.80 &  $<$ 2.55 &  $<$ 2.55 \\
$[$C I] $^3$P$_2 \rightarrow ^3$P$_1$ & 809.9 & 370.42 & 43.07 & 31.56 & 10.39 & 8.74 & 8.98 & 4.91 \\
$^{13}$CO  \jj{8}{7} & 881.5 & 340.18 &  $<$ 3.15 & 4.42 &  $<$ 2.61 &  $<$ 2.61 &  $<$ 2.61 &  $<$ 2.61 \\
CO  \jj{8}{7} & 921.8 & 325.23 & 138.10 & 162.27 & 28.94 &  $<$ 9.06 &  $<$ 9.06 &  $<$ 9.06  \\
CO \jj{9}{8} & 1036.9 & 289.12 & 141.13 & 101.74 &  $<$ 5.13 &  $<$ 5.13 &  $<$ 5.13 &  $<$ 5.13\\
CO \jj{10}{9} & 1152.0 & 260.24 & 158.23 & 76.49 & 32.30 &  $<$ 18.18 &  $<$ 18.18  &  $<$ 18.18 \\
CO  \jj{11}{10} & 1267.0 & 236.61 & 129.02 & 53.28 & 28.30 &  $<$ 20.19 &  $<$ 20.19 &  $<$ 20.19 \\
CO  \jj{12}{11} & 1382.0 & 216.93 & 106.28 & 13.17 &  $<$ 15.61 &  $<$ 15.61 &  $<$ 15.61 &  $<$ 15.61 \\
$[$N II] $^3$P$_1 \rightarrow ^3$P$_0$ & 1461.2 & 205.18 & 132.77 &  $<$ 39.55 & 119.38 & 69.09 & 289.03 &  $<$ 39.55 \\
CO  \jj{13}{12} & 1496.9 & 200.27 & 88.12 &  $<$ 40.94 &  $<$ 40.94 &  $<$ 40.94 &  $<$ 40.94 &  $<$ 40.94 \\
\hline
CO  \jj{14}{13} & 1611.7 &186.01 & 81.56 &  $<$ 28.74 & 37.50 &  $<$ 28.74 &  $<$ 28.74 &  $<$ 28.74\\
o-H$_2$O 3$_{03} \rightarrow 2_{12}$ & 1716.7 & 174.63  &   $<$ 15.75  &  $<$ 15.75  & 42.04 &  $<$ 15.75 &  $<$ 15.75 &  $<$ 15.75 \\
CO  \jj{15}{14} & 1726.6 & 173.63 & 82.01 & 50.79 & 64.06 &  $<$ 15.75 &  $<$ 15.75 &  $<$ 15.75 \\
CO  \jj{16}{15} & 1841.4 & 162.81 & 54.03 &  $<$ 27.66 & 46.75 &  $<$ 27.66 &  $<$ 27.66 &  $<$ 27.66 \\
$[$C II] $^2$P$_{3/2} \rightarrow ^2$P$_{1/2}$ & 1900.6 & 157.74 &  $<$ 22.44 & 137.33 &   $<$ 22.44 &  25.57 & 471.85 &  $<$ 22.44 \\
CO  \jj{17}{16} & 1956.0 & 153.27 & 48.31 &  $<$ 41.64 & 53.02 & $<$ 41.64 &  $<$ 41.64 &  $<$ 41.64 \\
$[$O I] $^3$P$_0 \rightarrow ^3$P$_1$ & 2060.0 & 145.53 & 39.62 & 38.13 &  $<$ 24.22 &  $<$ 24.22 &  $<$ 24.22 &  $<$ 24.22 \\
CO  \jj{18}{17} & 2070.7 & 144.78 & 29.20 &  $<$ 24.22 & 62.14 &  $<$ 20.67 &  $<$ 20.67 &  $<$ 20.67 \\
CO  \jj{19}{18} & 2185.1 & 137.20 & 41.15 &  $<$ 15.06 & 41.24 &  $<$ 15.06 &  $<$ 15.06 &  $<$ 15.06 \\
CO  \jj{20}{19} & 2299.6 & 130.37 & 26.11 &  $<$ 11.85 & 38.01 &  $<$ 11.85 &  $<$ 11.85 &  $<$ 11.85 \\
CO  \jj{21}{20} & 2414.0 & 124.19 &  $<$ 33.72  &  $<$ 33.72 & 60.58 &  $<$ 33.72 &  $<$ 33.72 &  $<$ 33.72 \\
$[$N II] $^3$P$_2 \rightarrow ^3$P$_1$ & 2459.3 & 121.90 & $<$ 17.67 & $<$ 17.67 & $<$ 17.67 & $<$ 17.67 & 89.11 & $<$ 17.67 \\
CO  \jj{22}{21} & 2528.2 & 118.58 &  $<$ 16.92 &  $<$ 16.92 & 42.94 &  $<$ 16.92 &  $<$ 16.92 &  $<$ 16.92 \\
CO  \jj{23}{22} & 2642.3 & 113.46 &  $<$ 42.44 &  $<$ 42.44 & 96.10 &  $<$ 42.44 &  $<$ 42.44 &  $<$ 42.44 \\
OH 3/2 $\rightarrow$ 3/2 ($7/2+ \rightarrow 5/2-$) & 3551.6 & 84.41 &$<$ 39.11 & $<$ 39.11 & 70.95 & $<$ 39.11 & $<$ 39.11 & $<$ 39.11 \\
$[$O I] $^3$P$_1 \rightarrow ^3$P$_2$ & 4745.1 & 63.18 & 837.08 & 417.61 & 355.54 & 38.76 & 31.51 & 183.45 \\
\enddata
\tablecomments{Spectral lines detected with SPIRE and PACS.  Where no 3$\sigma$ detection is found, we provide 3$\sigma$ upper limits from local continuum.  All line fluxes for HBC 722 are attributed primarily to contamination, with the exception of [O I] 63 $\mu$m; we update the line fluxes from \citet{green11b}.  Line fluxes for V1735 Cyg are partially contaminated by V1735 Cyg SM1. CO \jj{7}{6} and [C I] 371 $\mu$m are partially blended. The CO \jj{23}{22} may be blended with the H$_2$O 113.53 $\mu$m line.\\
$^1$Although not detected with SPIRE, we detect this line with HIFI, at an integrated line flux of  2.95 $\times$ 10$^{-18}$ W m$^{-2}$.
}
\end{deluxetable}
\end{center}

\clearpage

\begin{figure}
\begin{center}
\includegraphics[scale=0.7,angle=0]{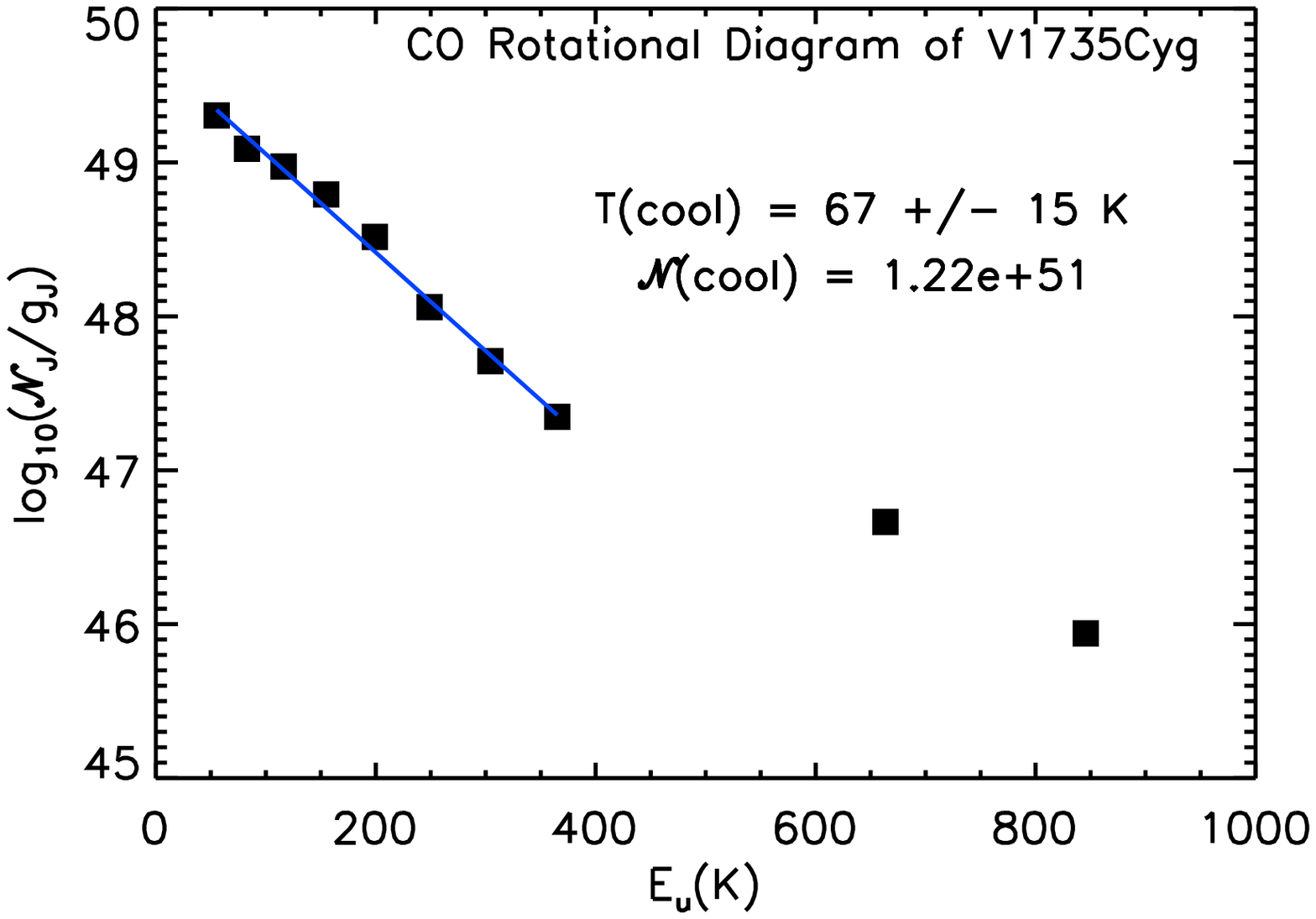}
\includegraphics[scale=0.7,angle=0]{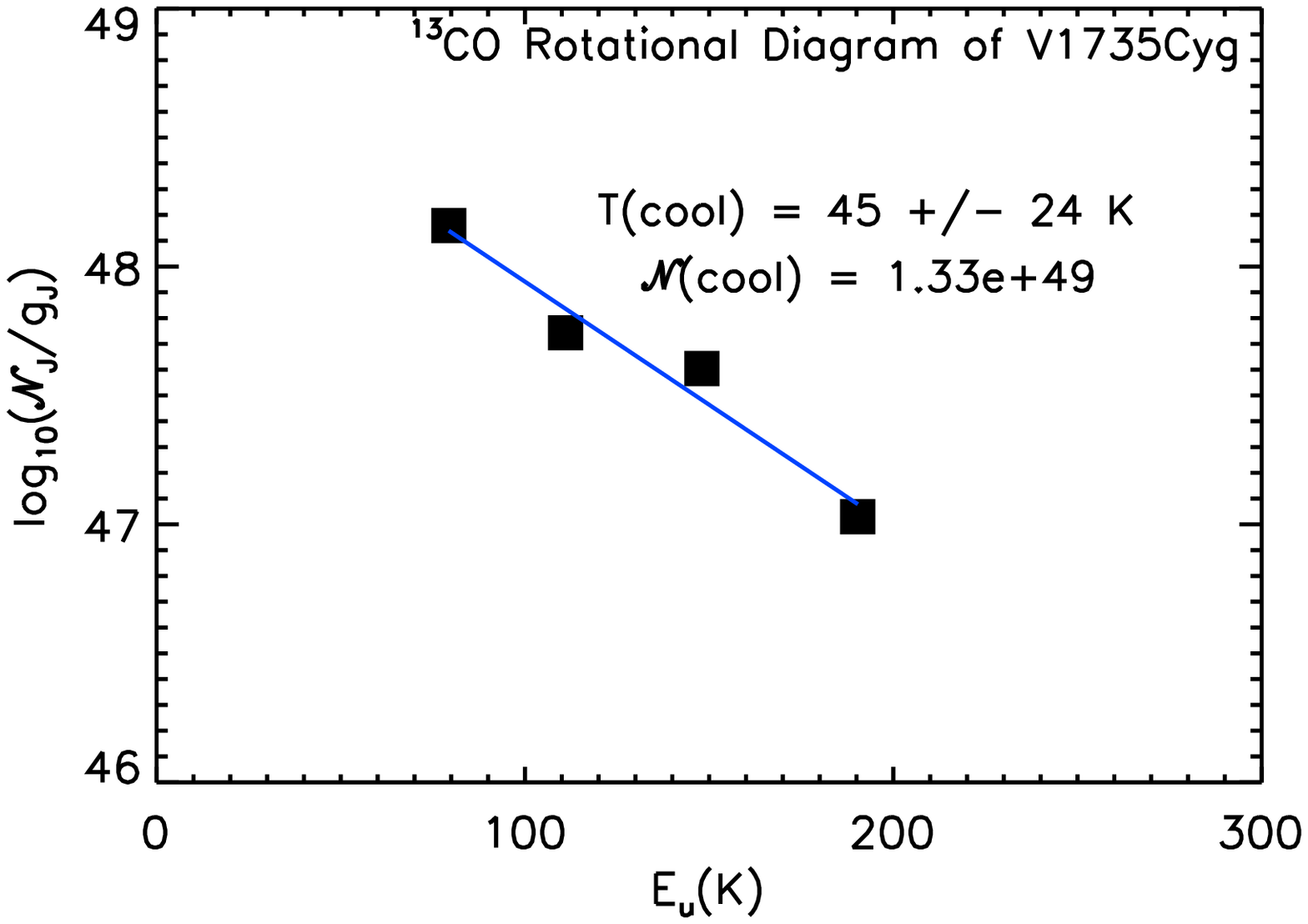}
\caption{CO (top) and $^{13}$CO (bottom) rotational diagram for the contaminated V1735 Cyg SPIRE spectrum.}
\label{rot13}
\end{center}
\end{figure}

\clearpage

\begin{figure}
\includegraphics[scale=0.43]{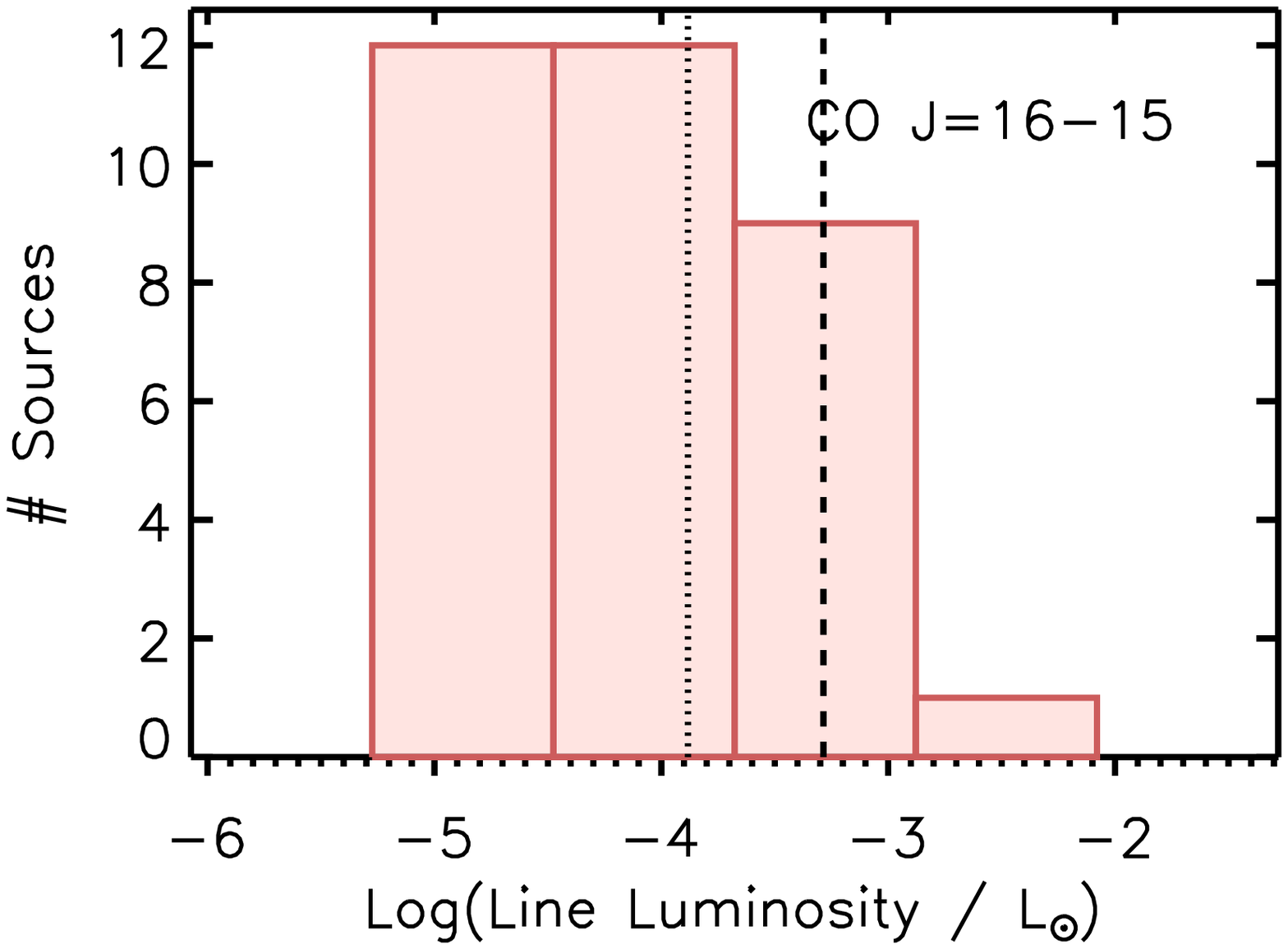}
\includegraphics[scale=0.43]{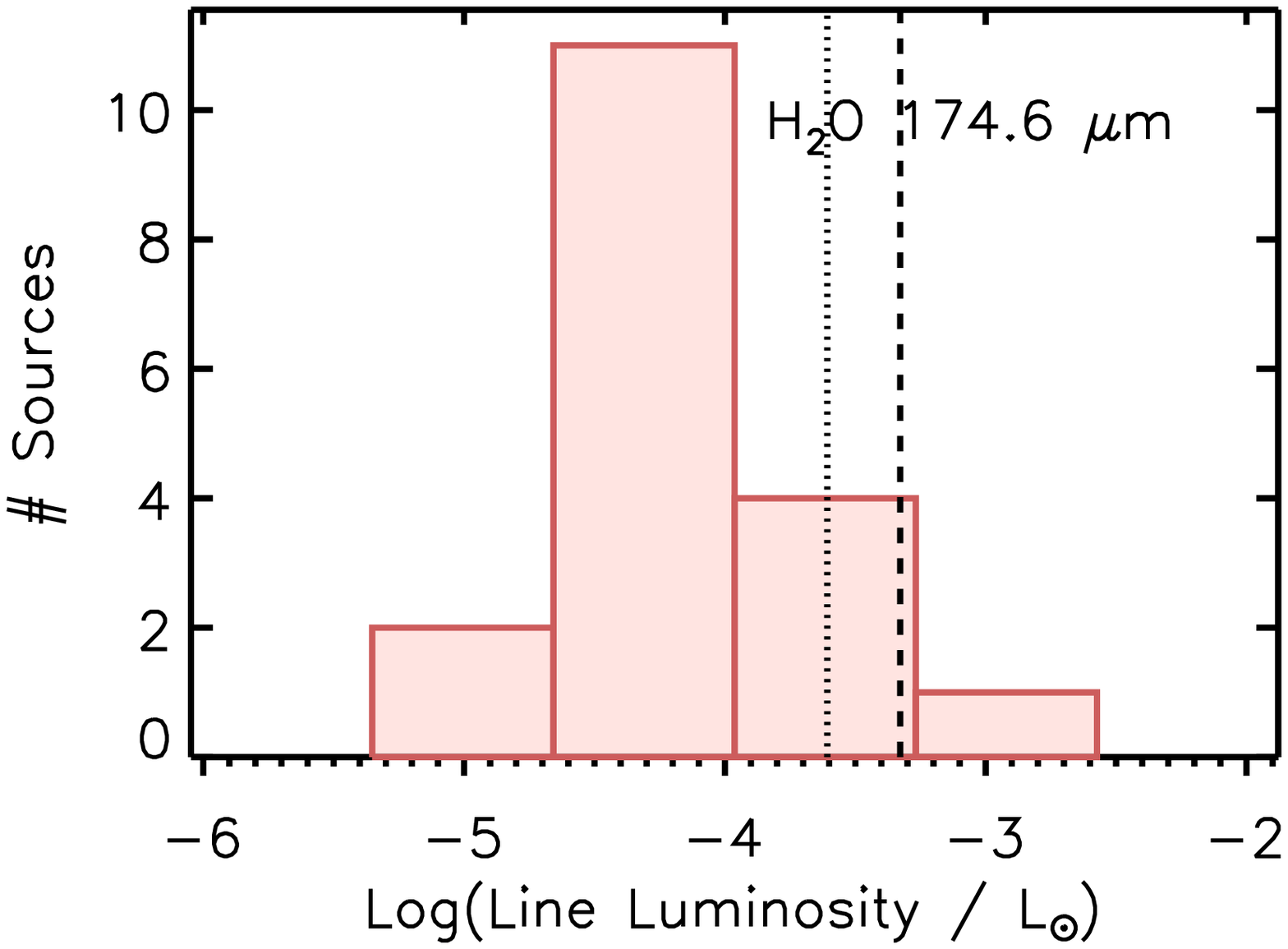}
\caption{{\bf Left:} 
Histogram of luminosities of the CO \jj{16}{15} line for the DIGIT sample from \citep{green13b}.  The vertical dashed line indicates the line strength for V1057 Cyg.  The vertical dotted line marks the 3$\sigma$ upper limit for sources with non-detections in our sample.
 {\bf Right:} The same analysis for the H$_2$O 174.63 $\mu$m line, detected in V1057 Cyg.}
\label{gas}
\end{figure}

\end{document}